\documentclass[journal,comsoc,finalcls,twocolumn,10pt,twoside]{IEEEtran}
\usepackage[T1]{fontenc}

\usepackage{cite}

\ifCLASSINFOpdf
   \usepackage[pdftex]{graphicx}
\else
   \usepackage[dvips]{graphicx}
\fi
\usepackage{amsmath}
\usepackage[cmintegrals]{newtxmath}
\usepackage{array}

\ifCLASSOPTIONcompsoc
  \usepackage[caption=false,font=normalsize,labelfont=sf,textfont=sf]{subfig}
\else
  \usepackage[caption=false,font=footnotesize,labelformat=simple]{subfig}
\fi
 
\usepackage{fixltx2e}
\usepackage{dblfloatfix}

\usepackage{multirow}
\usepackage{diagbox}
\usepackage{epstopdf}
\usepackage{xtab,afterpage}
\usepackage{comment}
\usepackage{color}
\definecolor{light-gray}{gray}{0.5}
\usepackage{pdflscape}
\usepackage{lscape}
\usepackage{rotating}
\usepackage{array}
\usepackage{multirow}
\usepackage{hhline}
\usepackage[table]{xcolor}

\newcolumntype{L}[1]{>{\raggedright\let\newline\\\arraybackslash\hspace{0pt}}m{#1}}
\newcolumntype{C}[1]{>{\centering\let\newline\\\arraybackslash\hspace{0pt}}m{#1}}
\newcolumntype{R}[1]{>{\raggedleft\let\newline\\\arraybackslash\hspace{0pt}}m{#1}}

\usepackage{url}

\newcounter{mysubtable}
\newcommand\modcounter{%
  \refstepcounter{mysubtable}%
  \renewcommand{\thetable}{\Roman{table}-\Alph{mysubtable}}%
}

\begin{document}
%
\title{Survey of Spectrum Sharing for\\ Inter-Technology Coexistence}
%
%
%

\author{Andra~M.~Voicu, Ljiljana Simi\'c and Marina Petrova
\thanks{A.~M.~Voicu and L.~Simi\'c are with the Institute for Networked Systems, RWTH Aachen University (e-mail: avo@inets.rwth-aachen.de; lsi@inets.rwth-aachen.de). M.~Petrova is with the School of Electrical Engineering and Computer Science, KTH Royal Institute of Technology (e-mail: petrovam@kth.se).}}

\maketitle

\begin{abstract}
Increasing capacity demands in emerging wireless technologies are expected to be met by network densification and spectrum bands open to multiple technologies. These will, in turn, increase the level of interference and also result in more complex inter-technology interactions, which will need to be managed through spectrum sharing mechanisms. 
Consequently, novel spectrum sharing mechanisms should be designed to allow spectrum access for multiple technologies, while efficiently utilizing the spectrum resources overall. 
Importantly, it is not trivial to design such efficient mechanisms, not only due to technical aspects, but also due to regulatory and business model constraints. 
In this survey we address spectrum sharing mechanisms for wireless inter-technology coexistence by means of a \emph{technology circle} that incorporates in a unified, system-level view the technical and non-technical aspects. 
We thus systematically explore the spectrum sharing design space consisting of parameters at different layers.
Using this framework, we present a literature review on inter-technology coexistence with a focus on wireless technologies with equal spectrum access rights, i.e. (i)~primary/primary, (ii)~secondary/secondary, and (iii)~technologies operating in a spectrum commons. 
Moreover, we reflect on our literature review to identify possible spectrum sharing design solutions and performance evaluation approaches useful for future coexistence cases.
Finally, we discuss spectrum sharing design challenges and suggest future research directions.

\end{abstract}


%
\IEEEpeerreviewmaketitle

\section{Introduction}
%
%
%
%

In order to cope with the growing wireless traffic volume demands, significant changes in wireless technology deployments are expected in the near future. 
Two important trends can be distinguished: (i)~the already ubiquitous wireless networks are predicted to undergo extreme densification~\cite{Nokia2016}, and (ii)~an increasing number of spectrum bands are being targeted by multiple wireless technologies, e.g. LTE was recently proposed to operate in the 5 GHz unlicensed band~\cite{3GPP2016, 3GPP2016a, Forum2015}, the 3.5~GHz Citizens Broadband Radio Service (CBRS) band in the U.S. is under discussion for being open to more technologies~\cite{FCC15-472015}.
These trends will, in turn, increase the level of interference and the complexity of wireless inter-technology interactions, which have to be managed through efficient spectrum sharing mechanisms. 

Traditionally, wireless technologies have operated in either licensed, or unlicensed bands. Licensed bands are granted by spectrum regulators to single entities, e.g. cellular operators, which then individually deploy and manage their networks in dedicated spectrum bands. Consequently, inter-technology coexistence has not been an issue in these bands. 
By contrast, in the unlicensed bands any technology and device has equal rights to access the spectrum, as long as basic regulatory restrictions are met, e.g. maximum transmit power. As such, mutual interference among different technologies is inherent to the unlicensed bands and has typically been managed by rather simple distributed spectrum sharing schemes, e.g. between \mbox{Wi-Fi} and Bluetooth.
 
Recently, due to the growing need for higher network capacity, several regulatory and technical changes have been introduced for wireless technologies.  
Firstly, spectrum regulators have opened an increasing number of bands to \emph{multiple} technologies, and have authorised novel access right frameworks, other than pure exclusive use or equal rights, i.e. different variants of primary/secondary access. 
Some examples of bands where such frameworks exist are: TV white space (TVWS)~\cite{Ofcom2015, FCC2010}, the recently proposed 3.5~GHz Citizens Broadband Radio Service (CBRS) band in the U.S.~\cite{FCC15-472015}, and the 2.3--2.4~GHz band in Europe, where recent coexistence trials under Licensed Shared Access (LSA) have been conducted~\cite{Guiducci2017}. 

New challenging coexistence cases are also expected in the unlicensed bands, where LTE has recently been proposed and standardized to operate in the unlicensed 5~GHz band~\cite{3GPP2016, 3GPP2016a, Forum2015}, where it must coexist with \mbox{Wi-Fi}.
As both LTE and \mbox{Wi-Fi} are broadband technologies designed to carry high traffic loads, this is different to prior inter-technology coexistence cases in unlicensed bands (\emph{cf.} \mbox{Wi-Fi}/Bluetooth coexistence).
Furthermore, a \emph{second} technology, i.e. NB-IoT (Narrowband Internet of Things)~\cite{3GPP2016a}, has been recently designed to coexist with LTE in the same licensed cellular bands where LTE used to operate exclusively.    

As demonstrated by these examples, a significant number of \emph{heterogeneous} wireless devices, in terms of technologies and traffic requirements, is expected to be deployed in shared spectrum bands. It follows that new inter-technology interactions are currently emerging, and they are too complex to be efficiently managed by traditional spectrum sharing mechanisms designed for either licensed cellular bands, or unlicensed bands with low to moderate traffic volumes. 
It is thus crucial to design novel inter-technology spectrum sharing mechanisms that: (i)~allow multiple devices and technologies to access the spectrum; and (ii)~facilitate an efficient overall use of the spectrum, while fulfilling the requirements of each device/technology.    

\begin{table*}[t!]
\begin{center}
\caption{Classification of inter-technology surveys in the literature. This survey addresses the categories shaded in green.
\label{table_0} }
\begin{tabular}{|l|l|l|l|c|}
\cline{5-5}
 \multicolumn{4}{c|}{} & \textsc{Prior Surveys}\\
\hline 
\multirow{2}{*}{\rotatebox[origin=c]{90}{\parbox{3.5cm}{\centering\textsc{Inter-technology \\Interactions}}}} 
 
  & \multirow{6}{*}{\parbox{3.5cm}{\textbf{Inter-technology coexistence} \\\textbf{(in shared spectrum bands)}}} 
  & \multirow{3}{*}{\parbox{1.5cm}{\textbf{hierarchical} \\\textbf{regulatory} \\\textbf{framework}}} 
  & \parbox{3cm}{\textbf{different access rights} \\ (i.e. \emph{primary/secondary})} & 
                   \begin{tabular}{p{6.5cm}}  dynamic spectrum access (DSA), e.g.~\cite{Paisana2014, Tehrani2016, Akyildiz2006, Akyildiz2008, Zhao2007, Wang2008, Yucek2009, Gavrilovska2014, Ren2012} 
                   \end{tabular}
  \\
 		\hhline{|~|~|~|--}
 			 & & & {\cellcolor{green!25}}\parbox{3cm}{\textbf{equal access rights} \\(i.e. \emph{primary/primary}, \\\emph{secondary/secondary})} &
 			       \begin{tabular}{p{6.5cm}} converged heterogeneous mobile networks with focus on M2M \emph{integration}~\cite{Jo2014} 
                   \end{tabular}
 \\ 
 \hhline{|~|~|---}   
 & &  \multicolumn{2}{l|}{\cellcolor{green!25}\parbox{4.5cm}{\textbf{flat regulatory framework with equal access rights} (i.e. \emph{spectrum commons})}} 
      &  \begin{tabular}{p{6.5cm}}  converged heterogeneous mobile networks with focus on M2M \emph{integration}~\cite{Jo2014};\\
                             early literature on \mbox{Wi-Fi}/LTE coexistence in the unlicensed bands~\cite{Ho2017}
         \end{tabular}             
 \\
\hhline{|~|====}
 & \multicolumn{3}{l|}{\textbf{Integration of technologies operating in different spectrum bands}} 
 &  \begin{tabular}{p{6.5cm}} mobile cellular and vehicular communications~\cite{Zheng2015};\\
                            interworking architectures for wireless technologies~\cite{Atayero2012} \end{tabular}
 \\
\hline
\end{tabular}
\end{center}
\end{table*}

Furthermore, the design of inter-technology spectrum sharing mechanisms does not only depend on purely technical aspects, but also on regulatory constraints, business models, and social practices. For instance, the regulators impose limits on the spectrum access rights for different devices/networks and in some cases even on the spectrum sharing mechanisms, e.g. listen-before-talk (LBT) being mandatory for the 5~GHz unlicensed band in Europe~\cite{ETSI2015}.
Business models and social practices affect the design of spectrum sharing mechanisms, as the most efficient mechanisms from a technical perspective may not be practically feasible due to e.g. lack of agreements among the involved network managers/device owners. 

Two important questions arise, pertinent to designing future wireless technologies: \textbf{(i)~how to design in a systematic manner efficient spectrum sharing mechanisms especially for inter-technology coexistence, by taking into account technical and non-technical parameters}; and \textbf{(ii)~how to evaluate their coexistence performance, with respect to a given technology itself, and its impact on other coexisting technologies}. 

In this survey we explore the first question by means of a multi-layer \emph{technology circle} that incorporates in a system-level view all relevant technical and non-technical aspects of a wireless technology. 
The technology circle, as proposed in~\cite{Mahonen2012} and illustrated in Fig.~\ref{fig_techCircle}, includes the seven layers of the OSI stack and introduces the regulatory framework at Layer~0, and business models and social practices at Layer~8. The technology circle thus represents a unified design space for spectrum sharing, consisting of parameters at different layers. 
Next, we identify the layers at which spectrum sharing is implemented, and the layers that impose constraints. We then discuss the individual effect of each layer on spectrum sharing and the feasibility of different design parameter combinations at different layers. 
To this end we present a classification of the literature on inter-technology coexistence, based on individual spectrum sharing design parameters at different layers.
We focus on coexistence under a regulatory framework with \emph{equal} spectrum access rights and especially on a spectrum commons. Importantly, these are the most challenging coexistence cases, as the limitations imposed by regulators tend to be more relaxed, but multiple, diverse technologies may share the same band, so that the design of spectrum sharing mechanism must take into account interactions with a wide range of other technologies. 

We address the second posed question by discussing the choice of performance evaluation methods and metrics in the literature on inter-technology coexistence with equal spectrum access rights. 
Finally, we reflect on the reviewed literature to determine suitable design approaches for future wireless technologies and we identify challenges and possible research directions.

\subsection{Related Surveys in the Literature} 
Earlier surveys addressing inter-technology spectrum sharing~\cite{Paisana2014, Tehrani2016, Akyildiz2006, Akyildiz2008, Zhao2007, Wang2008, Yucek2009, Gavrilovska2014, Ren2012, Zheng2015, Jo2014, Atayero2012, Ho2017} focused on only specific coexistence cases and did not present a comprehensive view of inter-technology wireless coexistence in general, as summarized in Table~\ref{table_0}. 
These surveys considered: spectrum that is shared in a primary/secondary manner, i.e. through dynamic spectrum access (DSA) techniques, e.g.~\cite{Paisana2014, Tehrani2016, Akyildiz2006, Akyildiz2008, Zhao2007, Wang2008, Yucek2009, Gavrilovska2014, Ren2012}; coexistence solutions in the form of integrated, coordinated technologies, e.g. converged heterogeneous mobile networks operating in shared spectrum bands~\cite{Jo2014}; or early literature on  \mbox{Wi-Fi}/LTE coexistence in the unlicensed bands~\cite{Ho2017}. 
Other surveys addressed inter-technology interactions for integrated technologies operating in different spectrum bands, e.g. mobile cellular and vehicular communications~\cite{Zheng2015}, and interworking architectures for wireless technologies~\cite{Atayero2012}. 
Therefore, the existing literature lacks a general and comprehensive view of inter-technology coexistence, which is especially important for the most challenging coexistence case, i.e. spectrum sharing when multiple technologies have the same rights to access the spectrum. 
\textbf{Our survey, instead, presents inter-technology coexistence from a unified, system-level perspective}, which is essential for answering the two posed research questions on systematically designing efficient spectrum sharing mechanisms and evaluating their coexistence performance. 
Moreover, we focus on technologies with equal spectrum access rights and especially on coexistence in a spectrum commons, which we expect to be of high practical relevance in the near future.

\begin{table*}[t!]
\caption{Interference classification and terminology based on the relative shift between the spectrum portions where the interferer transmitter and victim receiver operate}
\label{table_1}
\centering
\begin{tabular}{|p{3.5cm}|p{3.3cm}|c|}
\hline
	\diagbox[width=3.9cm, height=1cm]{\textbf{Terminology scope}}{\textbf{Spectrum used}} 
	& \centering Same frequency for interferer Tx and victim Rx 
	& Different frequencies for interferer Tx and victim Rx\\
\hline    
	Generic & \parbox{3cm}{\textbf{\emph{in-band interference}}~\cite{AgilentTechnologies2011}} 
	& \begin{tabular}{p{9cm}} $\bullet$ \textbf{\emph{out-of-band emissions}}~\cite{ITU-R2015a} (also \textbf{\emph{out-of-band interference}}~\cite{FCCTechnologicalAdvisoryCouncil2015}) -- due to Tx\\
	                     $\bullet$  \textbf{\emph{spurious emissions}}~\cite{ITU-R2015a} (sometimes included in out-of-band interference~\cite{AgilentTechnologies2011}) -- due to Tx  \\
	                     $\bullet$  \textbf{\emph{adjacent band interference}}~\cite{FCCTechnologicalAdvisoryCouncil2015} -- due to Rx
	   \end{tabular}\\
\hline  
    Technology-oriented & \parbox{3cm}{\textbf{\emph{co-channel}} \\ \textbf{\emph{interference}} \cite{FCCTechnologicalAdvisoryCouncil2015, AgilentTechnologies2011}}
    & \begin{tabular}{p{9cm}} \textbf{\emph{adjacent channel interference}}~\cite{3GPP2015, AgilentTechnologies2011}:\\
	                     $\bullet$  \textbf{\emph{adjacent channel leakage}}~\cite{3GPP2015} -- due to Tx  \\
	                     $\bullet$  \textbf{\emph{adjacent channel selectivity}}~\cite{3GPP2015}/ \textbf{\emph{rejection}}~\cite{IEEE2016} -- due to Rx
	   \end{tabular}\\ 
\hline 
\end{tabular}
\end{table*}

\subsection{Survey Structure}
The rest of this survey is structured as follows. 
In Section~\ref{interf_tax} we define the inter-technology coexistence problem in terms of interference and we present an interference taxonomy.
In Section~\ref{tech_circle} we present the technology circle and we discuss the impact of different layers on spectrum sharing mechanisms in general.
Section~\ref{litHierFr} presents our literature review of inter-technology coexistence within a hierarchical regulatory framework with a focus on technologies with the same spectrum access rights, i.e. primary/primary and secondary/secondary.
Section~\ref{litSpecComm} presents our literature review of inter-technology coexistence in a spectrum commons.
In Section~\ref{sec_discussion} we discuss the main findings of this survey and we identify challenges and potential future research directions.
Section~\ref{sec_conclusions} concludes the survey.

\section{Interference Taxonomy \& Problem Statement}
\label{interf_tax}

In this section we present an interference taxonomy and we define our problem statement for wireless inter-technology coexistence in terms of interference types.
We also present the spectrum management terminology used in this survey.

\textbf{\emph{Interference}} consists of perturbing signals that arrive at a receiver at the same time as the signal of interest. Consequently, the signal-to-interference-and-noise ratio (SINR) is decreased at the victim receiver, such that decoding the useful signal becomes more difficult.
Spectrum regulators consider interference when establishing operational bounds for devices, or technologies. 
From an engineering perspective, interference is important for determining the achievable data rates, depending on the capabilities of the radio hardware (e.g. filter characteristics, receiver noise figure).
Increased interference thus decreases the link capacity, which in turn affects the overall network capacity. 
We note that although interference fundamentally occurs at the Physical (PHY) layer, interference mitigation techniques are also implemented at other layers, especially at the MAC. The final link capacity is thus affected by such techniques, as well.

\begin{figure}[!t]
\centering
\subfloat[]{\includegraphics[width=1\columnwidth]{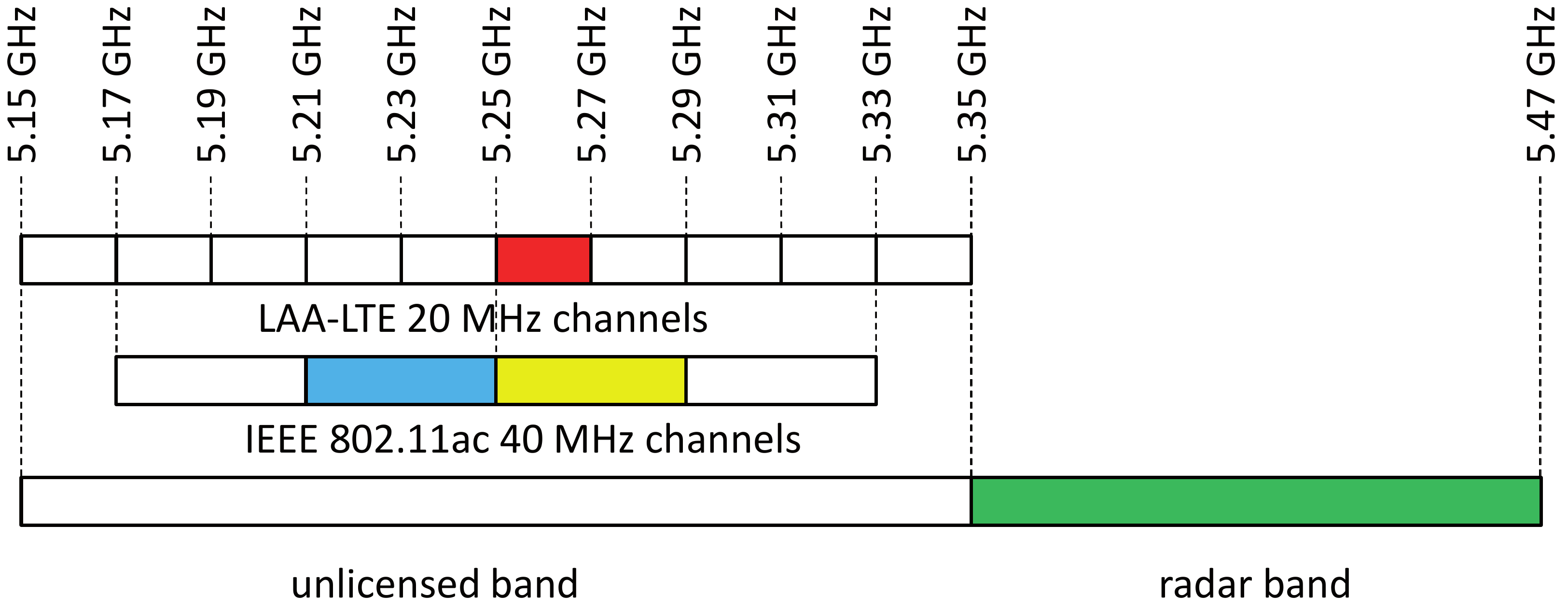} \label{fig_1a}}
\\
\subfloat[]{\includegraphics[width=1\columnwidth]{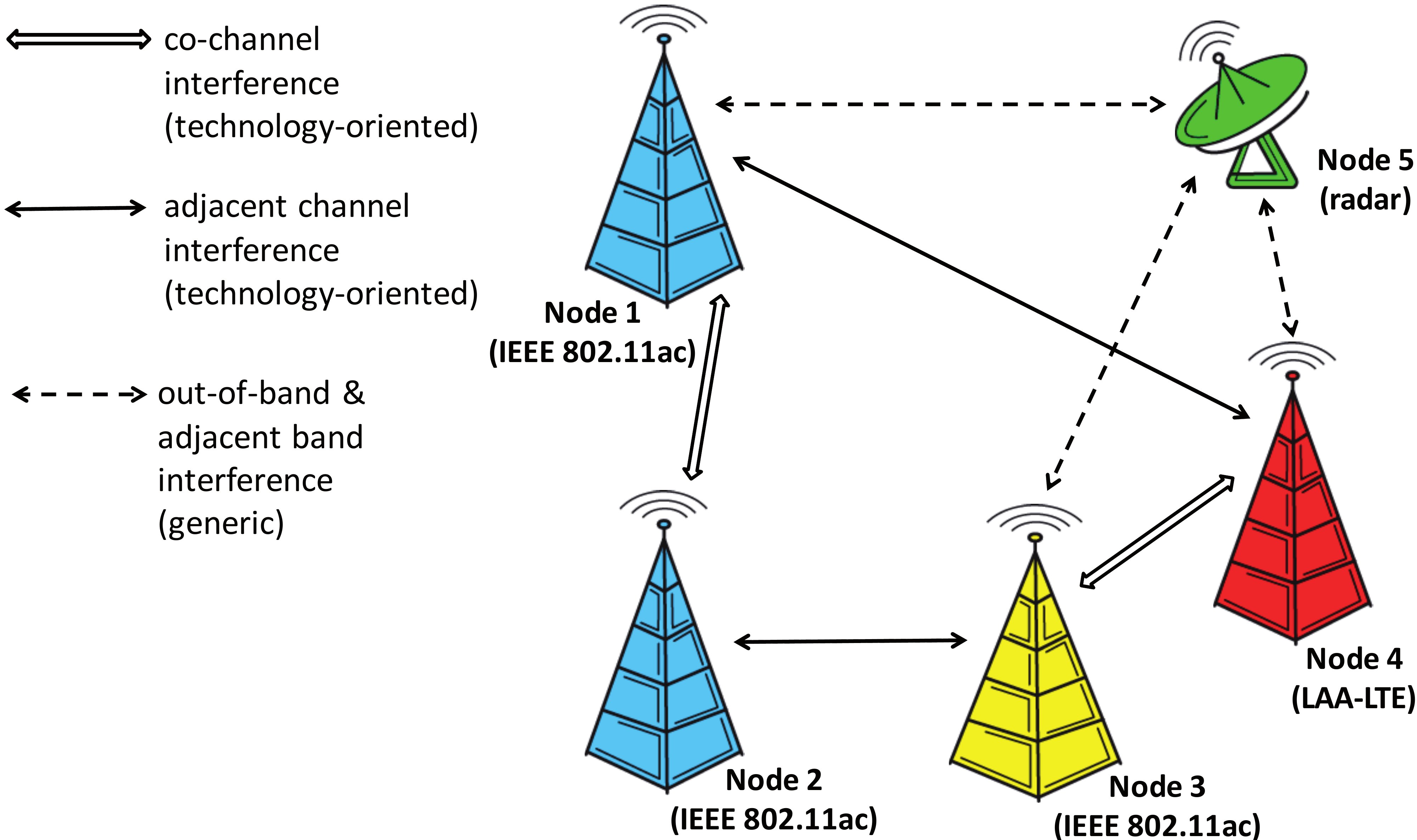} \label{fig_1b}}
\caption{Example of (a) two spectrum bands, where one is used as unlicensed by IEEE 802.11ac \mbox{Wi-Fi}~\cite{IEEE2016} and LAA-LTE~\cite{3GPP2016a}, and the other is allocated to radar services in Europe~\cite{ECC2016}; and (b) types of mutual interference occurring between different nodes operating in these bands: Nodes 1, 2, and 3 are IEEE 802.11ac \mbox{Wi-Fi} nodes operating on the 40 MHz channels in the same respective colour (blue and yellow), Node 4 is an LAA-LTE node operating on a 20 MHz channel (red), and Node 5 is a radar node operating on a channel in the radar band (green).}
\label{fig_1}
\end{figure}

Interference can be classified according to the imperfections of the transmitter and receiver, and the relative portions of the spectrum where the interfering transmitter and the victim receiver operate. 
There are a few terms widely used in this context, but their meaning is sometimes loosely defined, as summarized in Table~\ref{table_1}. We identify two types of terminology used to refer to interference, based on the scope: (i)~\textbf{generic} terms typically used in the regulatory domain, which is concerned with interference from another frequency band and limits imposed on the transmitters; and (ii)~\textbf{technology-oriented} terms that refer mostly to interference among devices within given technologies with further channel partitioning of the same spectrum band, where each device is allowed to access any of these channels. Fig.~\ref{fig_1} shows examples of different types of interference among IEEE~802.11ac~\mbox{Wi-Fi}, Licensed Assisted Access (LAA) LTE, and radars operating in the 5 GHz band.

Interference from a \textbf{generic} perspective can be \textbf{\emph{in-band}}, if both interfering transmitter and victim receiver operate in the same spectrum band~\cite{AgilentTechnologies2011}. 
In case the transmitter and receiver do not operate in the same band, the interference can be in the form of: \textbf{\emph{out-of-band emissions/interference}}, \textbf{\emph{spurious emissions}}, or \textbf{\emph{adjacent band interference}}~\cite{ITU-R2015a, FCCTechnologicalAdvisoryCouncil2015, AgilentTechnologies2011}. 
Out-of-band and spurious emissions refer to interference caused by the imperfection in the filters of the transmitter. 
We note that spectrum regulators are typically concerned with these kinds of interference, since regulation traditionally imposes operational limits on the transmitters and not the receivers (see e.g.~\cite{Vries2013}). 
Adjacent band interference was used in~\cite{FCCTechnologicalAdvisoryCouncil2015} to refer to the interference experienced by the receiver due to its own inability to perfectly filter out the received power in a band adjacent to the one it operates on.

\begin{figure}[!t]
\centering
\includegraphics[width=1\columnwidth]{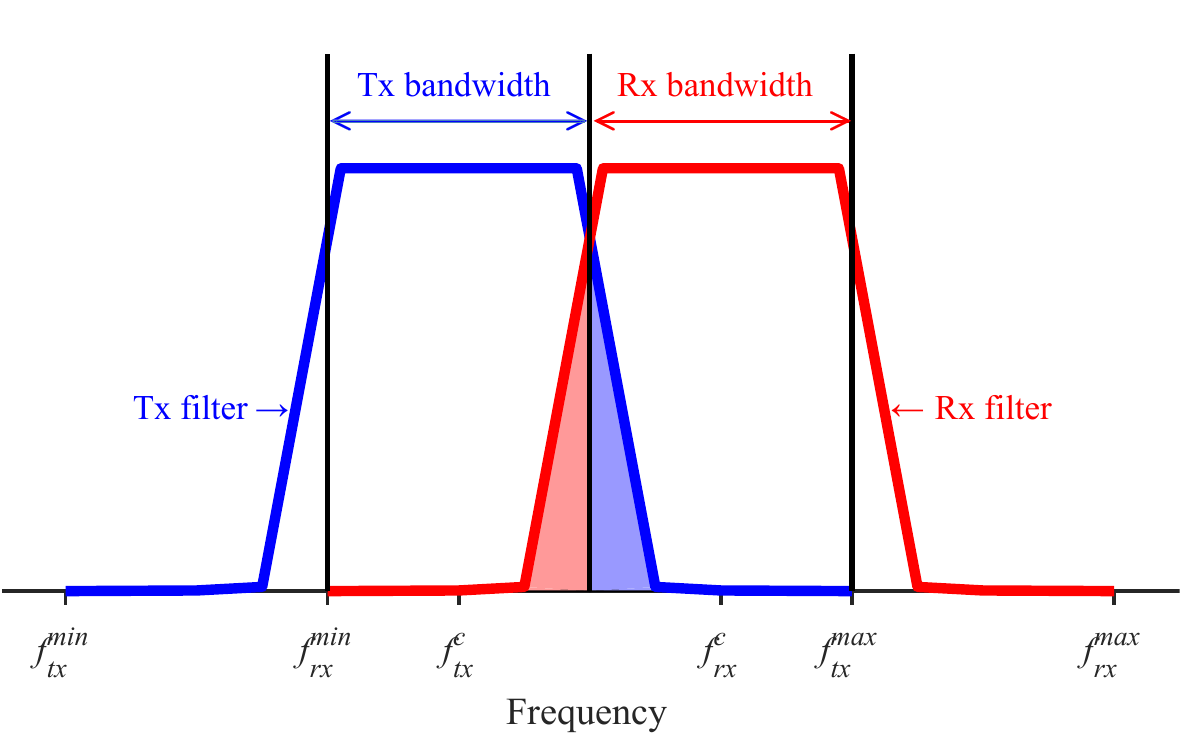}
\caption{Illustration of ACI as determined by the filters of the interferer transmitter (blue)
and victim receiver (red). The ACI caused by the power leaked by the transmitter
is shown as the area coloured in light blue. The ACI due to imperfect receiver filtering is
shown as the area coloured in light red.}
\label{fig_ACI_survey}
\end{figure}

From a \textbf{technology-oriented} perspective, where several channels are defined within a given band, the interference is defined with respect to the channel, not the band. We thus identify \textbf{\emph{co-channel interference}} for operation over the same channel~\cite{AgilentTechnologies2011, FCCTechnologicalAdvisoryCouncil2015} and \textbf{\emph{adjacent channel interference}} (ACI) for operation on adjacent channels~\cite{AgilentTechnologies2011, 3GPP2015}. 
We note that co- and adjacent channel interference can occur both among devices of the same technology, and among devices of different technologies (\emph{cf.} Fig.~\ref{fig_1}). 
Furthermore, it is important to distinguish between ACI caused by the imperfections of the interferer transmitter and imperfections of the victim receiver -- as shown in Fig.~\ref{fig_ACI_survey} -- as the performance of a technology in terms of link-level data rates depends on both. 
For instance, 3GPP~\cite{3GPP2015} distinguishes, in case of LTE, between \textbf{\emph{adjacent channel leakage}} (at the transmitter) and \textbf{\emph{adjacent channel selectivity}} (at the receiver). 
The IEEE 802.11 standard~\cite{IEEE2016} defines a similar concept to the receiver selectivity, i.e. \textbf{\emph{adjacent channel rejection}}, and specifies the transmitter spectrum mask as an equivalent of the allowed adjacent channel leakage.

\subsection*{Problem Statement}
In the context of multiple wireless technologies operating in the same spectrum band, an important aspect is achieving \textbf{\emph{inter-technology coexistence}}, which refers to the ability of two or more co-located technologies to carry out their communication tasks without significant negative impact on their performance. A consistent informal definition is reported in~\cite{Cypher2000}. 
We note that the definition of coexistence that we adopt in this survey is intentionally broad, in order to span the wide range of interpretations in the literature: some works use specific coexistence goals and metrics (e.g. achieving a minimum throughput value), whereas others study the coexistence impact on the performance of each technology in terms of various metrics (e.g. throughput, delay, packet collision probability, etc.), but do not target a specific coexistence goal. We discuss this further in our literature review in Section~\ref{litSpecComm} (see especially Tables~\ref{table_review_3b}, \ref{table_review_4b}, \ref{table_review_4c}, \ref{table_review_4d}).   

Wireless inter-technology coexistence can be achieved by mitigating \emph{co- and adjacent channel interference}, as these types of interference occur when multiple devices of different technologies share the same spectrum band. 
In order to mitigate this inter-technology interference and allow access to the spectrum for multiple devices, spectrum sharing mechanisms are typically implemented at Layer~2, in a similar manner as for traditional MAC schemes mitigating intra-technology interference.  
Such solutions allow each device to use only a portion of the spectrum resources, e.g. in time or frequency, while experiencing lower levels of interference.
It follows that each device will still experience a decrease in capacity when other portions of the spectrum resources are occupied by other devices.
For example, a single device is allowed to transmit for a shorter time duration (e.g. time division multiple access -- TDMA -- in cellular networks, carrier sense multiple access with collision avoidance -- CSMA/CA -- in \mbox{Wi-Fi}), or over a portion of the frequency band (e.g. frequency division multiple access -- FDMA -- in cellular networks, channel selection in \mbox{Wi-Fi}). 
We discuss spectrum sharing mechanisms at Layer~2 further in Section~\ref{spec_sharing}.

\begin{figure*}[!t]
\centering
\includegraphics[width=0.7\linewidth]{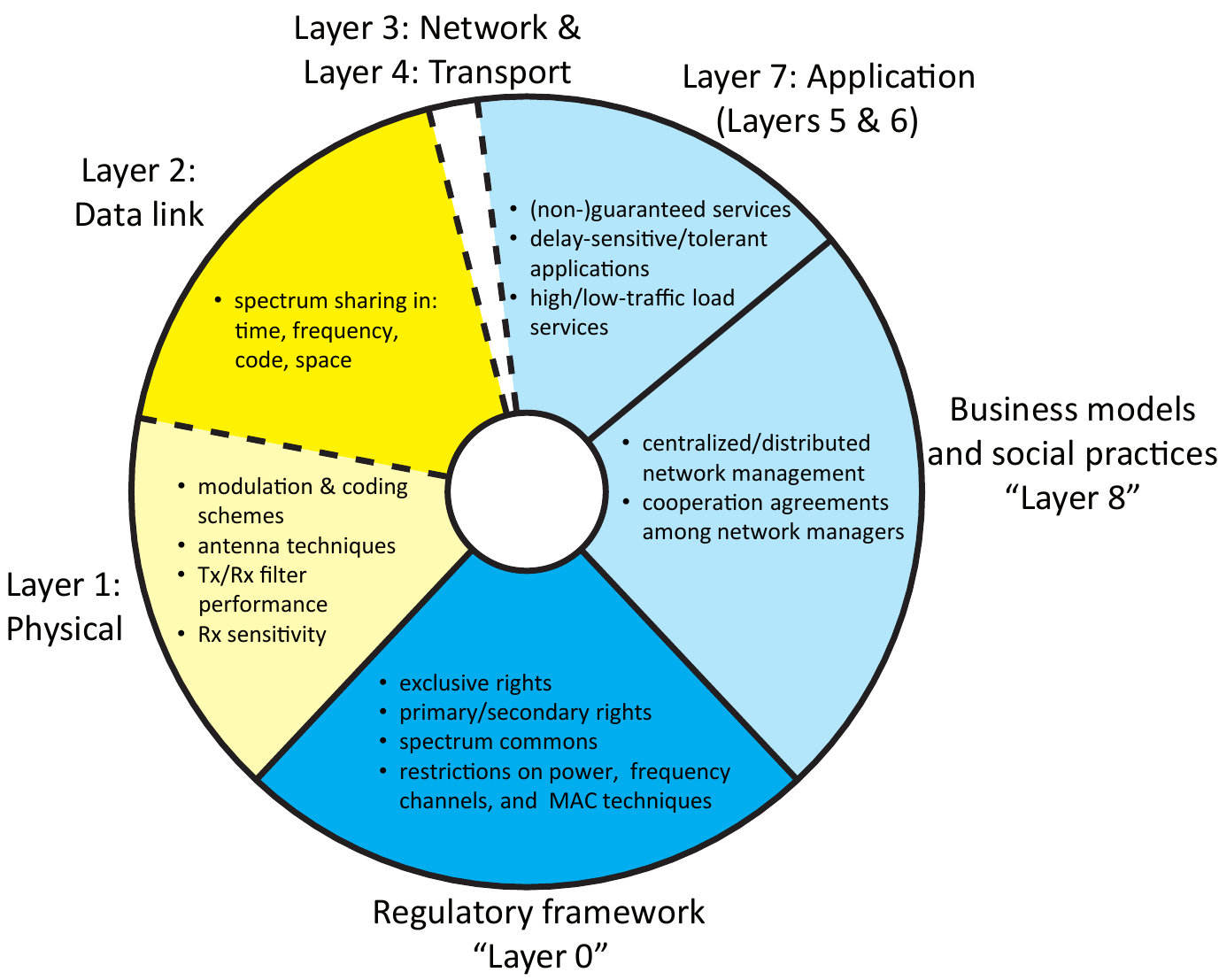}
\caption{Technology circle~\cite{Mahonen2012} as a general system-level framework for considering the design space of inter-technology spectrum sharing. Most of the spectrum sharing mechanisms (yellow) are implemented at Layer~2 and a few at Layer~1. The main constraints (blue) on spectrum sharing design are found at Layer~0 and some at Layers~7~and~8. The main features of each layer are summarized in the figure. This classification is further used in Tables~\ref{table_review_1}, \ref{table_review_2}, \ref{table_review_3}, and~\ref{table_review_4} in our literature review.}
\label{fig_techCircle}
\end{figure*}

\begin{table*}[t!]
\caption{General spectrum sharing taxonomy based on the technology circle. Specific mechanisms considered in the literature for inter-technology coexistence are further presented in Tables~\ref{table_review_1}, \ref{table_review_2}, \ref{table_review_3}, and~\ref{table_review_4}.}
\label{table_2}
\centering
\begin{tabular}{|c|p{1.5cm}|p{1cm}|p{9cm}|c|}
\hline
	\multicolumn{2}{|c|}{\textbf{Scope}} & \multicolumn{2}{c|}{\textbf{Spectrum sharing techniques}} & \textbf{Layer} \\
\hline    
	\multirow{11}{*}{Intra-technology} & \multirow{3}{*}{link level} & \multicolumn{2}{l|}{in time: TDD} & 0 \\
\cline{3-5}	
	                    			             & & \multicolumn{2}{l|}{in frequency: FDD} & 0 \\
\cline{3-5}	
	                    			             & & \multicolumn{2}{l|}{full duplex} & 1 \\ 
\cline{2-5}	                              
	                                   &  \multirow{9}{*}{network level}& \multicolumn{2}{l|}{in frequency: FDMA, OFDMA (and NC-OFDMA), channel selection, frequency reuse} & 2 \\
\cline{3-5}	
	                    			   &  & \multirow{5}{*}{in time} & periodic transmissions: TDMA, (adaptive) duty cycle & 2 \\
\cline{4-5}	                    			        
	                    			        &      &     & random access: without spectrum sensing, e.g. ALOHA, slotted ALOHA; LBT and no random backoff, e.g. ETSI frame based equipment (FBE); LBT with random backoff and fixed contention window (CW), e.g. ETSI load based equipment (LBE) B; LBT with random backoff and adaptive CW, e.g. CSMA/CA, ETSI LBE A & \multirow{4}{*}{2}\\
\cline{3-5}		                    			        
	                    			   
	                    			        &  & \multicolumn{2}{l|}{in code: CDMA} & 2 \\
\cline{3-5}		                    			        
	                    			        &  & \multicolumn{2}{l|}{in space: SDMA} & 2 \\
\cline{3-5}		                    			        
	                    			        &  & \multicolumn{2}{l|}{other: power control}  & 2\\                            
\hline 
    \multicolumn{2}{|c|}{\multirow{5}{*}{Inter-technology}} & \multicolumn{2}{l|}{in frequency: distributed channel selection, DSA techniques (database, spectrum sensing)} & 0, 2\\
\cline{3-5}    
	 \multicolumn{2}{|c|}{}                  			& \multicolumn{2}{l|}{in time: random access, distributed periodic, DSA techniques (database, spectrum sensing)} & 0, 2 \\
\cline{3-5}	                    			
	  \multicolumn{2}{|c|}{}                  		   & \multicolumn{2}{l|}{in code: FHSS, DSSS} & 1\\
\cline{3-5}	                    			
	  \multicolumn{2}{|c|}{}                 			&  \multicolumn{2}{l|}{in space: geolocation \& DSA techniques (database, spectrum sensing)} & 0, 2 \\
\cline{3-5}	                    			
	  \multicolumn{2}{|c|}{}                  			& \multicolumn{2}{l|}{other: power control -- distributed, DSA techniques (database, spectrum sensing)} & 0, 2   \\
  
\hline    
\end{tabular}
\end{table*}

\subsection*{Spectrum Management Terminology}
\label{terms}

Spectrum management refers to the manner in which the spectrum is used in general, in order to facilitate wireless communication among different devices~\cite{ECC2014}. We identify the most important terms describing aspects of spectrum management as follows: spectrum rights, spectrum allocation, spectrum sharing. 

\textbf{\emph{Spectrum rights}} is typically used by spectrum regulators to describe
under which conditions a party can use a spectrum band and what entitlement it has, e.g. how long, with what power level, and whether it has priority over other spectrum users when transmitting in this band. 
Spectrum rights are also relevant for engineers, who have to design and deploy technologies and devices that use the spectrum within the limits set by the spectrum regulators.

\textbf{\emph{Spectrum allocation/assignment}} is used in a regulatory context, to express that the spectrum regulator grants a certain party some rights to use a particular portion of the spectrum~\cite{ECC2014}, e.g. bands that are allocated to individual cellular operators. A related term that is widely used, but does not have a regulatory connotation, is \textbf{\emph{channel allocation}}, i.e. the channels on which different devices operate within a band, as configured by network managers. 

\textbf{\emph{Spectrum sharing}} is broadly defined by ECC ``as common usage of the same spectrum resource by more than one user. Sharing can be performed with respect to
all three domains: frequency, time and place."~\cite{ECC2014}
In this survey we adopt a similar definition as the one given by ECC, but we also include spectrum sharing via coding.

\section{Spectrum Sharing: A System-Level View}
\label{tech_circle}

The design, implementation, and performance of spectrum sharing schemes is determined by a multitude of inter-related factors beyond the pure technical approach. As such, in order to maximize the spectrum utility for individual devices and/or networks, spectrum sharing should be analysed from a system-level perspective that takes into account the technical, regulatory, and business aspects of wireless technologies.
In this section we present such a system-level perspective and some general classifications for spectrum sharing that can be further applied for inter-technology coexistence with equal spectrum access rights, as in Sections~\ref{litHierFr} and~\ref{litSpecComm}. 

In~\cite{Mahonen2012} the technical and non-technical aspects of wireless technologies were identified and grouped into nine layers forming a \emph{technology circle}, as shown in Fig.~\ref{fig_techCircle}. Layers~\mbox{1--7} are the technical layers of the OSI stack (i.e. Physical, Data Link, Network, Transport, Session, Presentation, Application\footnote{In real implementations (e.g. TCP/IP stack), the functionality of Layers~5--6 is integrated in Layer~7; we thus discuss only the \emph{Application} layer.}), whereas Layers~0~and~8 model the regulatory, and business and social aspects, respectively. As the circular representation suggests, there is an inter-dependence between all these layers, which together form a large design parameter space that determines the candidate spectrum sharing mechanisms: 
some layers correspond to the actual implementation of these mechanisms, whereas other layers impose design constraints. 
Specifically, the major spectrum sharing mechanisms are implemented at Layer~2, and some at Layer~1, as summarized in Table~\ref{table_2}.
Nonetheless, there are exceptions where sharing mechanisms are implemented at other layers, e.g. duplexing and DSA databases at Layer~0.
Most of the design constraints for spectrum sharing are specified at Layer~0, but also at Layers~7~and~8. 
We note that Layers~3~and~4 may have an indirect influence on the efficiency of inter-technology spectrum sharing mechanisms, by e.g. limiting the size of
the packets transmitted through fragmentation, or varying the data rate of the traffic flow; however, this is outside the scope of this survey. 

Importantly, not all combinations of technical and non-technical parameters at different layers are feasible when designing spectrum sharing mechanisms for inter-technology coexistence, and out of those that are feasible, some may be preferred over others. 
For instance, when deploying traditional cellular networks, each operator has exclusive rights to access the spectrum at Layer~0. This case is thus suitable for implementing spectrum sharing mechanisms at Layer~2 that are centrally coordinated, as a single operator manages the entire network at Layer~8. 
Consequently, cellular networks are ideally suited to carry delay-sensitive traffic such as voice (at Layer~7), as the performance of the centrally-managed network can be readily predicted and optimized. 
We note that, for this example, inter-technology coexistence only occurs for multiple integrated technologies (e.g. LTE and NB-IoT), which are deployed by the same operator (at Layer~8). 

Let us now consider a spectrum band where different networks operate based on primary/secondary spectrum rights at Layer~0. The primary network can then implement coordinated spectrum sharing mechanisms at Layer~2 as a result of typically having a single network manager at Layer~8 (i.e. similarly to cellular networks). The operation of the secondary networks is strictly limited at Layer~0 to ensure primary protection, e.g. by specifying a maximum allowable interference power from the secondary networks to the primary. 
As such, the access of the secondary networks to the spectrum can be coordinated at Layer~0 through a reliable database operated by a third party at Layer~8, e.g. for TVWS. 
By contrast, interactions among secondary devices can be managed by distributed spectrum sharing mechanisms at Layer~2, as they do not have the right to any protection at Layer~0; it follows that it is not straightforward to guarantee the quality of the services offered by these secondary networks at Layer~7~\cite{Akyildiz2006, Liang2011}. 

Lastly, in a spectrum commons like the unlicensed bands, where various technologies coexist and have the same spectrum access rights at Layer~0, distributed spectrum sharing mechanisms have been a popular choice at Layer~2. Fully centralized coordination is typically not feasible for this example, due to the lack of business agreements among numerous networks managers at Layer~8. 

As illustrated by these examples, there is a tight interconnection between the technical and non-technical design parameters and constraints at different layers. It is critical to consider these interconnections in a unified system-level view, as different parameter combinations result in specific inter-technology interactions. 
Correctly identifying and evaluating these interactions lays the foundation for a robust development framework for new wireless technologies that result in efficient spectrum use.

In this survey we adopt the technology circle proposed in~\cite{Mahonen2012} as a framework to facilitate our system-level analysis of inter-technology spectrum sharing.
In the following we briefly describe each layer of the technology circle and we highlight its impact on the design of spectrum sharing mechanisms in general.
We first discuss Layer~0, which specifies the main constraints on spectrum sharing, but which also includes a few sharing mechanisms; we then present Layer~2 where most of the spectrum sharing mechanisms are implemented; subsequently, further sharing mechanisms at Layer~1 are presented; lastly, we discuss further constraints at Layers~7~and~8.
In Sections~\ref{litHierFr} and~\ref{litSpecComm} we then apply these general spectrum sharing classifications to our literature review on inter-technology coexistence with equal access rights.

\subsection{Regulatory Framework Constraints \& Spectrum Sharing at Layer~0}
\label{reg_framework}

Layer~0 primarily defines regulatory constraints for spectrum sharing mechanisms at Layers~2~and~1. However, a few spectrum sharing mechanisms are actually implemented at this layer. In this section we first discuss the regulatory constraints and then spectrum sharing at Layer~0.

\subsubsection{Constraints at Layer~0}
The regulatory framework consists of the regulatory limitations imposed on the use of spectrum. These determine who is allowed to use the spectrum, for how long,
and within which technical parameter constraints, e.g. transmit power. Consequently, spectrum sharing mechanisms have to be designed and optimized under these constraints.

\begin{figure}[!t]
\centering
\includegraphics[width=1\columnwidth]{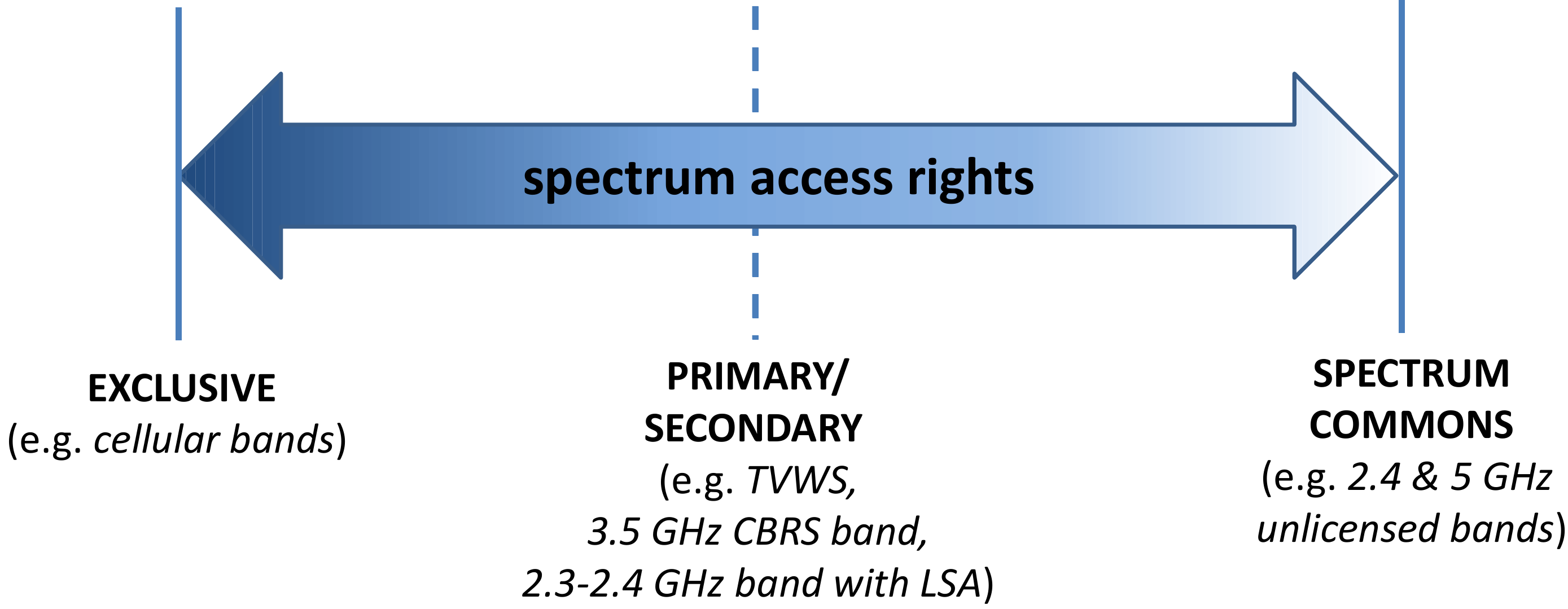}
\caption{Spectrum access rights based on the regulatory framework.}
\label{fig_regFramework}
\end{figure}

As shown in Fig.~\ref{fig_regFramework}, spectrum access rights span a continuum of access models, from  exclusive use  of  spectrum, i.e.  exclusive spectrum  access  rights  for  a  single network or technology, to a spectrum commons, where  all devices/networks/technologies have the same rights to access the spectrum. 
Spectrum access rights between these extremes include the primary/secondary spectrum use model, where secondary networks must give priority to the dominant primary network. 
We note that the vast regulatory literature on spectrum access rights is out of the scope of this survey and we instead refer the interested reader to~\cite{Peha2005, Zhao2007, ElectronicCommunicationsCommitte(ECC)2009, Buddhikot2007}. 

Traditionally, the spectrum access rights applied in practice have been at the two extremes in Fig.~\ref{fig_regFramework}.
On the one hand, exclusive rights to access the spectrum have been granted to e.g. mobile cellular networks, where each operator buys a license for a given spectrum band. Since there is a single operator deploying and managing the network, the regulators do not need to impose rules on the spectrum sharing techniques; the regulatory restrictions instead largely focus on transmit power levels and filter masks, in order to limit the interference towards other out-of-band networks/services.

On the other hand, in the unlicensed bands -- an example of a spectrum commons -- any device/technology/network has the same rights to access the spectrum (e.g. the 2.4~GHz and the 5~GHz unlicensed bands). Since such bands are open in principle to any technology, the spectrum regulators may decide to impose some restrictions also on the spectrum sharing mechanisms at Layer~2, such that multiple coexisting technologies have the opportunity to access the spectrum. For instance, in Europe ETSI requires devices to implement LBT at the MAC layer, where each device must sense and detect the medium free from other transmissions before starting its own transmission~\cite{ETSI2015}. Additionally, for the channels in the 5~GHz unlicensed bands where radar systems operate, mechanisms like dynamic frequency selection (DFS) and transmit power control (TPC) are specified in regulation~\cite{ETSI2015, ECC2004, 47CFR15.4072016}, in order to protect radar operations. 

Over the last fifteen years, several measurement studies have investigated how efficiently spectrum is used~\cite{FederalCommunicationsCommissionSpectrumPolicyTaskForce2002, Palaios2013, McHenry2006, Valenta2009, Islam2008}. The main findings revealed that some of the allocated spectrum with exclusive rights is not used to its full capacity. Consequently, other models for spectrum access rights have emerged, with the general aim of allowing more dynamic access to the spectrum, based on demand. However, incumbent technologies operating in these bands still have priority when accessing the spectrum, such that hierarchical primary/secondary regulatory frameworks are needed. 

Three recent examples where hierarchical regulatory frameworks are applicable are: TVWS, the CBRS band, and other bands granted through LSA, e.g. 2.3--2.4~GHz~\cite{Guiducci2017}.     
TVWS refers to the spectrum initially allocated for TV broadcasting, the coverage of which is not uniform, such that in particular locations the spectrum could be reused by other technologies with secondary access rights~\cite{Ofcom2015, FCC2010}, e.g. IEEE~802.11af \mbox{Wi-Fi}~\cite{IEEE2016}, LTE~\cite{ETSI2011}, IEEE 802.19.1~\cite{IEEE2014}, or IEEE 802.22~\cite{2011}. 
We note that this spectrum access framework has only recently been adopted by a few regulators, i.e. FCC in the U.S. and Ofcom in the U.K., and practical deployments are still in their infancy~\cite{Matsumura2017}.  
A three-layer hierarchical regulatory model is currently under discussion for the 3.5~GHz CBRS band in the U.S.~\cite{FCC15-472015} with: incumbent access, priority access, and general authorized access (GAA). The spectrum access system (SAS) manages the spectrum access of the secondary systems, corresponding to the two latter spectrum access layers.
LSA is specified by the ECC in Europe~\cite{ECC2014} and is primarily intended for mobile broadband operators that are willing to share spectrum with existing incumbents. 
We note that other models for spectrum access rights have also been proposed in the literature~\cite{Tehrani2016}, but have thus far largely not been adopted in practice.

Importantly, inter-technology coexistence can occur for any model of spectrum access rights. However, the most challenging coexistence cases are expected in the unlicensed bands as an example of a spectrum commons, where any technology is allowed to transmit while complying with rather relaxed rules. As such, there is also a growing tendency to extensively use unlicensed bands by different technologies.
One example trend is to aggregate unlicensed spectrum, e.g. LTE in the licensed bands aggregates carriers in the unlicensed bands (as carrier \mbox{Wi-Fi}~\cite{3gpp2014wifi}, LAA~\cite{3GPP2016, 3GPP2016a}, or LTE-U~\cite{Forum2015}); and \mbox{Wi-Fi} aggregates multiple 20~MHz channels in the 5 GHz unlicensed band~\cite{IEEE2016}. 
Moreover, both future 5G cellular technologies~\cite{3GPP2017} and IEEE 802.11ad \mbox{Wi-Fi}~\cite{IEEE2016} aim at extending their operation to the unlicensed 60~GHz band.

\subsubsection{Spectrum Sharing at Layer~0}
In hierarchical regulatory frameworks, spectrum sharing between primary/secondary networks is implemented through DSA mechanisms, where secondary networks access the spectrum opportunistically, whenever it is not occupied by primary networks.
In such deployments where the primary users are protected from the secondary users, e.g. TVWS, the secondary users typically acquire knowledge on the availability of channels from a database operated by a third party~\cite{Ofcom2015, FCC2010}.
In fact, there is a strong inter-connection among spectrum sharing in \textbf{frequency}, \textbf{time}, \textbf{space}, and \textbf{power} in such networks, i.e. the DSA database is a central coordinator that gives information on the availability of the channels in certain locations and imposes limits on the transmit power and duration of use for the secondary networks.
We consider these to be fundamental constraints imposed by the database on how the secondary networks access the spectrum and we include such spectrum sharing mechanisms at Layer~0.
We note that primary/secondary spectrum sharing could also be implemented in a solely distributed manner using spectrum sensing, or spectrum sensing could be used as additional input for DSA databases, but we consider such techniques as belonging to Layer~2, similarly to other sensing-based spectrum access mechanisms, e.g. CSMA/CA. 
Notably, many DSA and supporting cognitive radio techniques have been proposed in the literature~\cite{Ren2012, Gavrilovska2014, Yucek2009, Wang2008, Akyildiz2006, Liang2011}, but have not yet been implemented in commercial deployments.  

Finally, duplexing can be considered a spectrum sharing mechanism between the two directions of a single link, that is implemented at Layer~0 through regulatory and technical restrictions on channelization. Here we can distinguish frequency division duplexing (FDD) and time division duplexing (TDD), as shown in Table~\ref{table_2}.

\begin{figure*}[!t]
\centering
\includegraphics[width=0.7\linewidth]{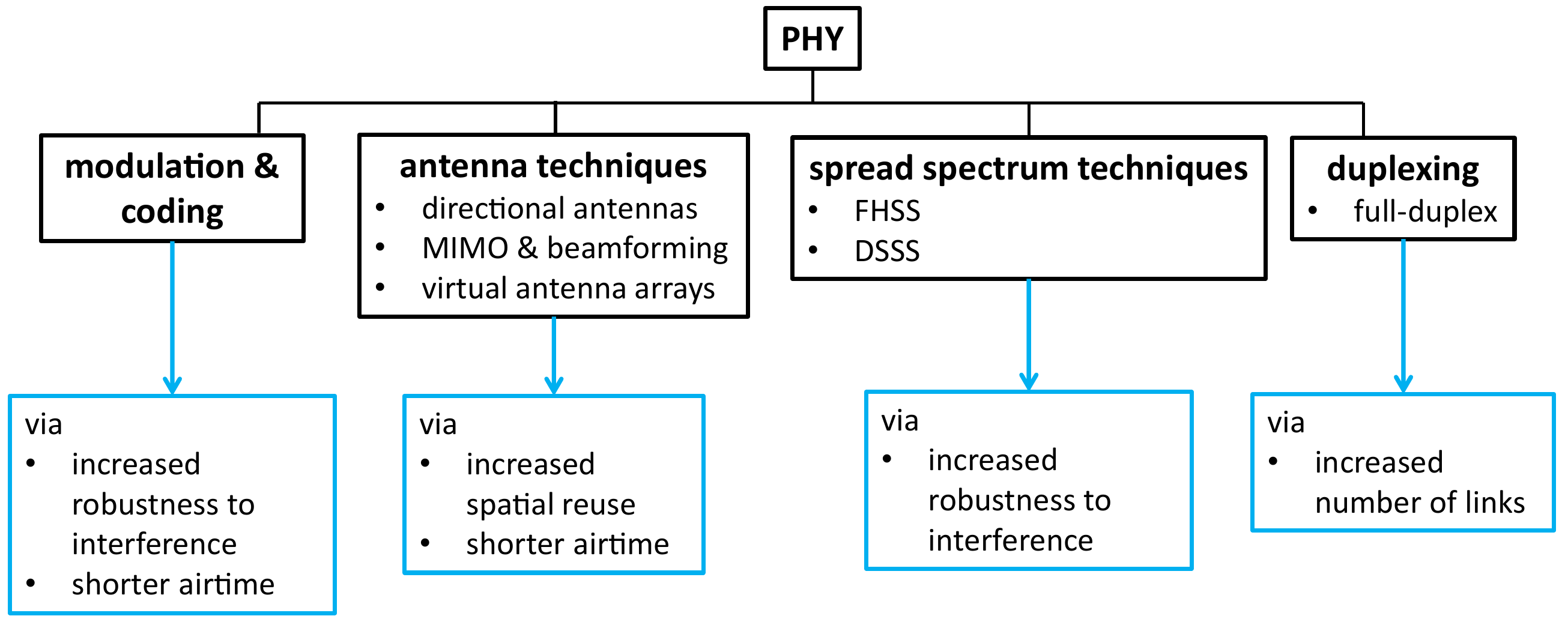}
\caption{General classification of Layer~1 techniques that can be used for facilitating wireless inter-technology coexistence. Specific Layer~1 techniques considered in the reviewed literature for inter-technology coexistence are presented in Tables~\ref{table_review_1}, \ref{table_review_2}, \ref{table_review_3}, and~\ref{table_review_4}.}
\label{fig_layer1}
\end{figure*}

\subsection{Spectrum Sharing at Layer~2}
\label{spec_sharing}

The majority of spectrum sharing mechanisms are implemented at Layer~2 of the technology circle. Although the focus of this survey is on inter-technology spectrum sharing, here we also present and discuss a taxonomy of intra-technology spectrum sharing, since the mechanisms implemented by devices within a technology can also affect the interactions with other technologies.   

\subsubsection{Intra-Technology Spectrum Sharing}

From an intra-technology network-level perspective, multiple devices within the same network have to access the same spectrum. In this context spectrum sharing is performed by the MAC sub-layer of Layer~2. Spectrum sharing in such a case can be performed in: \textbf{(i)~frequency}; \textbf{(ii)~time}; \textbf{(iii)~code}; or \textbf{(iv)~space}.

\paragraph*{Spectrum sharing in frequency}
The traditional technique is frequency division multiple access (FDMA), which divides the allocated band into multiple sub-carriers, which are then allocated to different users, e.g. in GSM. A similar concept, but with a finer frequency division granularity is orthogonal frequency division multiple access (OFDMA), which divides the band into closely-spaced orthogonal sub-carriers, e.g. in LTE and WiMAX.
Furthermore, frequency division can be used as a spectrum sharing mechanism between devices, without necessarily being implemented as a MAC protocol, e.g. channel selection/allocation for \mbox{Wi-Fi}, which can increase capacity and reduce interference among \mbox{Wi-Fi} devices~\cite{SurachaiChieochan2010}.   
Frequency reuse techniques have been applied analogously for cellular networks~\cite{Damnjanovic2011, Saquib2012, Katzela1996, Hamza2013}.
We note that, for modern and emerging wireless networks, implementing channel selection for interference management may not be straightforward, due to advanced features like channel bonding (in e.g. IEEE~802.11n/ac \mbox{Wi-Fi}, and LTE), where several channels are dynamically merged to form larger-bandwidth channels~\cite{Bukhari2016}. Consequently, partially overlapping channels of different widths may be used and reconfigured dynamically by different coexisting devices, which increases the complexity of network-wide interference interactions.

\paragraph*{Spectrum sharing in time}
This has traditionally been implemented among users in cellular networks through scheduled time division multiple access (TDMA), which is an instance of periodic transmissions that are centrally coordinated. 
A more general concept is duty cycling, which also refers to non-coordinated or only locally-coordinated periodic transmissions. Originally, duty cycling was proposed for sensor networks~\cite{Carrano2014} with the aim of reducing energy consumption. Recently, it has also been adopted by broadband technologies such as LTE-U, which implements adaptive duty cycling~\cite{Forum2015}.
A fundamentally different approach is random access in time, e.g. ALOHA and its variant slotted ALOHA, where each device transmits whenever there is traffic to be sent from the upper layers. Also random, but implementing carrier sensing, are LBT mechanisms, where each device first listens to the channel and transmits only if no other ongoing transmission is detected, e.g. CSMA/CA for \mbox{Wi-Fi} and several other LBT variants specified by ETSI~\cite{ETSI2015}, \emph{cf.} Table~\ref{table_2}. 
We note that, in order to reduce the number of colliding transmissions from different devices, some LBT mechanisms vary the sensing time that a device has to listen to the channel for, based on a random backoff, which is selected by each device randomly within a given interval, e.g. [0, CW], where CW (contention window) is a design parameter. Furthermore, the CW itself can be adapted, e.g. for CSMA/CA in IEEE~802.11 the CW is doubled every time that a collision occurs (i.e. binary exponential random backoff).

\paragraph*{Spectrum sharing via coding}
For multi-user networks this is known as code division multiple access (CDMA) and it is based on spread spectrum techniques at Layer~1. CDMA is implemented by allocating a unique code for each user and allowing all users to transmit over the same wide bandwidth. This was implemented in 3G systems like UMTS and CDMA2000, based on direct sequence spread spectrum (DSSS) at Layer~1. 

\begin{table*}[t!]
\caption{General classification of applications based on user requirements. Requirements for traffic volume are further used to classify existing literature on inter-technology coexistence in a spectrum commons in Section V.}
\label{table_app}
\centering
\begin{tabular}{|p{2cm}|p{2cm}|p{12cm}|}
\hline
	\centering\textbf{Requirement} & \centering\textbf{Classification} & \centering\let\newline\\\arraybackslash\hspace{0pt} \textbf{Examples} \\
\hline    
	\multirow{2}{*}{traffic volume} & high traffic load (broadband) & file sharing, video steaming, and video conferencing through cellular broadband (e.g. LTE, 5G), and IEEE 802.11 \mbox{Wi-Fi}; some industrial applications for sensor networks with need for high sampling rate \\
\cline{2-3}	
	               & low traffic load & home and industrial applications for sensor networks (e.g. IEEE 802.15.4, ZigBee, NB-IoT, Bluetooth), M2M applications  \\
\hline  
    \multirow{2}{*}{delay}          & delay-tolerant & web browsing, file transfer, email, some sensor applications \\
\cline{2-3}	    
	               & delay-sensitive & voice calls, streaming, some industrial IoT applications~\cite{Schulz2017}  \\
\hline
    \multirow{2}{*}{target end-user} &  human & web browsing, video conferencing \\
\cline{2-3}	    
	                &  machine & IoT, D2D, M2M  \\
\hline 
\end{tabular}
\end{table*}

\paragraph*{Spectrum sharing in space}
This is based on antenna directivity at Layer~1.
Deploying directional antennas facilitates e.g. sectorization in cellular networks, and thus interference reduction and more aggressive frequency reuse~\cite{Chan1992}.
As such, sectorization in cellular networks is used for combined spectrum sharing in space and frequency.   
A more recent multiple access technique is space division multiple access (SDMA) which emerged together with advanced antenna techniques at Layer~1. SDMA is based on using narrow beams pointed in the direction of the desired receiver, such that interference in other directions is reduced, which allows a higher number of simultaneous links over the same area, i.e. increases spatial reuse. We note that although the underlying Layer~1 techniques of sectorization and beamforming are similar, beamforming is a dynamic mechanism, whereas sectorization assumes a static antenna configuration.  
Multi-user multiple-input-multiple-output (MU-MIMO), i.e. an example of SDMA, has been standardized as an option in IEEE~802.11ac \mbox{Wi-Fi}~\cite{IEEE2016}, LTE~\cite{3GPP2017b}, IEEE 802.16 WiMAX~\cite{IEEE802.16-2012}. Example MU-MIMO MAC protocols for \mbox{Wi-Fi} were reviewed in~\cite{Liao2016}. 

Another method to share the spectrum, but not a MAC protocol, is transmit \textbf{power} control, which determines the transmission and interference range, and thus affects spatial reuse. Many technologies implement it as a mandatory or an optional feature, e.g. UMTS, LTE, \mbox{Wi-Fi}, sensor networks.

\subsubsection{Inter-Technology Spectrum Sharing}
For distributed spectrum access, the intra-technology spectrum sharing mechanisms can coincide with the inter-technology mechanisms at Layer~2, especially in the unlicensed bands, but also for secondary/secondary inter-technology coexistence in hierarchical regulatory models. 
Consequently, inter-technology spectrum sharing mechanisms can be implemented through (but are not restricted to) MAC protocols. 

Inter-technology spectrum sharing can be performed in \textbf{frequency} through channel selection schemes. An example is LTE-U/LAA performing channel selection to avoid co-channel \mbox{Wi-Fi} devices~\cite{Forum2015, 3GPP2016, 3GPP2016a}. 
Inter-technology spectrum sharing in \textbf{time} can be implemented in distributed networks at the MAC layer, through duty cycle transmissions (e.g. LTE-U) or through LBT MAC protocols (e.g. CSMA/CA for \mbox{Wi-Fi} and LBT for LAA). These mechanisms share the spectrum both within and among technologies.     
Another mechanism that facilitates both intra- and inter-technology coexistence for distributed networks is \textbf{power control}, which affects spatial reuse within and among technologies. This is considered for e.g. LAA~\cite{3GPP2015}, and the upcoming \mbox{Wi-Fi} amendment IEEE~802.11ax~\cite{Omar2016, Bellalta2016}. 

Importantly, most current technologies implement more than one spectrum sharing mechanism at Layer~2 to facilitate (both intra- and inter-technology) coexistence. Examples include GSM (FDMA and TDMA), LTE (OFDMA and TDMA), \mbox{Wi-Fi} (CSMA/CA, and optionally channel selection and SDMA), LTE in the unlicensed bands (duty cycle or LBT, and channel selection). We note that most technologies implement a variant of spectrum sharing in time and frequency, which suggests that these mechanisms are able to efficiently mitigate interference.

\subsection{Spectrum Sharing and Interference Mitigation at Layer~1}
\label{phy}

The PHY layer can affect inter-technology coexistence through techniques that influence the design and performance of spectrum sharing mechanisms, as shown in Fig.~\ref{fig_layer1}. Furthermore, some of these techniques, i.e. spread spectrum techniques and full-duplex, can be seen as spectrum sharing mechanisms implemented directly at Layer~1, as summarized in Table~\ref{table_2}. We briefly discuss the PHY techniques in Fig.~\ref{fig_layer1} in the following.

The PHY layer determines the manner in which the data is sent over the wireless channels, primarily through modulation and coding. Different combinations of modulation and coding schemes affect the spectrum reuse, since they may provide increased robustness to interference, so that the number of links that can be simultaneously active is increased.

Other mechanisms at Layer~1 that affect spectrum sharing are antenna techniques. 
Deploying directional antennas facilitates sectorization in cellular networks and beamforming based on multiple antennas supports SDMA mechanisms at Layer~2, as discussed in Section~\ref{spec_sharing}. 
Also, using multiple-input-multiple-output (MIMO) antenna systems enables multiple data streams per link, which can increase the link capacity and reduce the effect of channel quality fluctuations through spatial diversity.
Spatial diversity is also exploited through cooperative communication, which proposes virtual antenna arrays built with single-antenna devices. The impact of such techniques on the MAC in general is surveyed in~\cite{Sami2016}. 

Spread spectrum techniques have been used at Layer~1 to increase robustness against interference in intra- and inter-technology coexistence scenarios. Frequency hopping spread spectrum (FHSS) allows rather low data rates and is thus implemented by technologies like Bluetooth~\cite{Bluetooth2016}, which transports lower volumes of traffic. Direct sequence spread spectrum (DSSS) was used instead for technologies that transport moderate to high traffic volumes, e.g. IEEE 802.11b and code division multiple access (CDMA) systems like UMTS and CDMA2000. 

\begin{figure*}[!t]
\centering
\includegraphics[width=0.65\linewidth]{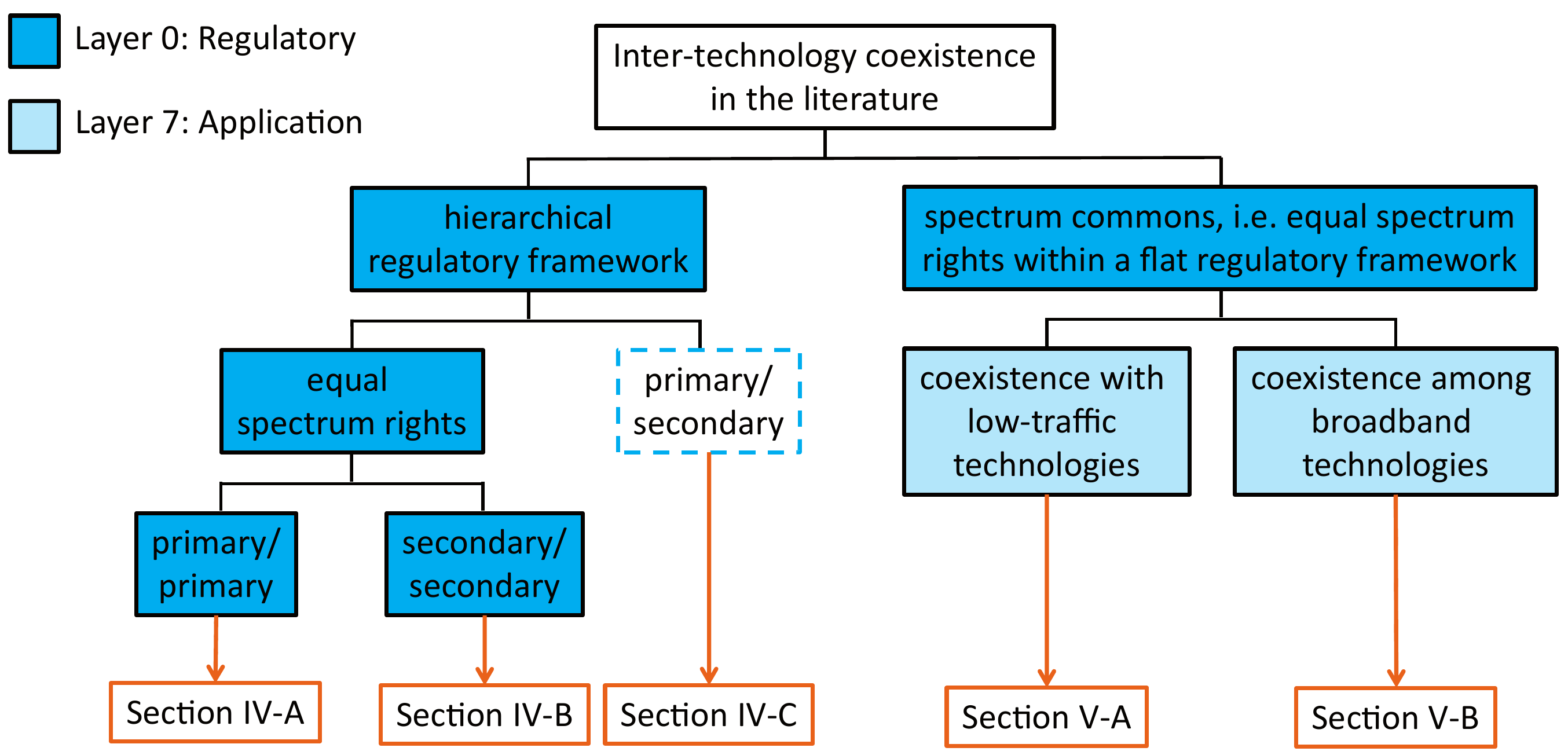}
\caption{Classification of research work in our literature review in Sections~\ref{litHierFr}~and~\ref{litSpecComm}, where we focus on inter-technology coexistence with \emph{equal} spectrum access rights within a hierarchical regulatory framework (i.e. primary/primary and secondary/secondary) and a flat regulatory framework (i.e. spectrum commons).}
\label{fig_classifLit}
\end{figure*}

Finally, recent interference cancellation techniques at Layer~1, which allow full-duplex communication, i.e. bidirectional for the same link at the same time, are a promising solution to increase spectrum utilization efficiency (see, e.g.~\cite{Kim2015} for full-duplex from the perspective of the PHY and MAC layers, and~\cite{Amjad2017} for full-duplex with cognitive radios). Full-duplex would impact spectrum sharing techniques at Layer~2, which would have to be redesigned (e.g. CSMA/CA for \mbox{Wi-Fi}~\cite{Xie2014}).
We note that full-duplex can be considered a spectrum sharing technique at Layer~1, since it refers to sharing spectrum resources at the link level, by means of PHY techniques. By contrast, other duplexing techniques like FDD and TDD share the resources as determined by regulations at Layer~0.

\subsection{Constraints of Applications at Layer~7}

This layer can have a major impact on the design and performance of spectrum sharing mechanisms, since the specific requirements for each target application in a given network should be reflected in the choice of spectrum sharing technique. 
The applications can be grouped in different categories, according to their requirements in terms of traffic volume, delay, and target end-users, as shown in Table~\ref{table_app}. 
The application type affects the selected environment where the networks are deployed, their mobility patterns, and thus the interference characteristics. Coexistence in these specific conditions has to be managed by the spectrum sharing mechanisms.

\subsection{Constraints of Business Models and Social Practices at Layer~8}
\label{business_models}

Business models and social practices affect the network deployment likelihood, topology, ownership, and level of coordination. These result in different interference
conditions. A taxonomy of business models is outside the scope of this survey, but we provide some examples to illustrate such interactions between technical and non-technical requirements.

Outdoor public cellular networks are owned and managed by mobile operators, as they have the financial resources to acquire a license for the cellular
bands. Consequently, the spectrum sharing techniques can be centrally coordinated. However, outdoor base station (BS) deployments in private locations, e.g. on top of buildings, are restricted by the existence of an agreement with the building owners. The optimization of the spectrum sharing parameters depends, in turn, on the physical locations of BSs and resulting propagation conditions.

By contrast, in private deployments, e.g. indoor residential \mbox{Wi-Fi}, there are multiple distributed networks, which operate individually, often with the default configuration. 
In \mbox{Wi-Fi} business deployments, a higher level of coordination is expected than in private deployments, e.g. for channel allocation or client-access point (AP) association, as there is a single manager configuring the network. A similar example are hotspot deployments. However, it may also occur that multiple hotspots from different operators transmit over the same spectrum, such that coordination can be achieved within a network managed by a single operator, but not among networks.

Finally, based on Layer~8 considerations we identify inter-technology interactions of two types: (i)~integration and (ii)~competition.
Inter-technology \emph{integration} refers to different technologies that interconnect, in order to increase capacity, or extend the range of the offered services, e.g. carrier \mbox{Wi-Fi} (i.e. integration of \mbox{Wi-Fi} into the 3GPP cellular networks for data offloading purposes~\cite{He2016, Rebecchi2015}); standardization of \mbox{LAA-LTE} operating in the 5~GHz unlicensed band for capacity increase; and NB-IoT in LTE Advanced Pro (i.e. Release 13) for supporting device-to-device (D2D) IoT applications.
Inter-technology \emph{competition} occurs among different technologies that share the same spectrum, but for their individual offered services, e.g. secondary technologies operating within hierarchical regulatory frameworks; IEEE 802.15.4, Bluetooth, and \mbox{Wi-Fi} sharing the 2.4 GHz unlicensed band; LTE-U, LAA, and \mbox{Wi-Fi} sharing the 5~GHz unlicensed band.     
Importantly, interactions of the competition type lead to the most challenging inter-technology coexistence cases, where optimizing the overall spectrum utility is not trivial, due to the potentially greedy or conflicting individual goals for each technology.

\subsection{Literature Review Structure}

In the following sections we present a review of the literature addressing inter-technology coexistence with equal spectrum access rights and we classify the work according to different layers of the technology circle, as shown in Fig.~\ref{fig_classifLit}. We first differentiate the work based on the regulatory framework at Layer~0, i.e. hierarchical in Section~\ref{litHierFr}, and flat in Section~\ref{litSpecComm}. For the hierarchical regulatory framework, we distinguish coexistence between \emph{primary/primary} (Section~\ref{litPrimPrim}) and \emph{secondary/secondary} (Section~\ref{litSecSec}) technologies. Although this is not the focus of our survey, in Section~\ref{litPrimSec} we also give some illustrative examples of \emph{primary/secondary} coexistence from the recent literature, in order to show how increasingly more bands are being considered for operation of multiple technologies.   
For coexistence within a flat regulatory framework, where all technologies have the same rights (i.e. \emph{spectrum commons}), we classify the work further based on the Application Layer~7, i.e. \emph{low-traffic} and \emph{broadband} technologies. For each of the identified categories we review the spectrum sharing mechanisms at Layers~2~and~1.

\section{Literature Review of Inter-Technology Spectrum Sharing within a Hierarchical Regulatory Framework}
\label{litHierFr}

This section focuses on inter-technology coexistence within a hierarchical regulatory framework. We first review existing work on inter-technology coexistence with equal spectrum rights, i.e. primary/primary in Section~\ref{litPrimPrim} and secondary/secondary in Section~\ref{litSecSec}. Then we give some examples of work on primary/secondary coexistence in Section~\ref{litPrimSec}.  
We note that primary/secondary inter-technology coexistence has already been extensively addressed in previous surveys, e.g.~\cite{Paisana2014, Tehrani2016, Akyildiz2006, Akyildiz2008, Zhao2007, Wang2008, Yucek2009, Gavrilovska2014, Ren2012}, and so we give only a few representative examples from the recent literature, in order to show that there is an increasing number of bands considered for operation of multiple technologies, which may open the possibility to also accommodate inter-technology coexistence with equal spectrum rights in the future.
Finally, in Section~\ref{hier_frame_sum} we summarize and discuss the main findings.

\begin{table*}[!t]
\caption{Literature review of inter-technology spectrum sharing with equal rights, within a hierarchical regulatory framework}
\label{table_review_1}
\centering
\begin{tabular}{|p{2.2cm}|p{2cm}|p{0.61cm}|p{5cm}|p{2.3cm}|p{3.2cm}|}
\hline
 \centering \textbf{Spectrum Rights at Layer~0} 
& \centering \textbf{Technologies}
& \centering \textbf{Ref.}  
& \centering \textbf{Coexistence at Layer~2} 
& \centering \textbf{Coexistence at Layer~1}
& \centering\let\newline\\\arraybackslash\hspace{0pt} \textbf{Coordination at Layer~2 based on constraints at Layer~8} \\ 
\hline
    \multirow{2}{*}{primary/primary} & \multirow{2}{2cm}{LTE/NB-IoT} & \cite{Mangalvedhe2016} & \textbf{LTE}: resource blanking in time and frequency; \textbf{NB-IoT}: power boosting & -- & centralized\\
\cline{3-6}
                    &            & \cite{Wang2016} & \textbf{both technologies}: adjacent frequencies & -- & -- \\
\hline
    \multirow{7}{*}{secondary/secondary} & GAA users/GAA users in the CBRS band & \cite{Sahoo2017, Ying2017} & \textbf{both technologies}: channel allocation through SAS & -- & centralized \\
\cline{2-6}  
	                    & \multirow{2}{2cm}{IEEE 802.22/802.11af in TVWS} & \cite{Kang2011} & \textbf{802.11af}: likely CSMA/CA & -- & --\\
\cline{3-6} 
                        &                         & \cite{Feng2013} & \textbf{802.11af}: CSMA/CA; \textbf{802.22}: busy tone  & \textbf{802.11af}: signal pattern comparison & distributed \\	
\cline{2-6}
                        & \multirow{3}{2cm}{\mbox{Wi-Fi}/LTE in TVWS} & \cite{Cavalcante2013, Paiva2013} & \textbf{\mbox{Wi-Fi}}: CSMA/CA~\cite{Cavalcante2013}; \textbf{LTE}: none & -- & distributed \\                                             
\cline{3-6} 
                        &           & \cite{Almeida2013} & \textbf{\mbox{Wi-Fi}}: CSMA/CA; \textbf{LTE}: fixed duty cycle (0--80\%) with different subframe blanking patterns & -- & distributed \\
\cline{3-6}
                        &           & \cite{Beluri2012} & \textbf{\mbox{Wi-Fi}}: CSMA/CA; \textbf{LTE}: fixed and adaptive duty cycle, LBT, channel selection & -- & distributed \\                      
\hline           
\end{tabular}
\end{table*}

\subsection{Primary/Primary Coexistence}
\label{litPrimPrim}

This section reviews the literature on primary/primary inter-technology coexistence, as summarized in Table~\ref{table_review_1}. 
We first present a literature overview in Section~\ref{litprimprim}. We then review in detail the work on LTE/NB-IoT coexistence in Section~\ref{ltenbiot}.

\subsubsection{\textbf{Literature Overview}} 
\label{litprimprim}
Primary/primary inter-technology coexistence was considered in the literature for different technologies that are integrated, such that exclusive spectrum access rights at Layer~0 are assigned to a single entity that deploys and manages at Layer~8 a multi-technology network in the same spectrum band, e.g. cellular networks that incorporate LTE and NB-IoT. 
As such, designing inter-technology spectrum sharing mechanisms is less challenging and only a few papers addressed this by considering centralized mechanisms specific to single-technology cellular networks, i.e. channel allocation, power control, resource blanking.

\subsubsection{\textbf{LTE/NB-IoT}}
\label{ltenbiot}
The authors in~\cite{Mangalvedhe2016} identified interference problems occurring when LTE coexists with an in-band \mbox{NB-IoT} deployment (i.e. both technologies use the same subcarriers) of the same operator, for the case where only some of the BSs are \mbox{NB-IoT-capable}. NB-IoT devices could thus associate to only some BSs, such that they may suffer from strong interference from BSs that are only LTE-capable. 
As coexistence solutions, the authors investigated power boosting, i.e. increasing the downlink power for the NB-IoT resource blocks compared to that for LTE resource blocks; and resource blanking, i.e. not scheduling LTE transmissions on resource blocks that are used for NB-IoT by neighbouring BSs. 
The simulation results in~\cite{Mangalvedhe2016} showed that LTE resource blanking was an efficient method to avoid co-channel interference for NB-IoT users. 
We note that this technique is essentially a dynamic variant of spectrum sharing in time and frequency among different BSs.  

By contrast, the authors in~\cite{Wang2016} considered LTE/NB-IoT coexistence for the complementary case, where the two technologies transmitted on different frequency channels. The effects of ACI were evaluated for different filter capabilities of the transmitter (i.e. ACLR) and of the receiver (i.e. ACS). The authors found through simulations that the effect of ACI on the LTE and NB-IoT networks was in general negligible.

\subsection{Secondary/Secondary Coexistence}
\label{litSecSec}

This section reviews the literature on secondary/secondary inter-technology coexistence, as summarized in Table~\ref{table_review_1}. We first present a literature overview in Section~\ref{lit_ov_ss}. We then review in detail the work on: (i)~the newly-available CBRS band in the U.S. in Section~\ref{CBRS_review}; and (ii)~TVWS in Section~\ref{TVWS_review}.

\subsubsection{\textbf{Literature Overview}} 
\label{lit_ov_ss}
Only a few works have addressed secondary/secondary inter-technology coexistence in the CBRS band and they considered centralized channel allocation through a database. However, they only presented preliminary results and the addressed allocation issue is similar in any other centrally managed network.
Most of the work focusing on secondary/secondary inter-technology coexistence in TVWS assumed that protection of the primary technology had been met, such that the addressed coexistence issues are in fact equivalent to those in a spectrum commons.
We thus emphasize that secondary/secondary inter-technology coexistence mechanisms in the literature are similar to either those used for primary/primary or spectrum commons coexistence.

\subsubsection{\textbf{CBRS band}}
\label{CBRS_review}
As discussed in Section~\ref{reg_framework}, spectrum access in the CBRS band is managed through SASs, where there is a three-layer hierarchical regulatory framework for access rights. An example of coexisting secondary technologies is coexistence among GAA users of different technologies. This was briefly addressed in~\cite{Sahoo2017, Ying2017}, which are short poster papers that provide at most preliminary results.
The authors in~\cite{Sahoo2017} proposed schemes for fair allocation of the channels among GAA users managed by a SAS, i.e. static and max-min fair allocations. 
In~\cite{Ying2017} another scheme was proposed for the SAS to allocate channels dynamically to coexisting GAA devices, but this required the devices to perform carrier sensing and was based on graph theory and the transmission activity of each device. Examples of such users included \mbox{Wi-Fi}, \mbox{LTE-U}, LAA.
Since channel allocation is centrally performed by the SAS, we emphasize that such allocation problems are similar in any other centralized networks, e.g. cellular networks.

\subsubsection{\textbf{TVWS}}
\label{TVWS_review}
The authors in~\cite{Kang2011, Feng2013} addressed coexistence between IEEE 802.11af and IEEE 802.22 in TVWS. We note that IEEE 802.11af accesses the spectrum based on CSMA/CA, whereas IEEE 802.22 implements scheduled transmissions.
In~\cite{Kang2011} an evaluation of co-channel interference from IEEE 802.11af to 802.22 was presented, where no additional inter-technology coexistence mechanism was implemented, and it was found via OPNET simulations that the IEEE 802.22 upstream throughput was severely degraded. No results were presented for IEEE 802.11af. Also, it is not clear to which extent CSMA/CA for 802.11af was implemented in~\cite{Kang2011}. 
The authors in~\cite{Feng2013} proposed implementing a busy tone by the IEEE 802.22 nodes, in order to avoid 802.11af hidden nodes. Additionally, IEEE 802.11af compared the signal pattern of the busy tone and the 802.22 signal, in order to detect 802.22 exposed terminals. The proposed scheme was shown via simulations to provide an increase in the aggregate throughput over the case without busy tone, especially for high traffic loads.

\begin{table*}[!t]
\caption{Examples from the literature on inter-technology coexistence with primary/secondary spectrum access rights}
\label{table_review_2}
\centering
\begin{tabular}{|p{3cm}|c|p{4cm}|p{4cm}|p{3cm}|}
\hline
	\centering \textbf{Technologies} 
	& \centering \textbf{Ref.} 
	& \centering \textbf{Coexistence for Secondary Technology at Layer~0}
	& \centering \textbf{Coexistence for Secondary Technology at Layer~2} 
	& \centering\let\newline\\\arraybackslash\hspace{0pt} \textbf{Coordination based on constraints at Layer~8}\\
\hline
    shipborne radars/CBRS devices & \cite{Nguyen2017}, \cite{Palola2017} & in frequency and space based on SAS database and additional sensing network & -- & centralized\\
\hline
   (non-)governmental incumbents/LTE & \cite{Guiducci2017} & time, frequency, and space based on an LSA system & -- & centralized\\
\hline
	radar/LTE & \cite{Labib2016} &  channels restricted for indoor use & LBT, TPC, DFS & distributed\\
\hline
	radar/IoT & \cite{Khan2017} & frequency, time, and space based on REM and SA database & -- & centralized\\
\hline
	\multirow{3}{3cm}{IEEE 802.11p DSRC (ITS)/\mbox{Wi-Fi}} & \cite{Lansford2013} & -- & standardized CSMA/CA & distributed\\
\cline{2-5}
               & \cite{Naik2017} & -- & real-time channelization & distributed\\
\cline{2-5}
               & \cite{Khan2017a} & -- & LBT with lower priority; reduced \mbox{Wi-Fi} transmit power & distributed \\              	                  
\hline
	TV/LTE & \cite{Holland2014}, \cite{Ibuka2015} & frequency, time, space through database & -- & centralized\\
\hline	
	TV/IEEE 802.11af & \cite{Mizutani2015, Holland2014} & frequency, time, space through database & -- & centralized\\
\hline
    TV/next-generation cognitive radio TV (ATSC 3.0) & \cite{Rempe2017} & time, frequency, space based on database & spectrum sensing & centralized \\	                  
\hline
\end{tabular}
\end{table*}

A number of papers addressed \mbox{Wi-Fi}/LTE coexistence in TVWS~\cite{Beluri2012, Almeida2013, Cavalcante2013, Paiva2013}. Importantly, these papers did not consider the incumbent TV transmissions, so the addressed coexistence problem and the proposed solutions are the same as for \mbox{Wi-Fi}/LTE coexistence in the 5~GHz unlicensed band, which has been extensively studied in the literature and is reviewed in Section~\ref{litrev_hightraffic} and summarized in Table~\ref{table_review_4}.
In \cite{Cavalcante2013} an  evaluation of the impact of \mbox{Wi-Fi}/LTE coexistence in TVWS was presented, where LTE did not implement any inter-technology coexistence mechanism. The authors found via simulations that \mbox{Wi-Fi} was severely affected, due to its CSMA/CA mechanism through which \mbox{Wi-Fi} deferred to LTE, whereas LTE transmitted almost continuously.
By contrast, \cite{Paiva2013} evaluated the mutual interference between \mbox{Wi-Fi}/LTE at the PHY layer only (i.e. CSMA/CA was not modelled) and found via simulations that the performance of both technologies was degraded. 
The authors in~\cite{Almeida2013} simulated blank subframe allocation for LTE to coexist with \mbox{Wi-Fi} in TVWS, with fixed duty cycle and different blank subframe patterns. Their main finding was that there was a tradeoff between \mbox{Wi-Fi} and LTE performance and that duty cycle tuning depended on deployment and requirements.
The authors in~\cite{Beluri2012} proposed fixed and adaptive duty cycle for LTE when coexisting with \mbox{Wi-Fi} and compared these schemes with LBT through simulations. Additionally, LTE could select less loaded channels. The authors found that LBT was more efficient than duty cycle for high traffic load, but claimed that LBT was not justified, given that LTE would likely avoid loaded channels.  
We note that the overall results reported for \mbox{Wi-Fi}/LTE coexistence in TVWS are consistent with those for \mbox{Wi-Fi}/LTE coexistence in the 5~GHz unlicensed band in Section~\ref{litrev_hightraffic}.

\subsection{Primary/Secondary Coexistence}
\label{litPrimSec}

This section presents some representative examples from the recent literature on primary/secondary inter-technology coexistence, as summarized in Table~\ref{table_review_2}, illustrating that an increasing number of bands are opening for multiple technologies. 
This trend is relevant to our survey, as it can also potentially lead to an increasing number of coexistence cases among technologies with the same spectrum access rights, i.e. secondary/secondary, or even to opening more bands as a spectrum commons in the future.
We note that an extensive survey of primary/secondary coexistence is not our focus, but the interested reader may refer to e.g.~\cite{Paisana2014, Tehrani2016, Akyildiz2006, Akyildiz2008, Zhao2007, Wang2008, Yucek2009, Gavrilovska2014, Ren2012}.

We first present a literature overview in Section~\ref{litovprimsec}.
We then present in detail works addressing coexistence in: the CBRS band in Section~\ref{revcbrs}; LSA bands in Section~\ref{revlsa}; radar bands in Section~\ref{revradar}; the 5.9~GHz band allocated to Dedicated Short-Range Communications (DSRC) in Section~\ref{revdsrc}; and TVWS in Section~\ref{revtvws}.

\subsubsection{\textbf{Literature Overview}} 
\label{litovprimsec}
Most of the proposed spectrum sharing mechanisms are based on central coordination, due to the constraints at Layers~0~and~8. We note that for most of the primary/secondary coexistence cases, spectrum sharing is implemented at Layer~0 through a database that imposes fundamental restrictions on the way that the secondary technologies access the spectrum. However, for some coexistence cases, e.g. DSRC/\mbox{Wi-Fi}, distributed mechanisms were implemented at Layer~2. We note, however, that opening the DSRC band for \mbox{Wi-Fi} is still under discussion~\cite{FCC2016}, so it has not yet been clarified through regulation which level of protection must be offered to DSRC.

\subsubsection{\textbf{CBRS Band}}
\label{revcbrs}
The authors in~\cite{Nguyen2017} considered via simulations the case of incumbent shipborne radars coexisting with secondary CBRS devices, for which an additional sensing network had to detect the incumbents and report their presence to the SAS.
Several algorithms were proposed for determining the sensing capabilities of these sensors and their placement.
The work in~\cite{Palola2017} addressed a similar coexistence case and experimentally evaluated the evacuation time and the reconfiguration performance in the CBRS band for an SAS, where the incumbents were shipborne radars and the secondary users implemented LTE.    

\subsubsection{\textbf{LSA Bands}}
\label{revlsa}
The first large-scale LSA implementation was presented in~\cite{Guiducci2017}, where an LTE deployment coexisted with several incumbents in the 2.3--2.4~GHz band (e.g. fixed services, Programme-Making and Special Events -- PMSE -- video links). Several drive tests and simulations were conducted for functional and regulatory compliance verification. 

\subsubsection{\textbf{Radar Bands}}
\label{revradar}
The authors in~\cite{Labib2016} reviewed the spectrum sharing techniques imposed by regulators for LTE (but also valid for any other technology) to coexist with radars in the 5~GHz band. 
The work in~\cite{Khan2017} proposed coexistence of IoT and rotating radars through an SAS with radio environmental maps, which shared the spectrum in frequency, time, and space. Results from a measurement campaign on spectrum usage by rotating radars was presented, in order to show the coexistence potential with IoT. 

\subsubsection{\textbf{DSRC/Wi-Fi}}
\label{revdsrc}
The authors in~\cite{Lansford2013, Naik2017, Khan2017a} addressed coexistence between DSRC (or Intelligent Transportation Systems -- ITS) devices and \mbox{Wi-Fi} in the 5.9 GHz band, which is currently under consideration for becoming open to \mbox{Wi-Fi} operations. 
In~\cite{Lansford2013} potential DSRC/\mbox{Wi-Fi} coexistence issues were discussed, if \mbox{Wi-Fi} implemented its original CSMA/CA coexistence mechanism. 
The authors in~\cite{Naik2017} proposed a real-time channelization algorithm for IEEE 802.11ac \mbox{Wi-Fi} to coexist with DSRC devices, where the \mbox{Wi-Fi} APs selected a primary channel and bandwidth, such that the \mbox{Wi-Fi} throughput was maximized. Both experimental and simulation results were presented, showing that the \mbox{Wi-Fi} throughput was increased via the proposed scheme compared to static channel allocation.
In \cite{Khan2017a} the performance of two mechanisms proposed in Europe for \mbox{Wi-Fi} to coexist with ITS was evaluated via simulations. Both mechanisms were based on LBT, as follows: the first mechanism used higher-duration sensing parameters when detecting ITS; the second mechanism probed for hidden ITS stations, and was able to vacate the channel. The authors found that there were three ITS transmitter-receiver distance ranges, corresponding to different coexistence characteristics: for short distances there were no coexistence problems; for medium distances outdoor \mbox{Wi-Fi} coexisted better than indoor \mbox{Wi-Fi}; and for long distances the ITS packet loss was high, but this was not considered problematic for safety applications.

\subsubsection{\textbf{TVWS}}
\label{revtvws}
The work in~\cite{Mizutani2015, Holland2014, Ibuka2015, Rempe2017} addressed coexistence with primary TV services, where spectrum resources were shared through a centralized database, in frequency, time, and space domains.
The authors in~\cite{Mizutani2015, Holland2014, Ibuka2015} experimentally verified the correct operation and performance of IEEE 802.11af and/or LTE in TVWS, through Ofcom's TVWS trial pilot program.
In~\cite{Rempe2017} a different topic was addressed, i.e. an implementation of a next-generation cognitive radio TV based on the ATSC 3.0 standard, coexisting with legacy TV devices. We note that~\cite{Rempe2017} is a short poster paper that did not present performance evaluation results. 

\subsection{Summary \& Insights}
\label{hier_frame_sum}

There are few works that have addressed inter-technology coexistence with equal spectrum access rights within a hierarchical regulatory framework, i.e. primary/primary and secondary/secondary. Specifically, for primary/primary coexistence only integrated LTE/NB-IoT deployments have been considered, where centralized spectrum sharing mechanisms were implemented similarly to single-technology cellular networks. Resource blanking, i.e. sharing in time and frequency with fine granularity, was found efficient. We note that this technique has the advantage of already being standardized for LTE. Furthermore, ACI from LTE had a negligible effect on the network performance. 

For secondary/secondary coexistence, the spectrum sharing mechanisms were proposed to be implemented either in a centralized manner via databases, or in a distributed manner as for a spectrum commons like the unlicensed bands. We note that centralized spectrum sharing can be applied in a straightforward way due to the requirement that secondary devices cooperate in any case with the database, in order to protect the primary. However, managing resource allocation also among secondary devices increases the computational effort for the database and is less dynamic with respect to the offered traffic, i.e. higher delays are expected due to the communication overhead between secondary devices and the database. 
By contrast, distributed spectrum sharing is implemented directly in the wireless secondary devices. It was found that sharing in frequency (i.e. channel selection) is an efficient way to protect different technologies from each other. For co-channel coexistence, LBT performed better than duty cycling for high traffic load, but duty cycling may be sufficient. We note that the choice of implementing LBT or duty cycling may also depend on the required changes for existing standards. For instance, \mbox{Wi-Fi} already implements CSMA/CA as an LBT variant, whereas for LTE rather complex changes were needed to implement LBT in LAA. Implementing duty cycling for LTE via the already standardized resource blanking is a more straightforward technical solution. 

Primary/secondary coexistence is not the focus of this survey, but a few examples from the literature were presented, in order to show the large number of bands that are being targeted by multiple technologies, where coexistence with the same access rights may also become an issue in the future. Such bands are the 3.5~GHz CBRS band in the U.S., the 2.3-2.4~GHz LSA band in Europe, 5~GHz radar bands, the 5.9 GHz DSRC band, and TVWS. Spectrum sharing for primary/secondary coexistence was implemented either in a centralized manner via databases, or via distributed sensing mechanisms. Sharing via databases is considered safer for protecting primary technologies in TVWS, the CBRS band, or the LSA bands, especially since different technologies may be deployed as secondary ones. By contrast, secondary \mbox{Wi-Fi} devices could protect primary DSRC devices by implementing distributed channel selection or sensing mechanisms, as \mbox{Wi-Fi} already implements variants of such mechanisms.

\begin{table*}[!t]
\modcounter
\caption{Literature review of inter-technology spectrum sharing with low-traffic technologies in a spectrum commons}
\label{table_review_3}
\centering
\begin{tabular}{|p{2.3cm}|p{1cm}|p{5cm}|p{4cm}|p{3cm}|}
\hline
     \centering \textbf{Technologies} 
     & \centering \textbf{Ref.} 
     & \centering \textbf{Coexistence at Layer~2} 
     & \centering \textbf{Coexistence at Layer~1}
     & \centering\let\newline\\\arraybackslash\hspace{0pt} \textbf{Coordination at Layer~2 based on constraints at Layer~8}\\
\hline     
    \multirow{6}{2.3cm}{\mbox{Wi-Fi}/ IEEE~802.15.4} & \cite{Pollin2008, Yuan2007, SikoraOttawa2005, Shuaib2006, PetrovaLasVegas2006} &  \textbf{\mbox{Wi-Fi}}: CSMA/CA; \textbf{802.15.4}: CSMA/CA & -- & distributed \\
\cline{2-5}
                    & \cite{Angrisani2008} &  \textbf{\mbox{Wi-Fi}}: CSMA/CA; \textbf{802.15.4}: CSMA/CA, polling & -- & distributed \\
\cline{2-5}
                    & \cite{Howitt2003} & \textbf{both}: frequency selection & -- & distributed or centralized \\ 
\cline{2-5}
                    & \cite{Won2005} & \textbf{\mbox{Wi-Fi}}: CSMA/CA; \textbf{802.15.4}: adaptive channel allocation & -- & local coordination for 802.15.4\\                    
\cline{2-5}
                    & \cite{Petrova2007} &  \textbf{both}: CSMA/CA, static channel allocation & \textbf{\mbox{Wi-Fi}}: beamforming & distributed \\
\cline{2-5}
                    & \cite{Hauer2009} &  \textbf{\mbox{Wi-Fi}}: CSMA/CA; \textbf{802.15.4}: CSMA/CA, adaptive power control & -- & distributed \\                                                                                	
\hline  
	\multirow{6}{2.3cm}{\mbox{Wi-Fi}/Bluetooth} & \cite{Shuaib2006, Sydanheimo2002, Lansford2001, Golmie2003a} & \textbf{\mbox{Wi-Fi}}: CSMA/CA & \textbf{Bluetooth}: FHSS & distributed \\
\cline{2-5} 
                     & \cite{Chiasserini2002} &  \textbf{\mbox{Wi-Fi}}: CSMA/CA; \textbf{both}: MAC traffic scheduling & \textbf{Bluetooth}: FHSS & collaborative or non-collaborative \\ 	                 
\cline{2-5}
                     & \cite{Golmie2003s} &  \textbf{\mbox{Wi-Fi}}: CSMA/CA; \textbf{Bluetooth}: scheduling & \textbf{Bluetooth}:  adaptive FHSS & Bluetooth local coordination \\ 	                 
\cline{2-5} 
                     & \cite{Park2003} & -- & \textbf{\mbox{Wi-Fi}}: coded OFDM; \textbf{Bluetooth}: FHSS & distributed \\  
\cline{2-5}
                     & \cite{Arumugam2003} & -- &  \textbf{\mbox{Wi-Fi}}: weighing sub-carriers; \textbf{Bluetooth}: FHSS, antenna diversity & distributed \\  
\cline{2-5}
                     & \cite{Ghosh2003} & -- & \textbf{\mbox{Wi-Fi}}: interference cancellation against Bluetooth & distributed \\ 
\hline
	IEEE~802.15.4/ Bluetooth & \cite{SikoraOttawa2005} & \textbf{802.15.4}: CSMA/CA & \textbf{Bluetooth}: FHSS & distributed \\
\hline
	IEEE 802.15.4/ microwave oven & \cite{SikoraOttawa2005} & \textbf{802.15.4}: CSMA/CA & -- & distributed \\
\hline
	Bluetooth/ \{WCAM, RFID, microwave oven\} & \cite{Sydanheimo2002} & -- & \textbf{Bluetooth}: FHSS & distributed \\
\hline	
    \mbox{Wi-Fi}/LTE D2D & \cite{Wu2016} & \textbf{LTE D2D}: LBT, interference avoidance routing, switch to licensed band & -- & distributed\\
\hline
    5G/IEEE~802.15.4 & \cite{Lackpour2017} & -- & \textbf{5G}: non-contiguous-OFDM, reconfigurable antennas & distributed \\ 
\hline         
    LTE/ZigBee & \cite{Parvez2016} & \textbf{LTE}: two 0.5 ms guard periods per frame; \textbf{802.15.4}: CSMA/CA  & -- & distributed \\                                
\hline
    IEEE 802.15.4/any interfering signal & \cite{Vermeulen2017} & -- & \textbf{802.15.4:} collision detection at transmitter with full duplex (self-interference cancellation) & distributed \\ 	                 
\hline 
\end{tabular}
\end{table*}

\begin{table*}[!t]
\addtocounter{table}{-1}
\modcounter
\caption{Inter-technology coexistence goals and performance for literature review of spectrum sharing with low-traffic technologies in a spectrum commons in Table~\ref{table_review_3}}
\label{table_review_3b}
\centering
\begin{tabular}{|p{2.3cm}|p{4cm}|p{2.2cm}|p{3cm}|p{3.5cm}|}
\hline
     \centering \multirow{2}{*}{\textbf{Technologies}} 
     & \centering \multirow{2}{*}{\textbf{Coexistence Goals}} 
     & \multicolumn{3}{c|}{\textbf{Performance Evaluation}}\\
\cline{3-5}
     &      
     & \centering \textbf{method} 
     & \centering \textbf{metric} 
     & \centering\let\newline\\\arraybackslash\hspace{0pt} \textbf{network size} \\    
\hline     
    \begin{tabular}{p{2cm}}
    \mbox{Wi-Fi}/ IEEE~802.15.4 \\ \\ \\ \\
    \scriptsize\cite{Pollin2008, Howitt2003, Yuan2007, SikoraOttawa2005, PetrovaLasVegas2006, Petrova2007, Hauer2009, Shuaib2006, Angrisani2008, Won2005}
    \end{tabular}
    & \begin{tabular}{p{3.3cm}}
      \emph{Impact on \mbox{Wi-Fi}}\\
      $-$\textbf{(implicitly) vs. standalone} \scriptsize\cite{Pollin2008, Howitt2003, Shuaib2006, Angrisani2008, PetrovaLasVegas2006} \\
      \hline
      \emph{Impact on IEEE 802.15.4}\\
      $-$\textbf{(implicitly) vs. standalone} \scriptsize\cite{Yuan2007, SikoraOttawa2005, PetrovaLasVegas2006, Petrova2007, Hauer2009, Shuaib2006, Angrisani2008, Won2005} \\
      \hline
      \emph{Other}\\
      $-$\textbf{\mbox{Wi-Fi} packet error rate below 8\%} \scriptsize\cite{Howitt2003} \\
      $-$\textbf{solve performance degradation of 802.15.4} \scriptsize\cite{Won2005}
      \end{tabular}
    & \begin{tabular}{p{2cm}}
      $-$\textbf{measurements} \scriptsize\cite{Pollin2008, SikoraOttawa2005, PetrovaLasVegas2006, Petrova2007, Hauer2009, Shuaib2006, Angrisani2008, Won2005}\\ \\
      $-$\textbf{analytical} \scriptsize\cite{Howitt2003, Yuan2007}\\ \\
      $-$\textbf{simulations} \scriptsize\cite{Yuan2007, Won2005}
      \end{tabular}
    & \begin{tabular}{p{2.8cm}}
      $-$\textbf{throughput} \scriptsize\cite{Pollin2008, Yuan2007, Shuaib2006} \\
      $-$\textbf{packet error rate/loss} \scriptsize\cite{Pollin2008, Howitt2003, SikoraOttawa2005, Hauer2009, Angrisani2008, PetrovaLasVegas2006}\\
      $-$\textbf{packet delivery ratio/success rate} \scriptsize\cite{Petrova2007, Hauer2009, Won2005}\\
      $-$\textbf{received power} \scriptsize\cite{Petrova2007}\\
      $-$\textbf{channel power} \scriptsize\cite{Angrisani2008}\\
      $-$\textbf{SIR} \scriptsize\cite{Angrisani2008}\\
      $-$\textbf{delay} \scriptsize\cite{Won2005}
      \end{tabular}
    & \begin{tabular}{p{3.3cm}}
      $-$~\textbf{1 link of each technology} \scriptsize\cite{Pollin2008, Yuan2007, SikoraOttawa2005, PetrovaLasVegas2006, Petrova2007, Hauer2009, Shuaib2006, Won2005} \\ \\
      $-$~\textbf{1 \mbox{Wi-Fi} link \& several 802.15.4 devices} \scriptsize\cite{Howitt2003, Angrisani2008}\\ \\
      $-$~\textbf{100 802.15.4 devices and abstract interference} \scriptsize\cite{Won2005}
      \end{tabular} \\
\hline  
	\begin{tabular}{p{2cm}} 
	\mbox{Wi-Fi}/Bluetooth \\ \\ \\ \\
	\scriptsize\cite{Shuaib2006, Sydanheimo2002, Lansford2001, Golmie2003a, Chiasserini2002, Golmie2003s, Park2003, Arumugam2003, Ghosh2003}
	\end{tabular}
	& \begin{tabular}{p{3.3cm}}
      \emph{Impact on \mbox{Wi-Fi}}\\
      $-$\textbf{vs. standalone} \scriptsize\cite{Shuaib2006, Sydanheimo2002, Lansford2001, Golmie2003a, Park2003, Arumugam2003} \\
      $-$\textbf{vs. coexistence without additional spectrum sharing mechanisms} \scriptsize\cite{Chiasserini2002, Golmie2003s, Ghosh2003}\\
      \hline
      \emph{Impact on Bluetooth}\\
      $-$\textbf{vs. standalone} \scriptsize\cite{Shuaib2006, Sydanheimo2002, Lansford2001, Golmie2003a, Arumugam2003} \\
      $-$\textbf{vs. coexistence without additional spectrum sharing mechanisms} \scriptsize\cite{Chiasserini2002, Golmie2003s} \\
      $-$\textbf{vs. other coexistence mechanisms} \scriptsize\cite{Golmie2003s}\\
      \end{tabular}
    & \begin{tabular}{p{2cm}}
      $-$\textbf{measurements} \scriptsize\cite{Shuaib2006, Sydanheimo2002, Lansford2001}\\ \\
      $-$\textbf{analytical} \scriptsize\cite{Park2003}\\ \\
      $-$\textbf{simulations} \scriptsize\cite{Lansford2001, Golmie2003a, Chiasserini2002, Golmie2003s, Arumugam2003, Ghosh2003}
      \end{tabular}
    & \begin{tabular}{p{2.8cm}}
      $-$\textbf{throughput} \scriptsize\cite{Shuaib2006, Sydanheimo2002, Lansford2001} \\
      $-$\textbf{packet error rate/loss} \scriptsize\cite{Sydanheimo2002, Golmie2003a, Golmie2003s, Arumugam2003, Ghosh2003}\\
      $-$\textbf{delay} \scriptsize\cite{Chiasserini2002, Golmie2003s}\\
      $-$\textbf{jitter} \scriptsize\cite{Golmie2003s} \\
      $-$\textbf{goodput} \scriptsize\cite{Chiasserini2002, Golmie2003s} \\
      $-$\textbf{bit error probability} \scriptsize\cite{Park2003}
      \end{tabular}
    & \begin{tabular}{p{3.3cm}}
    $-$~\textbf{1 link of each technology} \scriptsize\cite{Shuaib2006, Sydanheimo2002, Lansford2001, Park2003, Arumugam2003, Ghosh2003}\\ \\
    $-$~\textbf{1 Bluetooth link \& up to 2 \mbox{Wi-Fi} links} \scriptsize\cite{Golmie2003s}\\ \\
    $-$~\textbf{up to 10 \mbox{Wi-Fi} devices and several Bluetooth links} \scriptsize\cite{Golmie2003a, Chiasserini2002}
      \end{tabular} \\
\hline
	\begin{tabular}{p{2cm}} 
	IEEE~802.15.4/ Bluetooth \\
	\scriptsize\cite{SikoraOttawa2005}
	\end{tabular}
	 & study mutual impact on both technologies (implicitly vs. standalone) & measurements & packet loss & two Bluetooth links and one 802.15.4 link \\
\hline
	\begin{tabular}{p{2cm}}
	IEEE 802.15.4/ microwave oven \\ 
	\scriptsize\cite{SikoraOttawa2005} 
	\end{tabular} 
	& study impact on 802.15.4 (implicitly vs. standalone) & measurements & packet loss & one 802.15.4 link and one microwave oven \\
\hline
	\begin{tabular}{p{2cm}}
	Bluetooth/ \{WCAM, RFID, microwave oven\} \\
	 \scriptsize\cite{Sydanheimo2002} 
	 \end{tabular}
	 & study impact on Bluetooth vs. standalone & measurements & data rate, packet error rate & one Bluetooth link and one interferer of another technology \\
\hline	
    \begin{tabular}{p{2cm}}
    \mbox{Wi-Fi}/LTE D2D \\
    \scriptsize\cite{Wu2016}
    \end{tabular}
    & increase D2D throughput vs. different licensed/unlicensed spectrum use strategies & simulations & throughput & one \mbox{Wi-Fi} link and one multi-hop D2D flow \\
\hline
    \begin{tabular}{p{2cm}}5G/IEEE~802.15.4 \\
    \scriptsize\cite{Lackpour2017} 
    \end{tabular}
    & mitigate mutual interference vs. standalone \& vs. coexistence with 5G without spectrum sharing mechanisms & simulations & throughput & one ZigBee and one 5G link\\ 
\hline         
    \begin{tabular}{p{2cm}}LTE/ZigBee \\ 
    \scriptsize\cite{Parvez2016} 
    \end{tabular}
    & study mutual impact between LTE and ZigBee vs. standalone & simulations & throughput, SINR & 18 LTE BSs and 54 ZigBee APs\\                                
\hline
    \begin{tabular}{p{2cm}}
    IEEE 802.15.4/any interfering signal \\
    \scriptsize\cite{Vermeulen2017} 
    \end{tabular}
     & detect collisions while transmitting & measurements & detection and false alarm probabilities & one 802.15.4 link and one 802.15.4 interferer\\ 	                 
\hline 
\end{tabular}
\end{table*}

\section{Literature Review of Inter-Technology Coexistence in a Spectrum Commons}
\label{litSpecComm}

This section presents a review of the literature addressing inter-technology coexistence in a spectrum commons. We focus on the unlicensed bands as an example of a spectrum commons where the most diverse interactions between technologies occur, due to the largely technology-agnostic regulatory framework that allows any technology to operate in these bands without license costs, provided that the technical regulatory constraints at Layer~0 are met. 
We classify the abundant literature on this topic according to the Application-layer criteria in Table~\ref{table_app}, i.e. work that addresses: (i)~coexistence with low traffic technologies in Section~\ref{litrev_lowtraffic}, Tables~\ref{table_review_3} to \ref{table_review_3b}; and (ii)~coexistence among high traffic technologies in Section~\ref{litrev_hightraffic}, Tables~\ref{table_review_4} to \ref{table_review_4d}.  
In Section~\ref{sumSpecComm} we summarize and discuss the main findings.

\subsection{Coexistence with Low Traffic Technologies}
\label{litrev_lowtraffic}
We first present in Section~\ref{over_lowtraffic} an overview of our literature review on coexistence with low traffic technologies.
We then review in detail work on: (i)~IEEE~802.11 \mbox{Wi-Fi}/IEEE 802.15.4 in Section~\ref{litrev_802.15.4}; (ii)~IEEE~802.11 \mbox{Wi-Fi}/Bluetooth in Section~\ref{rev_bluetooth}; and (iii)~other technologies in Section~\ref{rev_other}.
Table~\ref{table_review_3} summarizes the spectrum sharing mechanisms and Table~\ref{table_review_3b} summarizes coexistence performance evaluation aspects, where \emph{standalone} is sometimes considered as a baseline case where there is no other coexisting technology present.    

\subsubsection{\textbf{Literature Overview}}
\label{over_lowtraffic}
For coexistence with low traffic technologies in a spectrum commons, a similar number of works considered spectrum sharing mechanisms at Layer~2 as at Layer~1 (\emph{cf.} Table~\ref{table_review_3}). This shows the importance of Layer~1 techniques for mitigating interference, especially for coexistence cases where at least one technology carries a low traffic volume.
Furthermore, most of the work assumed distributed spectrum sharing mechanisms at Layer~2 as influenced by ownership at Layer~8, as expected in a spectrum commons.
In terms of coexistence goals (\emph{cf.} Table~\ref{table_review_3b}), most of the works compared the coexistence performance with either the standalone case, or coexistence without additional spectrum sharing mechanisms. We note that such an approach does not facilitate the performance comparison of different mechanisms among themselves, so that selecting an efficient mechanism for future coexistence cases is not straightforward.
The preferred performance evaluation methods were measurements and simulations. We emphasize that conducting measurements was facilitated by the existence of commercially available hardware (for e.g. Bluetooth, \mbox{Wi-Fi}, and IEEE 802.15.4), especially for works that did not propose new coexistence mechanisms. However, most of the work based on measurements considered very simplistic deployments of one link for each technology.

\subsubsection{\textbf{IEEE 802.11 \mbox{Wi-Fi}/IEEE 802.15.4}}
\label{litrev_802.15.4}
Coexistence between these technologies was addressed in~\cite{Won2005, Yuan2007, Pollin2008, SikoraOttawa2005, Angrisani2008, PetrovaLasVegas2006, Petrova2007, Shuaib2006, Hauer2009, Howitt2003}.
The authors in~\cite{Pollin2008, Yuan2007, SikoraOttawa2005, Shuaib2006,  PetrovaLasVegas2006} addressed coexistence for basic standardized specifications, whereas~\cite{Howitt2003, Angrisani2008, Won2005, Petrova2007, Hauer2009} evaluated or proposed more advanced features to mitigate interference.

Specifically, the authors in~\cite{Pollin2008, PetrovaLasVegas2006} measured the impact of IEEE 802.15.4 on \mbox{Wi-Fi} performance. In~\cite{Pollin2008} it was found that the \mbox{Wi-Fi} throughput significantly decreased when the IEEE 802.15.4 transmitter was located close to the \mbox{Wi-Fi} receiver, due to the slow responsiveness of IEEE 802.15.4 when sensing the channel, which resulted in collisions. For other location configurations, both~\cite{Pollin2008, PetrovaLasVegas2006} found that the \mbox{Wi-Fi} packet loss was only marginally increased by coexistence, due to the much higher transmit power of \mbox{Wi-Fi} vs. IEEE 802.15.4. 

The works in~\cite{Yuan2007, SikoraOttawa2005, Shuaib2006, PetrovaLasVegas2006, Angrisani2008} reported complementary results, i.e. that the IEEE 802.15.4 performance in terms of throughput and packet loss rate degraded significantly when coexisting with \mbox{Wi-Fi}, especially for high \mbox{Wi-Fi} load. This was explained by the higher transmit power, higher sensing threshold, and shorter backoff time slot for \mbox{Wi-Fi} vs. IEEE 802.15.4. 
Also, \cite{Shuaib2006} reported that the \mbox{Wi-Fi} performance was affected by Bluetooth more than by IEEE~802.15.4. Although this effect was not explained in~\cite{Shuaib2006}, it was likely caused by Bluetooth frequency hopping, \emph{cf.} Section~\ref{rev_bluetooth}.

Two solutions were evaluated in~\cite{Angrisani2008} to improve the performance of IEEE~802.15.4 when coexisting with \mbox{Wi-Fi}: reducing the \mbox{Wi-Fi} duty cycle (i.e. the duration of a frame vs. total time between two frames) by reducing the \mbox{Wi-Fi} packet size, or increasing the time duration of the IEEE~802.15.4 polling window.
We note, however, that adjusting the \mbox{Wi-Fi} packet size is not a practical solution, especially since this also depends on the application type, which is not controlled by the network manager, but by the end user. As such, adjusting the IEEE~802.15.4 polling window could be more feasible for real deployments.  

The authors in~\cite{Howitt2003, Won2005, Petrova2007} considered different channel selection schemes for enabling coexistence.
Specifically, in~\cite{Howitt2003} the impact of IEEE 802.15.4 on 802.11b was evaluated with generic frequency management and it was reported that \mbox{Wi-Fi} was only marginally affected when the channels were allocated such that the inter-technology interference was reduced.
In~\cite{Won2005} an adaptive channel allocation scheme was proposed for multi-hop IEEE 802.15.4 networks, in order to protect them from IEEE 802.11b. The scheme required local coordination among IEEE 802.15.4 nodes, which temporarily formed a group and changed their channel if a high level of interference was detected. This scheme was found to be effective for improving IEEE 802.15.4 coexistence performance especially in large-scale networks. 
An experimental evaluation was presented in~\cite{Petrova2007}, which focused on the coexistence impact of IEEE 802.11g/n on 802.15.4 networks. Overlapping and non-overlapping channel configurations were considered and it was reported that the IEEE 802.15.4 network severely suffered in case of high co-channel \mbox{Wi-Fi} traffic load and that interference from adjacent channels may also be critical. 
This shows overall that spectrum sharing in frequency is efficient for enabling inter-technology coexistence, but this technique requires a larger portion of spectrum,  where multiple non-overlapping channels can be accommodated and where ACI from \mbox{Wi-Fi} is not negligible.  
Also, the extent to which \mbox{Wi-Fi} beamforming decreased the IEEE~802.15.4 packet delivery ratio differed greatly depending on the beam orientations~\cite{Petrova2007}. This suggests that SDMA via beamforming at PHY cannot be used as a stand-alone spectrum sharing technique, especially for wireless networks with mobile nodes, where different beams may be oriented in the same direction. However, beamforming can be used as an additional spectrum sharing technique to improve the coexistence performance for deployments with enough spatial separation between interfering devices.

Unlike previous work, \cite{Hauer2009} focused on the impact of IEEE~802.11b \mbox{Wi-Fi} on 802.15.4 body area networks and found that the 802.15.4 packet loss was significantly affected only for the very low power regime. Adaptive power control was suggested as a solution.   

\subsubsection{\textbf{IEEE 802.11 \mbox{Wi-Fi}/Bluetooth}}
\label{rev_bluetooth}

Coexistence between these technologies was addressed in~\cite{Sydanheimo2002, Lansford2001, Golmie2003a, Chiasserini2002, Golmie2003s, Shuaib2006, Park2003, Arumugam2003, Ghosh2003}. The authors in~\cite{Shuaib2006, Sydanheimo2002, Lansford2001, Golmie2003a} assumed standard specifications, whereas in~\cite{Chiasserini2002, Golmie2003s, Park2003, Arumugam2003, Ghosh2003} advanced features were proposed.

The authors in~\cite{Sydanheimo2002} measured the impact of mutual interference between IEEE 802.11b and Bluetooth. They found that the decrease in data rate was in general tolerable for both technologies. 
In~\cite{Lansford2001} it was found through simulations and measurements that Bluetooth was less affected by \mbox{Wi-Fi} than vice-versa, for closely spaced \mbox{Wi-Fi} and Bluetooth links. This showed that the FHSS technique implemented by Bluetooth is quite effective when the hopping channels cover a wider band than a \mbox{Wi-Fi} channel. Also, the CSMA/CA MAC was not as efficient at mitigating interference that occurred with a high hopping rate. 
Consistently, \cite{Golmie2003a} reported that a slower Bluetooth hopping rate caused less interference to \mbox{Wi-Fi}. Furthermore, increasing the \mbox{Wi-Fi} transmit power did not reduce the \mbox{Wi-Fi} packet loss, so lower transmit power was found to be desirable.   
 
The authors in~\cite{Chiasserini2002} proposed two MAC traffic scheduling algorithms to cope with the interference between DSSS-based IEEE 802.11 (i.e. IEEE 802.11b) and Bluetooth: the first algorithm scheduled and adjusted the \mbox{Wi-Fi} packets when coexisting with Bluetooth voice links, whereas the second one adjusted Bluetooth packets for data links when coexisting with \mbox{Wi-Fi}. 
Both schemes reportedly require only slight modifications of the IEEE 802.11 and Bluetooth standards. 
The simulation results showed a significant increase in goodput for both technologies.
However, these schemes require \mbox{Wi-Fi} and Bluetooth to have information about each other's traffic. Although~\cite{Chiasserini2002} suggested that both collaborative information exchange and non-collaborative sensing and interference pattern recognition are possible solutions, it may be difficult to implement either of them in practice, especially if multiple devices are active. 
Another scheduling scheme was considered in~\cite{Golmie2003s}, which postponed Bluetooth transmissions until a time slot associated with a good-quality frequency channel. This was compared with an adaptive frequency hopping mechanism for Bluetooth, which avoided channels used by \mbox{Wi-Fi}.
The proposed frequency hopping scheme required Bluetooth specification modifications and was found to be more suitable for environments where the interference conditions did not change fast, such that the same hopping sequence could be used for longer. By contrast, the scheduling scheme was found to be more suitable for the opposite case and did not require specification modifications.   

The authors in~\cite{Park2003, Arumugam2003, Ghosh2003} considered PHY techniques for coexistence between Bluetooth and OFDM-based \mbox{Wi-Fi}, i.e. IEEE~802.11g.
Specifically, \cite{Park2003} found through an analytical model that coding significantly decreased the bit error probability for \mbox{Wi-Fi}, when interference from Bluetooth occured.  
Also, \cite{Arumugam2003} found that the packet error rate could be decreased for both technologies through antenna diversity for Bluetooth, and through weighing of bits according to the interference level of the respective subcarriers for \mbox{Wi-Fi}.  
Finally, an interference cancellation technique was proposed for \mbox{Wi-Fi} in~\cite{Ghosh2003}, where the multipath channel and interference characteristics were estimated, in order to reduce the impact of interference from Bluetooth. A reported advantage was the potentially higher throughput compared to MAC schemes, since \mbox{Wi-Fi} could operate simultaneously with Bluetooth. However, the proposed PHY scheme was only evaluated for a single link of each technology, so it is unclear what its performance in realistic, larger deployments is.    
We note that features like coding and using antenna diversity are already a part of modern wireless communication standards, i.e. Wi-Fi and LTE.

\subsubsection{\textbf{Other Technologies}}
\label{rev_other}
Coexistence between other technologies where at least one of them is low-traffic was addressed in~\cite{Wu2016, Lackpour2017, Parvez2016, SikoraOttawa2005, Vermeulen2017, Sydanheimo2002}.

The authors in~\cite{SikoraOttawa2005} evaluated the performance of IEEE~802.15.4 when coexisting with Bluetooth or microwave ovens and reported that IEEE 802.15.4 was only marginally affected in terms of packet loss.
The work in~\cite{Sydanheimo2002} reported measurement results for Bluetooth coexisting with a wireless camera (WCAM), RFID, and a microwave oven and showed that the data rate of Bluetooth could be significantly reduced, especially for short distances between Bluetooth devices and coexisting devices of a different technology.
This shows that the spatial separation has a significant impact on the performance of low-power networks.

The routing performance of LTE-based multi-hop D2D communications coexisting with \mbox{Wi-Fi} in the unlicensed band was investigated in~\cite{Wu2016}. Three coexistence mechanisms were considered for D2D: LBT with sensing until the channel is available; interference avoidance routing (i.e. routing around \mbox{Wi-Fi}, so as to avoid contention); and switching to the licensed cellular band. The authors found that LTE-based D2D in the unlicensed band could increase the LTE network-wide capacity, but suggested that efficient algorithms to select the D2D transmission time are needed, as they may impact \mbox{Wi-Fi} negatively. 

The authors in~\cite{Lackpour2017, Vermeulen2017} proposed PHY techniques for coexistence with IEEE 802.15.4.
In~\cite{Lackpour2017} non-contiguous-OFDM and reconfigurable antennas were proposed for 5G to coexist with IEEE~802.15.4, whereas in~\cite{Vermeulen2017} self-interference cancellation with an in-band full duplex radio was proposed for IEEE 802.15.4 transmitters, in order to stop transmission in case of collision with any other signal and thus save energy.
We note, however, that \cite{Lackpour2017, Vermeulen2017} only considered one link of each coexisting technology, so it is unclear what the performance of these techniques is in realistic deployments with multiple active links. 
The coexistence performance of LTE and ZigBee (i.e. IEEE 802.15.4 at MAC and PHY layers) was evaluated in~\cite{Parvez2016}, for the 2.4~GHz band. Two guard periods were proposed in each LTE frame, so that ZigBee could access the channel. The authors found that ZigBee's performance was degraded more than that of LTE, but that the requirements for smart meter communications with ZigBee were still met. 
This shows the efficiency of time-sharing schemes.

\begin{table*}[!t]
\addtocounter{mysubtable}{-2}
\modcounter
\caption{Literature review of inter-technology spectrum sharing among broadband technologies in a spectrum commons}
\label{table_review_4}
\centering
\begin{tabular}{|p{2.2cm}|p{1.5cm}|p{7cm}|p{1.8cm}|p{3cm}|}
\hline
 \centering \textbf{Technologies} 
 & \centering \textbf{Ref.} 
 & \centering \textbf{Coexistence at Layer~2} 
 & \centering \textbf{Coexistence at Layer~1}
 & \centering\let\newline\\\arraybackslash\hspace{0pt} \textbf{Coordination at Layer~2 based on constraints at Layer~8}\\
\hline     
	\mbox{Wi-Fi} EDCA/DCF & \cite{Hwang2006, Bianchi2005} & \textbf{both}: CSMA/CA with different sensing time & -- & distributed \\
\hline
	\multirow{2}{2.2cm}{\mbox{Wi-Fi}/ IEEE~802.16} & \cite{Fu2007} & \textbf{\mbox{Wi-Fi}}: CSMA/CA, transmit power; \textbf{802.16}: transmit power & \textbf{both}: modulation & distributed \\
\cline{2-5}
                     & \cite{Berlemann2006} & \textbf{\mbox{Wi-Fi}}: CSMA/CA; \textbf{802.16}: channel blocking, ordering contention slots  & -- & distributed \\ 	                 
\hline
    \multirow{6}{2.2cm}{\mbox{Wi-Fi}/LTE} & \cite{Babaei2014, Jian2015, Capretti2016, Gomez-Miguelez2016, BhorkarNewOrleans2015, Jeon2014, SagariLondon2015, ChenGlasgowMay2015, Nhtilae2013, Rupasinghe2014, VoicuLondon2015, Sagari2015, FuadM.Abinader2014, Li2016a, Voicu2016, Jian2017, JiaLondon2015, Bhorkar2014} & \textbf{\mbox{Wi-Fi}}: CSMA/CA; \textbf{LTE}: none & -- & distributed \\	 
\cline{2-5}
                     & \cite{SagariLondon2015, VoicuLondon2015} & \textbf{\mbox{Wi-Fi}}: CSMA/CA; \textbf{both}: channel allocation (random~\cite{SagariLondon2015, VoicuLondon2015}; graph coloring~\cite{SagariLondon2015}; avoid occupied channels~\cite{VoicuLondon2015}) & -- & distributed\cite{SagariLondon2015, VoicuLondon2015}; coordinated\cite{SagariLondon2015} \\ 
\cline{2-5}
                     & \cite{Hajmohammad2013, Cai2016} & \textbf{both}: spectrum splitting between technologies (subcarrier granularity~\cite{Cai2016}) & -- & likely cooperative~\cite{Hajmohammad2013}; cooperative~\cite{Cai2016} \\
\cline{2-5}
                     & \cite{Chaves2013} & \textbf{\mbox{Wi-Fi}}: CSMA/CA; \textbf{LTE}: power control in the uplink & -- & distributed \\     
\cline{2-5}      
                     & \cite{Yun2015} & \textbf{\mbox{Wi-Fi}}: modified CSMA/CA & \textbf{both}: decoding & colocated LTE and \mbox{Wi-Fi} receivers \\  
\cline{2-5}
                     & \cite{Li2016c} & \textbf{\mbox{Wi-Fi}}: CSMA/CA & \textbf{LTE}: beamforming & distributed; LTE nodes also have 802.11 receivers \\                                   
\hline
	 \multirow{7}{2.2cm}{\mbox{Wi-Fi}/LBT-LTE} & \cite{Li2016, VoicuLondon2015, Cano2016a, Sandoval2016} & \textbf{\mbox{Wi-Fi}}: CSMA/CA; \textbf{LTE}: \emph{generic LBT} -- different ED thresholds~\cite{Li2016}, ideal MAC and different channel selection schemes (random, least interfered)~\cite{VoicuLondon2015}, ETSI LBE~\cite{Cano2016a} & optimized topologies~\cite{Sandoval2016} & distributed \cite{Li2016, VoicuLondon2015, Cano2016a}, likely centralized~\cite{Sandoval2016} \\
\cline{2-5}	 
	               & \cite{JiaLondon2015, Zhang2015a, Xiao2016, Song2016} & \textbf{\mbox{Wi-Fi}}: CSMA/CA; \textbf{LTE}: \emph{LBT without random backoff} --  \cite{JiaLondon2015} two time granularity levels; \cite{Zhang2015a} ETSI FBE; adaptive transmission duration~\cite{Xiao2016}; dynamic channel switch~\cite{Xiao2016} & -- & distributed~\cite{JiaLondon2015, Zhang2015a, Song2016}, centralized~\cite{Xiao2016} \\
\cline{2-5}
                   & \cite{Li2016a, Zhang2015c, Jeon2014, ChenGlasgowMay2015, Mushunuri2017, Bhorkar2014, LiHongKong2015, Song2016, Gao2016} & \textbf{\mbox{Wi-Fi}}: CSMA/CA; \textbf{LTE}: \emph{LBT with random backoff within fixed interval (or fixed CW)} -- different ED thresholds~\cite{Li2016a, Bhorkar2014}, adaptive ED threshold~\cite{LiHongKong2015}, variable transmission duration~\cite{Zhang2015c}, different channel selection schemes (random, least power at AP or UE)~\cite{Bhorkar2014} & -- & distributed \\  
\cline{2-5}	                 
	                 & \cite{Yoon2017, BhorkarNewOrleans2015, Ali2016, Falconetti2016, Simic2016, Voicu2016, Voicu2017, Voicu2017a, Li2016b, Gao2016} & \textbf{\mbox{Wi-Fi}}: CSMA/CA; \textbf{LTE}: \emph{LBT with binary exponential random backoff} -- backoff freeze~\cite{Falconetti2016}, ETSI LBE~\cite{Gao2016}, different ED thresholds~\cite{Falconetti2016}, different channel selection~\cite{Simic2016, Voicu2016, Voicu2017, Voicu2017a, Li2016b}, different transmit power~\cite{Simic2016, Voicu2016} & -- & distributed \\
\cline{2-5}
                     & \cite{TaoHongKong2015, Yin2015, Li2017, Hasan2016} & \textbf{\mbox{Wi-Fi}}: CSMA/CA; \textbf{LTE}: \emph{LBT with random backoff and adaptive contention window (other than binary exponential)} & -- & cooperative~\cite{TaoHongKong2015}, LTE coordination~\cite{Yin2015, Li2017, Hasan2016} \\ 
\hline
	 \multirow{4}{2.3cm}{\mbox{Wi-Fi}/ duty-cycle-LTE} & \cite{Chaves2013, Gomez-Miguelez2016, Nhtilae2013, Ali2016, RupasingheNewOrleans2015, Simic2016} & \textbf{\mbox{Wi-Fi}}: CSMA/CA; \textbf{LTE}: \emph{fixed duty cycle} -- 80\% with subframe granularity~\cite{Chaves2013}; 0-100\% with mean period 150~ms~\cite{Gomez-Miguelez2016}; 0-100\%~\cite{Nhtilae2013}; 50\% with period 80 ms and maximum 20 ms ON time~\cite{Ali2016}; 20-100\%\cite{RupasingheNewOrleans2015}; 50\%~\cite{Simic2016}; different channel selection \& Tx power~\cite{Simic2016} & -- & distributed \\
\cline{2-5}
                     & \cite{Abdelfattah2017, Capretti2016, Jeon2014, Li2016a, Voicu2016} & \textbf{\mbox{Wi-Fi}}: CSMA/CA; \textbf{LTE}: \emph{fixed duty cycle with different transmission patterns} -- 50\%~\cite{Abdelfattah2017} and 60\%~\cite{Capretti2016} consecutive/alternative active subframes; 50\% successive/alternative and synchronous/asynchronous~\cite{Jeon2014}; 50\% coordinated/uncoordinated~\cite{Voicu2016}; 33-67\% synchronous/asynchronous~\cite{Li2016a} & -- & distributed~\cite{Abdelfattah2017, Capretti2016, Jeon2014, Voicu2016, Li2016a}, LTE coordination~\cite{Voicu2016, Jeon2014, Li2016a} \\ 
\cline{2-5}
                     & \cite{Sadek2015, Voicu2017, Voicu2017a, Guan2016, Zhang2015a, Cano2016a, Cano2015, Jian2017, Sagari2015, Sriyananda2016, RupasingheNewOrleans2015, Simic2016, Voicu2016} & \textbf{\mbox{Wi-Fi}}: CSMA/CA; \textbf{LTE}: \emph{adaptive duty cycle} -- channel selection\cite{Sadek2015, Voicu2017, Voicu2017a, Guan2016, Simic2016, Voicu2016}; carrier aggregation (channel width)~\cite{Guan2016}; power control~\cite{Sagari2015, Sriyananda2016}; ideal TDMA (perfect scheduling)~\cite{Simic2016, Voicu2016}; different Tx power~\cite{Simic2016} & -- & distributed\cite{Sadek2015, Voicu2017, Voicu2017a, Zhang2015a, Cano2016a, Jian2017, Sriyananda2016, Simic2016, Voicu2016}; centralized~\cite{Sagari2015}; LTE coordination~\cite{Simic2016, Voicu2016, Guan2016, Cano2015} \\                                          
\hline
\end{tabular}
\end{table*}

\begin{table*}[!t]
\addtocounter{table}{-1}
\modcounter
\caption{Inter-technology coexistence goals and performance for literature review of spectrum sharing among broadband technologies in a spectrum commons in Table~\ref{table_review_4}: \mbox{Wi-Fi} EDCA/DCF, \mbox{Wi-Fi}/IEEE 802.16, \mbox{Wi-Fi}/LTE}
\label{table_review_4b}
\centering
\begin{tabular}{|p{1.7cm}|p{4cm}|p{3cm}|p{3cm}|p{3.5cm}|}
\hline
 \centering \multirow{2}{*}{\textbf{Technologies}} 
 & \centering \multirow{2}{*}{\textbf{Coexistence Goals}} 
 & \multicolumn{3}{c|}{\textbf{Performance Evaluation}}\\
\cline{3-5}
 &      
 & \centering \textbf{method} 
 & \centering \textbf{metric} 
 & \centering\let\newline\\\arraybackslash\hspace{0pt} \textbf{network size} \\ 
\hline     
	\mbox{Wi-Fi} EDCA/DCF \scriptsize\cite{Hwang2006, Bianchi2005} 
	& study mutual impact between technologies vs. each other 
	& \begin{tabular}{p{2.8cm}}
	  $-$\textbf{analytical} \scriptsize\cite{Hwang2006, Bianchi2005};\\
	  $-$\textbf{OPNET simulations} \scriptsize\cite{Hwang2006} 
	  \end{tabular}
	  & \begin{tabular}{p{2.8cm}}
	  $-$\textbf{throughput} \scriptsize\cite{Hwang2006, Bianchi2005}; \\
	  $-$\textbf{slot occupancy probability} \scriptsize\cite{Bianchi2005}
	  \end{tabular} 
	  & 20--30 stations of each technology \\
\hline
	\mbox{Wi-Fi}/ IEEE~802.16  \scriptsize\cite{Fu2007} 
	& study mutual impact between technologies vs. each other & analytical, simulations & bit error rate & one \mbox{Wi-Fi} and one 802.16 link \\
\hline
    \begin{tabular}{p{1.5cm}}
    \mbox{Wi-Fi}/LTE \\ \\ \\ \\
    \scriptsize\cite{Babaei2014, Jian2015, Capretti2016, Gomez-Miguelez2016, BhorkarNewOrleans2015, Jeon2014, SagariLondon2015, ChenGlasgowMay2015, Nhtilae2013, Rupasinghe2014, VoicuLondon2015, Sagari2015, FuadM.Abinader2014, Li2016a, Voicu2016, Jian2017, JiaLondon2015, Bhorkar2014, Chaves2013, Yun2015, Li2016c, Hajmohammad2013, Cai2016}
    \end{tabular}
    & \begin{tabular}{p{3.7cm}}
    \emph{Impact of coexistence with unmodified LTE on \mbox{Wi-Fi}} \\
    $-$\textbf{no baseline} \scriptsize\cite{Babaei2014, ChenGlasgowMay2015, Jian2017};\\
    $-$\textbf{vs. standalone} \scriptsize\cite{Jian2015, Capretti2016, Gomez-Miguelez2016, BhorkarNewOrleans2015, Jeon2014, SagariLondon2015, Nhtilae2013, Rupasinghe2014, VoicuLondon2015, Sagari2015, FuadM.Abinader2014, JiaLondon2015};\\
    $-$\textbf{vs. coexistence with itself} \scriptsize\cite{BhorkarNewOrleans2015, Sagari2015, Voicu2016, Li2016a, Bhorkar2014} \\
    \hline
    \emph{Impact of coexistence on unmodified LTE}\\
    $-$\textbf{no baseline} \scriptsize\cite{Capretti2016, ChenGlasgowMay2015, Sagari2015, Li2016a, Voicu2016, Jian2017};\\
    $-$\textbf{vs. standalone} \scriptsize\cite{Jeon2014, SagariLondon2015, Nhtilae2013, Rupasinghe2014, VoicuLondon2015, FuadM.Abinader2014, JiaLondon2015};\\
    $-$\textbf{vs. coexistence with itself} \scriptsize\cite{BhorkarNewOrleans2015, Bhorkar2014}\\
    \hline
    \emph{Other} \\
    $-$\textbf{increase aggregate throughput vs. coexistence with unmodified LTE} \scriptsize\cite{SagariLondon2015}\\
    $-$\textbf{mutual coexistence impact vs. channel selection and vs. LBT} \scriptsize\cite{VoicuLondon2015}\\
    $-$\textbf{mutual coexistence impact vs. standalone \& vs. duty cycle} \scriptsize\cite{Chaves2013}\\
    $-$\textbf{enable simultaneous \mbox{Wi-Fi} and LTE transmissions and compare aggregate throughput with time division} \scriptsize\cite{Yun2015}\\
    $-$\textbf{enable simultaneous LTE and \mbox{Wi-Fi} transmissions and compare with an LBT variant} \scriptsize\cite{Li2016c}\\
    $-$\textbf{maximize total capacity, ensure fairness and QoS for both technologies} \scriptsize\cite{Hajmohammad2013}\\
    $-$\textbf{maximize overall resource utilization vs. an LBT variant} \scriptsize\cite{Cai2016}
   \end{tabular}
   & \begin{tabular}{p{3cm}}
     $-$\textbf{simulations} \scriptsize\cite{BhorkarNewOrleans2015, Jeon2014, SagariLondon2015, Nhtilae2013, Rupasinghe2014, VoicuLondon2015, FuadM.Abinader2014, Li2016a, Voicu2016, Jian2017, JiaLondon2015, Bhorkar2014, Chaves2013, Yun2015, Cai2016}; \\ \\ \\
     $-$\textbf{analytical} \scriptsize\cite{Babaei2014, SagariLondon2015, ChenGlasgowMay2015, Sagari2015, Li2016a, Bhorkar2014, Hajmohammad2013}; \\ \\ \\
     $-$\textbf{measurements} \scriptsize\cite{Jian2015, Capretti2016, Gomez-Miguelez2016, Sagari2015, Yun2015}
     \end{tabular} 
    & \begin{tabular}{p{2.8cm}}
      $-$\textbf{throughput} \scriptsize\cite{Jian2015, Capretti2016, Gomez-Miguelez2016, BhorkarNewOrleans2015, Jeon2014, SagariLondon2015, Nhtilae2013, VoicuLondon2015, Sagari2015, FuadM.Abinader2014, Voicu2016, Jian2017, JiaLondon2015, Bhorkar2014, Chaves2013, Yun2015, Hajmohammad2013, Cai2016};\\
      $-$\textbf{no. transmitted packets} \scriptsize\cite{Jian2015}; \\
      $-$\textbf{channel/medium access probability} \scriptsize\cite{Babaei2014, ChenGlasgowMay2015, Li2016a}; \\
      $-$\textbf{number/ probability of successful transmissions/links} \scriptsize\cite{ChenGlasgowMay2015, Li2016a}; \\
      $-$\textbf{delay} \scriptsize\cite{Babaei2014}; \\
      $-$\textbf{jitter} \scriptsize\cite{Capretti2016};\\
      $-$\textbf{SINR}  \scriptsize\cite{SagariLondon2015, Rupasinghe2014, Sagari2015, Li2016a, Yun2015, Li2016c}; \\
      $-$\textbf{interference} \scriptsize\cite{Li2016a};\\
      $-$\textbf{coverage probability} \scriptsize\cite{Li2016a}; \\
      $-$\textbf{false sensing probability} \scriptsize\cite{Yun2015};\\
      $-$\textbf{mean square error} \scriptsize\cite{Yun2015};\\
      $-$\textbf{channel occupancy time} \scriptsize\cite{Li2016c};\\
      $-$\textbf{Jain's fairness index} \scriptsize\cite{Hajmohammad2013};\\
      $-$\textbf{utility} \scriptsize\cite{Cai2016}
      \end{tabular} 
    & \begin{tabular}{p{3.2cm}}
      $-$~\textbf{1 LTE link/eNB \& several \mbox{Wi-Fi} devices} \scriptsize\cite{Jian2015, Gomez-Miguelez2016, Sagari2015, Babaei2014, Capretti2016, Jian2017, Yun2015, Li2016c}; \\ \\ \\
      $-$~\textbf{$\leq$10 APs of each technology} \scriptsize\cite{Jeon2014, ChenGlasgowMay2015, Nhtilae2013, FuadM.Abinader2014, Voicu2016, Rupasinghe2014, JiaLondon2015,  Chaves2013, Cai2016};\\ \\ \\
      $-$~\textbf{$\leq$100 APs or 400--5000~APs/km\textsuperscript{2} of each technology, or 50 total links} \scriptsize\cite{BhorkarNewOrleans2015, Bhorkar2014, SagariLondon2015, Li2016a, VoicuLondon2015, Hajmohammad2013}; \\
      \end{tabular} \\
\hline
\end{tabular}
\end{table*}

\begin{table*}[!t]
\addtocounter{table}{-1}
\modcounter
\caption{Inter-technology coexistence goals and performance for literature review of spectrum sharing among broadband technologies in a spectrum commons in Table~\ref{table_review_4}: \mbox{Wi-Fi}/LBT-LTE}
\label{table_review_4c}
\centering
\begin{tabular}{|p{1.7cm}|p{6cm}|p{2cm}|p{3.5cm}|p{2cm}|}
\hline
 \centering \multirow{2}{*}{\textbf{Technologies}} 
 & \centering \multirow{2}{*}{\textbf{Coexistence Goals}} 
 & \multicolumn{3}{c|}{\textbf{Performance Evaluation}}\\
\cline{3-5}
 &      
 & \centering \textbf{method} 
 & \centering \textbf{metric} 
 & \centering\let\newline\\\arraybackslash\hspace{0pt} \textbf{network size} \\ 
\hline     
  \begin{tabular}{p{1.5cm}} \mbox{Wi-Fi}/ \mbox{LBT-LTE} \\ \\ \\ \\ \scriptsize\cite{Li2016, VoicuLondon2015, Cano2016a, Sandoval2016, Li2016a, Zhang2015c, JiaLondon2015, Zhang2015a, Xiao2016, Jeon2014, ChenGlasgowMay2015, Mushunuri2017, Bhorkar2014, LiHongKong2015, Song2016, Yoon2017, BhorkarNewOrleans2015, Ali2016, Falconetti2016, Simic2016, Voicu2016, Voicu2017, Voicu2017a, Li2016b, Gao2016, TaoHongKong2015, Yin2015, Li2017, Hasan2016} 
  \end{tabular}
	                                       &  \begin{tabular}{p{5.5cm}}
	                                           \emph{Impact on \mbox{Wi-Fi}} \\
	                                           $-$\textbf{no baseline}  \scriptsize\cite{Zhang2015c}; \\
	                                           $-$\textbf{vs. coexistence with itself}  \scriptsize\cite{Li2016, Li2016a, Mushunuri2017, Bhorkar2014, LiHongKong2015, Song2016, Yoon2017, BhorkarNewOrleans2015, Ali2016, Falconetti2016, Simic2016, Voicu2016, Voicu2017, Voicu2017a, Li2016b, Gao2016, TaoHongKong2015, Yin2015, Zhang2015a} \\
	                                           $-$\textbf{vs. plain coexistence}  \scriptsize\cite{Li2016, VoicuLondon2015, Li2016a, JiaLondon2015, Zhang2015a, Jeon2014, ChenGlasgowMay2015, Bhorkar2014, BhorkarNewOrleans2015, Simic2016, Voicu2016} \\
	                                           $-$\textbf{vs. standalone}  \scriptsize\cite{JiaLondon2015, Zhang2015a, Jeon2014, Voicu2017, Voicu2017a, Gao2016} \\
	                                           $-$\textbf{vs. channel selection}  \scriptsize\cite{VoicuLondon2015, Simic2016, Voicu2016, Voicu2017, Voicu2017a, Xiao2016} \\
	                                           $-$\textbf{vs. duty cycle variants}  \scriptsize\cite{Li2016a, Zhang2015a, Jeon2014, Ali2016, Simic2016, Voicu2016, Voicu2017, Voicu2017a, Xiao2016} \\
	                                           $-$\textbf{vs. other LBT variants}  \scriptsize\cite{Jeon2014, Mushunuri2017, Bhorkar2014, LiHongKong2015, Song2016, Yoon2017, Falconetti2016, Li2016b, Gao2016, TaoHongKong2015, Yin2015, Li2017, Hasan2016}\\
	                                           \hline
                                               \emph{Impact on LTE}  \\
                                               $-$\textbf{no baseline}  \scriptsize\cite{Zhang2015c} \\
                                               $-$\textbf{vs. plain coexistence}  \scriptsize\cite{Li2016, VoicuLondon2015, Li2016a, JiaLondon2015, Zhang2015a, Jeon2014, ChenGlasgowMay2015, Bhorkar2014, BhorkarNewOrleans2015, Simic2016, Voicu2016} \\
                                               $-$\textbf{vs. coexistence with itself}  \scriptsize\cite{Bhorkar2014, LiHongKong2015, BhorkarNewOrleans2015, Falconetti2016, Li2016b, TaoHongKong2015, Li2017} \\
                                               $-$\textbf{vs. standalone}  \scriptsize\cite{JiaLondon2015, Zhang2015a, Jeon2014, Voicu2017a} \\
                                               $-$\textbf{vs. channel selection}  \scriptsize\cite{VoicuLondon2015, Bhorkar2014, Voicu2016, Xiao2016, Voicu2017a} \\
                                               $-$\textbf{vs. duty cycle variants}  \scriptsize\cite{Li2016a, Zhang2015a, Jeon2014, Ali2016, Simic2016, Voicu2016, Xiao2016, Voicu2017a} \\
                                               $-$\textbf{vs. other LBT variants}  \scriptsize\cite{Jeon2014, Bhorkar2014, LiHongKong2015, Song2016, Yoon2017, Falconetti2016, Li2016b, Gao2016, TaoHongKong2015, Yin2015, Li2017, Hasan2016}\\
                                               \hline
                                               \emph{Other}  \\
                                               $-$\textbf{fairness (implicitly) for \mbox{Wi-Fi} vs. coexistence with itself}  \scriptsize\cite{Mushunuri2017, Bhorkar2014, LiHongKong2015, Song2016, Ali2016, Simic2016, Voicu2017, Voicu2017a, TaoHongKong2015}\\
                                               $-$\textbf{proportional fair rate allocation for \mbox{Wi-Fi} and LTE}  \scriptsize\cite{Cano2016a} \\
                                               $-$\textbf{fairness as same \mbox{Wi-Fi}/LTE airtime}  \scriptsize\cite{Yoon2017}\\
                                               $-$\textbf{proportional fair channel switch}  \scriptsize\cite{Li2016b} \\
                                               $-$\textbf{fairness as minimization of collision probability to \mbox{Wi-Fi}}  \scriptsize\cite{Yin2015} \\
                                               $-$\textbf{fairness as constant aggregate \mbox{Wi-Fi} throughput}  \scriptsize\cite{Li2017} \\
                                               $-$\textbf{airtime fairness for \mbox{Wi-Fi} based on altruistic gains}  \scriptsize\cite{Hasan2016}\\
                                               $-$\textbf{maximize aggregate LTE capacity in presence of \mbox{Wi-Fi}}  \scriptsize\cite{Sandoval2016} \\
                                               $-$\textbf{enable different levels of protection for \mbox{Wi-Fi}}  \scriptsize\cite{Zhang2015c} \\
                                               $-$\textbf{maximize total throughput given requirements of each technology}  \scriptsize\cite{Song2016}   \\
                                               \end{tabular}
	                                         & \begin{tabular}{p{1.5cm}}
	                                           $-$\textbf{simulations} \scriptsize\cite{VoicuLondon2015, Cano2016a, Sandoval2016, Li2016a, Zhang2015c, JiaLondon2015, Zhang2015a, Xiao2016, Jeon2014, Mushunuri2017, Bhorkar2014, LiHongKong2015, Yoon2017, BhorkarNewOrleans2015, Ali2016, Falconetti2016, Simic2016, Voicu2016, Voicu2017, Voicu2017a, Li2016b, TaoHongKong2015, Yin2015, Li2017, Hasan2016};\\ \\ \\ \\
	                                           $-$\textbf{analytical} \scriptsize\cite{Li2016, Sandoval2016, Li2016a, Zhang2015c, ChenGlasgowMay2015, Mushunuri2017, Bhorkar2014, Song2016, Ali2016, Gao2016, Yin2015, Li2017, Hasan2016}  \\
	                                           \end{tabular}
	                                         & \begin{tabular}{p{3cm}}
	                                           $-$\textbf{throughput} \scriptsize\cite{Li2016, VoicuLondon2015, Cano2016a, Sandoval2016, Zhang2015c, JiaLondon2015, Zhang2015a, Xiao2016, Jeon2014, Mushunuri2017, Bhorkar2014, LiHongKong2015, Song2016, Yoon2017, BhorkarNewOrleans2015, Ali2016, Falconetti2016, Simic2016, Voicu2016, Voicu2017,Voicu2017a,  Li2016b, Gao2016, TaoHongKong2015, Yin2015, Li2017};\\
	                                           $-$\textbf{delay} \scriptsize\cite{Cano2016a, Mushunuri2017, Ali2016, Gao2016, TaoHongKong2015};\\
	                                           $-$\textbf{coverage probability} \scriptsize\cite{Sandoval2016, Li2016a};\\
	                                           $-$\textbf{successful transmissions} \scriptsize\cite{Li2016a, ChenGlasgowMay2015}; \\
	                                           $-$\textbf{protection level} \scriptsize\cite{Zhang2015c}; \\
	                                           $-$\textbf{transmission duration} \scriptsize\cite{Zhang2015c} \\
	                                           $-$\textbf{channel access probability} \scriptsize\cite{ChenGlasgowMay2015, Mushunuri2017}\\
	                                           $-$\textbf{collision probability} \scriptsize\cite{Mushunuri2017, Yin2015}\\
	                                           $-$\textbf{SINR} \scriptsize\cite{LiHongKong2015}\\
	                                           $-$\textbf{airtime} \scriptsize\cite{Yoon2017, Hasan2016}\\
	                                           $-$\textbf{Jain's fairness} \scriptsize\cite{Yoon2017, Voicu2017, Voicu2017a, Hasan2016}\\
	                                           $-$\textbf{channel occupation} \scriptsize\cite{Yin2015}\\
	                                           $-$\textbf{utility} \scriptsize\cite{Li2017}\\
	                                           $-$\textbf{Q-value} \scriptsize\cite{Li2017}\\
	                                           $-$\textbf{entropy} \scriptsize\cite{Hasan2016}\\
	                                           $-$\textbf{risk-informed interference assessment} \scriptsize\cite{Voicu2017, Voicu2017a}\\
	                                           \end{tabular}
	                                         & \begin{tabular}{p{1.7cm}}
                                               $-$~\textbf{1 LTE link/AP \& several \mbox{Wi-Fi} devices} \scriptsize\cite{Yoon2017, Cano2016a, Zhang2015c, Mushunuri2017, Yin2015}\\ \\ \\
                                               $-$~\textbf{$\leq$10 APs, or 15 APs/km\textsuperscript{2} of each technology} \scriptsize\cite{Li2016, ChenGlasgowMay2015, Gao2016, Ali2016, Li2016b, Hasan2016, Falconetti2016, JiaLondon2015, Xiao2016, Jeon2014, Song2016, Li2017};\\ \\ \\
                                               $-$~\textbf{10--90 APs, or 400--5000 APs/km\textsuperscript{2} of each technology}: \scriptsize\cite{VoicuLondon2015, Simic2016, Voicu2016, Sandoval2016, Li2016a, Zhang2015a, Bhorkar2014, BhorkarNewOrleans2015, LiHongKong2015, TaoHongKong2015, Voicu2017, Voicu2017a}
                                               \end{tabular}
	                                                \\	
\hline
\end{tabular}
\end{table*}

\subsection{Coexistence among Broadband Technologies}
\label{litrev_hightraffic}

We first present in Section~\ref{ov_broadband} a literature overview of broadband technology coexistence, and then review in detail various strands of the work as follows.
We review the literature addressing coexistence among technologies of the IEEE 802.x standards in Section~\ref{litrev_802}. 
We then focus on IEEE 802.11 \mbox{Wi-Fi}/LTE coexistence in the unlicensed bands, which has been recently extensively investigated in light of the two main proposed LTE variants for the unlicensed bands, i.e. LAA~\cite{3GPP2015} and LTE-U~\cite{Forum2015}. We classify the existing literature based on the main Layer~2 coexistence approaches for LTE\footnote{We note that \mbox{Wi-Fi} always implements CSMA/CA at the MAC layer, i.e. LBT with binary exponential random backoff. Although different approximations were adopted for modelling CSMA/CA in different papers (e.g.~\cite{VoicuLondon2015} does not consider the MAC inefficiency due to sensing time, \cite{Li2016a, Bhorkar2014} assume random backoff with fixed CW, and \cite{Voicu2016} estimates the binary exponential random backoff by means of an analytical model), a detailed review of such modelling techniques is out of the scope of this survey.}:
 (i)~no MAC coexistence mechanism, i.e. LTE continuously transmits, in Section~\ref{wifiltecoex}; (ii)~LBT, i.e. the approach adopted by 3GPP for LAA~\cite{3GPP2016, 3GPP2016a}, in Section~\ref{litrev_lbt}; and (iii)~duty cycle, i.e. the approach adopted by the LTE-U Forum, in Section~\ref{litrev_dut}. 
Table~\ref{table_review_4} summarizes the spectrum sharing mechanisms in the reviewed literature and Tables~\ref{table_review_4b} to \ref{table_review_4d} summarize coexistence performance evaluation aspects, where \emph{standalone} refers to the baseline case with a single technology, i.e. no coexisting technology is considered, and \emph{plain coexistence} refers to \mbox{Wi-Fi}/LTE coexistence where no spectrum sharing mechanism is implemented for LTE. 
Furthermore, in Tables~\mbox{\ref{table_review_4b}} to \ref{table_review_4d} we group similar metrics in the literature under a few representative terms, e.g. \emph{throughput} also refers to goodput~\cite{Chiasserini2002}, offered/served load~\cite{Jeon2014}, capacity~\cite{Rupasinghe2014}, normalized throughput~\cite{FuadM.Abinader2014}, etc.   

\subsubsection{\textbf{Literature Overview}}
\label{ov_broadband}
Only a few works have addressed coexistence among IEEE 802.x standards, as the dominant IEEE standard in the unlicensed bands is 802.11 \mbox{Wi-Fi}, such that the devices implement similar spectrum sharing mechanisms.  
There is a large number of works that have addressed \mbox{Wi-Fi}/LTE coexistence in the unlicensed bands. Some of them consider LTE without any coexistence mechanism and identify the need to implement one, in order to allow \mbox{Wi-Fi} to access the spectum.
Most works consider different variants of either LBT-LTE, or duty-cycle-LTE and compare them only with standalone technologies, or with coexistence where LTE does not implement sharing mechanisms. We note that this approach does not facilitate a direct comparison between different mechanisms. A few works, however, considered both \mbox{Wi-Fi}/LBT-LTE and \mbox{Wi-Fi}/duty-cycle-LTE coexistence. 
The authors report in general that the adaptive sharing mechanisms at Layer~2 (either duty cycle or LBT) achieve the best coexistence performance. However, some of these mechanisms require information that is not trivial to obtain with distributed mechanisms (e.g. traffic requirements, number of nodes, etc.). We note that many works have considered fairness when evaluating the coexistence performance, but different fairness definitions were used (\emph{cf.} Tables~\ref{table_review_4c} and \mbox{\ref{table_review_4d}}). However, a significant number of papers have adopted the fairness criterion used by 3GPP, i.e. ``not impact \mbox{Wi-Fi} services more than an additional \mbox{Wi-Fi} network''~\cite{3GPP2015}. As such, some works found that the most fair coexistence performance was obtained when LTE implemented an LBT mechanism similar to \mbox{Wi-Fi}'s LBT.
Furthermore, there have been very few proposals for Layer~1 sharing mechanisms, which suggests that such techniques are not developed enough to mitigate interference for broadband technologies, such that the most efficient mechanisms are sharing in time and/or frequency at Layer~2.  
Finally, most of the works relied on simulations and analytical tools to evaluate the coexistence performance. Only few works have conducted basic experimental evaluations and only for duty-cycle-LTE. This shows the difficulty of obtaining such results due to the lack of devices that implement a fully functional open-source LTE stack, which could be modified in a straightforward manner for research purposes.

\subsubsection{\textbf{Coexistence among Broadband IEEE 802.x Technologies}}
\label{litrev_802}

The authors in~\cite{Hwang2006, Bianchi2005} addressed coexistence between legacy IEEE 802.11 devices implementing at the MAC the distributed coordination function (DCF) and new devices implementing enhanced distributed channel access (EDCA), i.e. different sensing durations, in order to grant different channel access priority levels for different traffic categories. The reported performance results validated the channel access priorities associated with different sensing durations~\cite{Hwang2006}. 
Additionally, EDCA had higher channel access priority than DCF, due to the different backoff counter decrement procedure, through which it gained one additional backoff slot~\cite{Bianchi2005}.  

Coexistence between IEEE 802.11a and 802.16 was addressed in~\cite{Fu2007, Berlemann2006}. In~\cite{Fu2007} the mutual interference was evaluated at the PHY layer, when transmissions from the two technologies overlapped in time and frequency. Furthermore, the authors suggested varying the transmit power and modulation scheme for coping with this interference. 
In~\cite{Berlemann2006} channel blocking and ordering of contention slots was proposed for IEEE 802.16, in order to reserve the channel before 802.11a and thus to guarantee QoS for 802.16. However, no performance evaluation results were presented.

\subsubsection{\textbf{\mbox{Wi-Fi}/LTE Coexistence}}
\label{wifiltecoex}

A number of papers investigated \mbox{Wi-Fi}/LTE coexistence performance when LTE does not implement any coexistence mechanism, e.g.~\cite{Babaei2014, Jian2015, Capretti2016, Gomez-Miguelez2016, BhorkarNewOrleans2015, Jeon2014, SagariLondon2015, ChenGlasgowMay2015, Nhtilae2013, Rupasinghe2014, VoicuLondon2015, Sagari2015, FuadM.Abinader2014, Li2016a, Voicu2016, Jian2017, JiaLondon2015, Bhorkar2014}, either as an individual coexistence case, or as a baseline for comparison with other mechanisms. They all reported that the \mbox{Wi-Fi} performance was severely degraded and that LTE should implement an inter-technology coexistence mechanism when operating in the unlicensed bands.

For an overview of the main coexistence approaches considered for LTE in the unlicensed bands we refer the reader to e.g.~\cite{Cano2016, FuadM.Abinader2014, Chaves2013, Cui2016, Ho2017, Zhang2015a, Chen2017, Kwon2017}, where~\cite{Ho2017} presented a survey of the early literature on \mbox{Wi-Fi}/LTE coexistence, and~\cite{Kwon2017} focused on LAA standardized by 3GPP. 

The authors in~\cite{SagariLondon2015, Chaves2013, Yun2015, Li2016c, Hajmohammad2013, Cai2016} proposed \mbox{Wi-Fi}/LTE coexistence solutions different than the MAC-based LBT and duty cycling.
The authors in~\cite{SagariLondon2015, Hajmohammad2013, Cai2016} focused on spectrum sharing in frequency.
In \cite{SagariLondon2015} it was found that even with random channel selection, a significant increase in network throughput can be achieved vs. co-channel deployments. Furthermore, two variants of a channel allocation scheme based on multigraph coloring were proposed, i.e. with intra- or inter-technology coordination. The inter-technology coordination did not improve the network throughput significantly compared to intra-technology coordination, but both were better than random channel selection. 
The authors in~\cite{Hajmohammad2013} proposed spectrum splitting between \mbox{Wi-Fi} and LTE and aimed to maximize the total \mbox{Wi-Fi} and LTE femtocell capacity, while taking into account fairness and QoS constrains. This scheme was shown to improve the capacity of the LTE femtocells, compared to licensed spectrum splitting between femtocells and macrocells. 
In~\cite{Cai2016} \mbox{Wi-Fi}/LTE coordinated spectrum splitting with subcarrier granularity was assumed. Some network controllers implemented decision trees and repeated games for spectrum splitting, in order to maximize their resource utilization. The scheme was shown to improve the throughput for both technologies compared to other LBT variants. 
Although the results in~\cite{SagariLondon2015, Hajmohammad2013, Cai2016} show overall that spectrum sharing in frequency is efficient for facilitating inter-technology coexistence, all the proposed mechanisms require intra- or inter-technology coordination, which cannot be easily achieved in distributed deployments, where the devices are owned and managed by different parties.

In~\cite{Chaves2013} an uplink power control mechanism was proposed for LTE users, which resulted in a similar or somewhat higher mean user throughput for both LTE and \mbox{Wi-Fi}, compared to LTE with a duty cycle of 80\%. 
However, selecting LTE with a duty cycle of 80\% as baseline does not prove the efficiency of LTE uplink power control overall, since it is expected that 80\% duty-cycling-LTE causes a significant level of interference, especially in dense deployments, and thus has a poor coexistence performance. Furthermore, for properly tuning the proposed LTE uplink power control, the network operator needs knowledge of the \mbox{Wi-Fi} network and its traffic. Finally, the proposed technique does not manage the interference caused by LTE downlink transmissions. As such, this power control mechanism could be used in conjunction with other spectrum sharing schemes, but is not sufficient as a standalone coexistence mechanism. 

Two different PHY-layer techniques were proposed in~\cite{Yun2015, Li2016c}. 
In~\cite{Yun2015} \mbox{Wi-Fi} and LTE could both transmit at the same time, on the same frequency, using a decoding method that enabled the separation of two overlapping OFDM signals (i.e. an interference cancellation technique). 
The authors in~\cite{Li2016c} proposed estimating the direction of arrival of \mbox{Wi-Fi} signals by LTE and then applying null steering, such that LTE does not cause interfere in the direction of \mbox{Wi-Fi} (i.e. a beamforming technique).
The techniques in~\cite{Yun2015, Li2016c} resulted in good coexistence performance, but they both required co-located LTE and \mbox{Wi-Fi} receivers and were evaluated for a single LTE link. Additionally,~\cite{Yun2015} also requires substantial changes to the CSMA/CA \mbox{Wi-Fi} mechanism.

\subsubsection{\textbf{\mbox{Wi-Fi}/LBT-LTE coexistence}}
\label{litrev_lbt}
The works~\cite{Li2016, VoicuLondon2015, Cano2016a, Sandoval2016, JiaLondon2015, Zhang2015a, Xiao2016, Li2016a, Zhang2015c, Jeon2014, ChenGlasgowMay2015, Mushunuri2017, Bhorkar2014, LiHongKong2015, Yoon2017, BhorkarNewOrleans2015, Ali2016, Falconetti2016, Simic2016, Voicu2016, Voicu2017, Song2016, Gao2016, TaoHongKong2015, Yin2015, Li2017, Hasan2016, Li2016b} addressed \mbox{Wi-Fi}/LBT-LTE coexistence.

\paragraph{Generic LBT}
The work in~\cite{Li2016, VoicuLondon2015, Cano2016a, Sandoval2016} assumed LBT models at a level of abstraction for which the specifics of the backoff type are irrelevant, so we refer to this as \emph{generic LBT}.
The authors in~\cite{Li2016} found that proper selection of the sensing threshold was beneficial for coexistence.
We note that the sensing threshold, which is an inherent parameter for LBT technologies, has a critical impact on how much a technology defers to another one and it is thus a natural parameter to configure for granting different priorities in accessing the channel. 
In~\cite{VoicuLondon2015} LBT-LTE was compared with different channel selection schemes for LTE, i.e. random or least-interfered channel. Channel selection was found to be more efficient than LBT at ensuring coexistence, which shows that spectrum sharing in frequency would be preferred over sharing in time in distributed deployments. However, this required a large number of channels, which may not always be available in practice. Furthermore, the rather large building shielding at 5~GHz contributed to reducing interference and ensuring harmonious coexistence.
In~\cite{Sandoval2016} a complementary solution to LBT was proposed, i.e. a framework that statistically optimizes the LTE network topology when coexisting with \mbox{Wi-Fi} in indoor scenarios, such that the aggregate LTE capacity is maximized and the required coverage achieved. However, this requires accurate models for radio propagation, service demand, load levels, and spatial distribution.
Overall,~\cite{Li2016, VoicuLondon2015, Sandoval2016} suggest that spectrum sharing in time, e.g. LBT, is required for \mbox{Wi-Fi}/LTE coexistence, but this can be complemented by other techniques like channel selection or topology optimization.

\paragraph{LBT without random backoff}
The work in~\cite{JiaLondon2015, Zhang2015a, Xiao2016} considered LBT-LTE without random backoff.
The authors in~\cite{JiaLondon2015} proposed two variants of LBT with fixed sensing duration, i.e. \emph{periodic} sensing with OFDM symbol granularity, and \emph{persistent} sensing with subframe granularity. In~\cite{Xiao2016} it was proposed that LTE directly transmits once the medium is sensed idle.
The results in~\cite{JiaLondon2015, Xiao2016} showed a satisfactory LTE and \mbox{Wi-Fi} user throughput, but both works implemented additional spectrum sharing techniques, i.e. \cite{JiaLondon2015} applied a much lower sensing threshold to defer to \mbox{Wi-Fi} than vice-versa, and in \cite{Xiao2016} LTE either dynamically switched the channel to allow \mbox{Wi-Fi} to transmit, or adaptively reserved some blank subframes for \mbox{Wi-Fi}.   
This suggests that implementing only LBT without additional configuration/adaptation of the sensing time cannot ensure coexistence among broadband technologies in a spectrum commons. Namely, LBT has to be enhanced by tuning further parameters, e.g. sensing threshold, random backoff, or by applying additional spectrum sharing mechanisms, e.g. sharing in frequency.

\paragraph{LBT with random backoff within fixed interval}
The work in~\cite{Li2016a, Zhang2015c, Jeon2014, ChenGlasgowMay2015, Mushunuri2017, Bhorkar2014, LiHongKong2015} addressed \mbox{Wi-Fi} coexistence with \mbox{LBT-LTE} with random backoff within a fixed interval (or with fixed CW).
In~\cite{ChenGlasgowMay2015} it was found that \mbox{Wi-Fi} performance was improved when coexisting with LBT-LTE with fixed CW compared to the case where it coexisted with LTE without any coexistence mechanism, as expected.
Furthermore, it is expected that LBT with random backoff and fixed CW can avoid collisions better than LBT without random backoff, especially for broadband technologies with high traffic load and dense deployments. 
Nonetheless, coexistence performance via LBT with random backoff and fixed CW was further improved with respect to a given coexistence goal by also tuning other parameters, e.g. sensing threshold~\cite{Bhorkar2014, LiHongKong2015}, channel selection schemes~\cite{Bhorkar2014}, transmission duration~\cite{Zhang2015c}. The results in~\cite{Bhorkar2014, Zhang2015c, LiHongKong2015} showed overall that different capacity gains and tradeoffs between \mbox{Wi-Fi} and LTE performance can be achieved. 
Furthermore, the authors in~\cite{Mushunuri2017, Song2016} evaluated coexistence for different fixed CW and found that \mbox{Wi-Fi} and the total system performance could be increased if the CW was properly selected.  

From the point of view of the resulting performance, tuning either of two different design parameters may be equivalent, but in practice the choice of parameter to adapt depends on the specific constraints at different layers of the technology circle, for a given deployment. For instance, sensing thresholds are lower-bounded by the minimum sensitivity of the receiver, whereas implementing channel selection requires that a sufficient number of channels are available.
The number of available channels is determined by regulatory constraints at Layer~0, whereas the receiver sensitivity is a PHY parameter, which is arguably in turn determined by equipment cost constraints at Layer~8.

\paragraph{LBT with binary exponential random backoff}
The authors in~\cite{Yoon2017, BhorkarNewOrleans2015, Ali2016, Falconetti2016, Simic2016, Voicu2016, Voicu2017, Li2016b, Gao2016} addressed \mbox{Wi-Fi} coexistence with LBT-LTE with binary exponential random backoff, which is one method to adapt the CW.
As \mbox{Wi-Fi} implements this method, this was also considered for LTE, in order to achieve the same behaviour when the two technologies share the spectrum and thus achieve fairness. For instance, \cite{Gao2016} found that a fixed CW was more beneficial for LTE instead of binary exponential random backoff, but at the same time degraded \mbox{Wi-Fi} performance more. We note that for LAA, binary exponential random backoff was eventually standardized in 3GPP Release 13.
In this context, further LTE parameters were either directly adopted from \mbox{Wi-Fi} (e.g. the sensing threshold~\cite{BhorkarNewOrleans2015}, varying the channel width by aggregating multiple channels~\cite{Falconetti2016}), or were adapted to match equivalent \mbox{Wi-Fi} parameters (e.g. the transmission time~\cite{Yoon2017}).
For other considered coexistence goals, \cite{BhorkarNewOrleans2015} reported that a suitable sensing threshold could improve the overall performance of \mbox{Wi-Fi} and LTE and in \cite{Li2016b} a proportional fair dynamic channel selection mechanism was proposed for LBT-LTE in order to coexist with \mbox{Wi-Fi}. A modification to binary exponential LBT was also introduced, i.e. a frozen period to ensure correct channel switching decision. The scheme was shown to be efficient especially for low traffic load.

\begin{table*}[!t]
\addtocounter{table}{-1}
\modcounter
\caption{Inter-technology coexistence goals and performance for literature review of spectrum sharing among broadband technologies in a spectrum commons in Table~\ref{table_review_4}: \mbox{Wi-Fi}/duty-cycle-LTE}
\label{table_review_4d}
\centering
\begin{tabular}{|p{1.7cm}|p{6cm}|p{2cm}|p{3.5cm}|p{2cm}|}
\hline
 \centering \multirow{2}{*}{\textbf{Technologies}} 
 & \centering \multirow{2}{*}{\textbf{Coexistence Goals}} 
 & \multicolumn{3}{c|}{\textbf{Performance Evaluation}}\\
\cline{3-5}
 &      
 & \centering \textbf{method} 
 & \centering \textbf{metric} 
 & \centering\let\newline\\\arraybackslash\hspace{0pt} \textbf{network size} \\  
\hline
	\begin{tabular}{p{1.5cm}} \mbox{Wi-Fi}/ duty-cycle-LTE \\ \\ \\ \\ \scriptsize\cite{Chaves2013, Gomez-Miguelez2016, Nhtilae2013, Ali2016, Abdelfattah2017, Capretti2016, Jeon2014, Li2016a, Sadek2015, Voicu2017, Voicu2017a, Guan2016, Zhang2015a, Cano2016a, Cano2015, Jian2017, Sagari2015, Sriyananda2016, RupasingheNewOrleans2015, Simic2016, Voicu2016}
	\end{tabular}
	                                            & \begin{tabular}{p{5.5cm}}
	                                              \emph{Impact on \mbox{Wi-Fi}} \\
	                                              $-$\textbf{no baseline} \scriptsize\cite{Cano2015}; \\
	                                              $-$\textbf{vs. coexistence with itself} \scriptsize\cite{Ali2016, Li2016a, Sadek2015, Voicu2017, Voicu2017a, Guan2016, Zhang2015a, Zhang2015a, Simic2016, Voicu2016}; \\
	                                              $-$\textbf{vs. plain coexistence} \scriptsize\cite{Chaves2013, Gomez-Miguelez2016, Nhtilae2013, Capretti2016, Jeon2014, Li2016a, Sadek2015, Zhang2015a, Jian2017, Sagari2015, RupasingheNewOrleans2015, Simic2016, Voicu2016};\\
	                                              $-$\textbf{vs. standalone} \scriptsize\cite{Chaves2013, Gomez-Miguelez2016, Nhtilae2013, Abdelfattah2017, Capretti2016, Jeon2014, Voicu2017, Voicu2017a, Zhang2015a, Sriyananda2016}; \\
	                                              $-$\textbf{vs. channel selection} \scriptsize\cite{Voicu2017, Voicu2017a, Guan2016, Simic2016, Voicu2016};\\
	                                              $-$\textbf{vs. other duty cycle variants} \scriptsize\cite{Nhtilae2013, Abdelfattah2017, Capretti2016, Jeon2014, Li2016a, Sriyananda2016, RupasingheNewOrleans2015, Simic2016, Voicu2016};\\
	                                              $-$\textbf{vs. LBT variants} \scriptsize\cite{Ali2016, Jeon2014, Li2016a, Voicu2017, Voicu2017a, Zhang2015a, Cano2016a, Simic2016, Voicu2016};\\
	                                              $-$\textbf{vs. power control} \scriptsize\cite{Chaves2013, Sagari2015}\\
	                                              \hline
                                                  \emph{Impact on LTE} \\
                                                  $-$\textbf{no baseline} \scriptsize\cite{Cano2015}; \\
                                                  $-$\textbf{vs. plain coexistence} \scriptsize\cite{Chaves2013, Gomez-Miguelez2016, Nhtilae2013, Capretti2016, Jeon2014, Li2016a, Sadek2015, Zhang2015a, Jian2017, Sagari2015, RupasingheNewOrleans2015, Simic2016, Voicu2016};\\
                                                  $-$\textbf{vs. coexistence with itself} \scriptsize\cite{Sadek2015};\\
                                                  $-$\textbf{vs. standalone} \scriptsize\cite{Chaves2013, Nhtilae2013, Jeon2014, Zhang2015a, Sriyananda2016, Voicu2017a};\\
                                                  $-$\textbf{vs. channel selection} \scriptsize\cite{Guan2016, Simic2016, Voicu2016, Voicu2017a};\\
                                                  $-$\textbf{vs. other duty cycle variants} \scriptsize\cite{Nhtilae2013, Capretti2016, Jeon2014, Li2016a, Sriyananda2016, RupasingheNewOrleans2015, Simic2016, Voicu2016}; \\
                                                  $-$\textbf{vs. LBT variants} \scriptsize\cite{Ali2016, Jeon2014, Li2016a, Zhang2015a, Cano2016a, Simic2016, Voicu2016, Voicu2017a}; \\
                                                  $-$\textbf{vs. power control} \scriptsize\cite{Chaves2013, Sagari2015}\\
                                                  \hline
                                                  \emph{Other} \\
                                                  $-$\textbf{fairness (implicitly) for \mbox{Wi-Fi} vs. coexistence with itself} \scriptsize\cite{Ali2016, Abdelfattah2017, Voicu2017, Voicu2017a, Simic2016}\\
                                                  $-$\textbf{fair coexistence for \mbox{Wi-Fi} as half the throughput of standalone \mbox{Wi-Fi}} \scriptsize\cite{Abdelfattah2017};\\
                                                  $-$\textbf{max. network utility with fairness for \mbox{Wi-Fi} as airtime vs. coexistence with itself} \scriptsize\cite{Guan2016};\\
                                                  $-$\textbf{proportional fair rate allocation for \mbox{Wi-Fi} and LTE} \scriptsize\cite{Cano2016a, Cano2015};\\
                                                  $-$\textbf{maximize overall throughput with fairness as same airtime for LTE \& \mbox{Wi-Fi}} \scriptsize\cite{Jian2017};\\
                                                  $-$\textbf{maximize capacity and minimize Tx power} \scriptsize\cite{Sriyananda2016}   \\
                                                  \end{tabular}
	                                            & \begin{tabular}{p{1.5cm}}
	                                              $-$\textbf{simulations} \scriptsize\cite{Chaves2013, Nhtilae2013, Ali2016, Abdelfattah2017, Jeon2014, Li2016a, Sadek2015, Voicu2017, Voicu2017a, Zhang2015a, Cano2016a, Jian2017, Sriyananda2016, RupasingheNewOrleans2015, Simic2016, Voicu2016};\\ \\ \\
	                                              $-$\textbf{analytical} \scriptsize\cite{Ali2016, Abdelfattah2017, Li2016a, Guan2016, Cano2015, Sagari2015, Sriyananda2016}  \\ \\ \\
	                                              $-$\textbf{measurements} \scriptsize\cite{Gomez-Miguelez2016, Capretti2016, Sadek2015, Sagari2015}  \\ 
	                                              \end{tabular}
	                                            & \begin{tabular}{p{3cm}}
	                                              $-$\textbf{throughput} \scriptsize\cite{Chaves2013, Gomez-Miguelez2016, Nhtilae2013, Ali2016, Abdelfattah2017, Capretti2016, Jeon2014, Sadek2015, Voicu2017, Voicu2017a, Zhang2015a, Cano2016a, Cano2015, Jian2017, Sagari2015, Sriyananda2016, RupasingheNewOrleans2015, Simic2016, Voicu2016};\\
	                                              $-$\textbf{jitter} \scriptsize\cite{Capretti2016}\\
	                                              $-$\textbf{delay} \scriptsize\cite{Ali2016, Cano2016a}\\
	                                              $-$\textbf{SINR} \scriptsize\cite{Abdelfattah2017, Sagari2015, RupasingheNewOrleans2015}\\
	                                              $-$\textbf{collision probability} \scriptsize\cite{Abdelfattah2017, Cano2015}\\
	                                              $-$\textbf{coverage probability} \scriptsize\cite{Li2016a}\\
	                                              $-$\textbf{successful links} \scriptsize\cite{Li2016a}\\
	                                              $-$\textbf{Jain's fairness index} \scriptsize\cite{Voicu2017, Voicu2017a, Guan2016}\\
	                                              $-$\textbf{airtime} \scriptsize\cite{Guan2016, Cano2015}\\
	                                              $-$\textbf{channel utilization} \scriptsize\cite{Jian2017}\\
	                                              $-$\textbf{energy efficiency} \scriptsize\cite{Sriyananda2016}\\
	                                              $-$\textbf{risk-informed interference assessment} \scriptsize\cite{Voicu2017, Voicu2017a}\\
	                                              \end{tabular}
	                                            & \begin{tabular}{p{1.7cm}}
                                                  $-$~\textbf{1 LTE link/AP \& several \mbox{Wi-Fi} devices} \scriptsize\cite{Gomez-Miguelez2016, Abdelfattah2017, Capretti2016, Cano2016a};\\ \\ \\
                                                  $-$~\textbf{$\leq$15 APs of each technology} \scriptsize\cite{Chaves2013, Nhtilae2013, Ali2016, Jeon2014, Guan2016, Cano2015, Jian2017, Sagari2015, Sriyananda2016, RupasingheNewOrleans2015}; \\ \\ \\
                                                  $-$~\textbf{10--30 APs, or up to 5000 APs/km\textsuperscript{2} of each technology} \scriptsize\cite{Voicu2017, Voicu2017a, Zhang2015a, Simic2016, Voicu2016, Li2016a, Sadek2015}\\
                                                  \end{tabular} \\
\hline
\end{tabular}
\end{table*}

\paragraph{LBT with random backoff and other adaptive CW} 
\label{lbtOptim}
The authors in~\cite{TaoHongKong2015, Yin2015, Li2017, Hasan2016} considered LBT-LTE with random backoff and contention window adaptation, other than binary exponential.
Specifically, the authors in~\cite{TaoHongKong2015} adapted the CW of LTE based on a target average transmission delay, but since cooperative information exchange among LTE via the X2 interface was required, the CW adaptation could be too slow in practice.
The other works, i.e.~\cite{Yin2015, Li2017, Hasan2016}, solved mathematical optimization problems and also required cooperation at least among LTE devices. In such cases, it is not clear how sensitive the proposed coexistence mechanisms are to conditions in real deployments, e.g. cooperation among only some LTE operators. 
In~\cite{Yin2015} the number of LTE users was maximized, while keeping the collision probability with \mbox{Wi-Fi} below a given threshold and in~\cite{Li2017} the LTE throughput was maximized, while keeping the \mbox{Wi-Fi} throughput constant via a genetic algorithm or multi-agent reinforcement learning.
In~\cite{Hasan2016} airtime fairness among LTE and \mbox{Wi-Fi} was considered and two mathematical approaches for characterizing fairness were compared, i.e. the Shapely value and proportional fairness.  
We note that estimating the number of \mbox{Wi-Fi} devices was required in all~\cite{Yin2015, Hasan2016, Li2017}, but it is not clear how efficiently the number of \mbox{Wi-Fi} devices can be estimated, especially in case of mixed \mbox{Wi-Fi} and LTE traffic sent over the same channel.
Furthermore, since in~\cite{Yin2015} only one LTE AP was assumed, it is not clear what the performance of the proposed mechanism is with multiple LTE APs, which are not necessarily coordinated. 
Finally, most works did not compare the performance of the proposed coordinated CW adaptation mechanism to that of an adaptive \emph{distributed} one with low computation complexity, e.g. binary exponential random backoff. Consequently, it is not clear whether these mathematical optimization approaches result in performance improvements over conventional CW adaptation approaches, especially in realistic deployments.

\subsubsection{\textbf{\mbox{Wi-Fi}/Duty-Cycle-LTE Coexistence}}
\label{litrev_dut}

The following work in the literature considered \mbox{Wi-Fi}/duty-cycle-LTE coexistence~\cite{Chaves2013, Gomez-Miguelez2016, Nhtilae2013, Ali2016, Abdelfattah2017, Capretti2016, Jeon2014, Li2016a, Sadek2015, Voicu2017, Guan2016, Zhang2015a, Cano2016a, Cano2015, Jian2017, Sagari2015, Sriyananda2016, RupasingheNewOrleans2015, Simic2016, Voicu2016}.

\paragraph{Fixed duty cycle}
The authors in~\cite{Chaves2013, Gomez-Miguelez2016, Nhtilae2013, Ali2016, RupasingheNewOrleans2015, Simic2016} considered LTE with fixed duty cycle.
We note that fixed duty cycling was initially proposed as a coexistence mechanism since it only required minimal modifications to the 3GPP LTE standard, i.e. it could be implemented based on subframe blanking. 
However, it was shown that the \mbox{Wi-Fi} performance was significantly affected when coexisting with LTE implementing fixed duty cycling, and that more sophisticated coexistence mechanisms were needed, e.g.~\cite{Nhtilae2013}.
Furthermore, the authors in~\cite{Gomez-Miguelez2016} varied experimentally the fixed duty cycle, the transmit power, and the LTE bandwidth and center frequency. It was found that the results are vendor-specific and fine tuning for fairness was difficult.

\paragraph{Fixed duty cycle with different transmission patterns}

The authors in~\cite{Abdelfattah2017, Capretti2016, Jeon2014, Li2016a, Voicu2016} further considered different transmission patterns for LTE with fixed duty cycle, 
in order to either study the coexistence performance of fixed duty cycling itself, or as a baseline for other coexistence mechanisms.
In general it was reported that, regardless of the transmission pattern, a fixed duty cycle for LTE could affect the \mbox{Wi-Fi} performance significantly.
For instance, the authors in~\cite{Abdelfattah2017} estimated the probability of collision and throughput for \mbox{Wi-Fi} via analytical models and ns-3 simulations,  for LTE with a fixed duty cycle of 50\% and different sub-frame transmission patterns. It was found that the \mbox{Wi-Fi} performance strongly depended on the packet size. Consequently, adjusting the duty cycle and duty cycle period was suggested, in order to improve \mbox{Wi-Fi} performance.
Furthermore, the authors in~\cite{Capretti2016} performed an empirical evaluation for a fixed duty cycle of 60\% and different consecutive/alternating sub-frame transmission patters. Importantly, they found that coexistence was possible, but that tuning the network parameters was non-trivial, especially since muting patterns that resulted in higher \mbox{Wi-Fi} throughput also resulted in higher \mbox{Wi-Fi} jitter.
These results suggest overall that fixed duty cycling for LTE is not sufficient to ensure \mbox{Wi-Fi}/LTE coexistence, but that adaptive duty cycling could be a feasible solution.

\paragraph{Adaptive duty cycle}
\label{dutOptim}
The authors in~\cite{Sadek2015, Voicu2017, Guan2016, Zhang2015a, Cano2016a, Cano2015, Jian2017, Sagari2015, Sriyananda2016, RupasingheNewOrleans2015, Simic2016, Voicu2016} considered LTE with adaptive duty cycle.
We note that, although duty cycling for LTE can be implemented based on the existing subframe blanking specifications, algorithms that adapt the duty cycle require more advanced features, e.g. monitoring the channel (potentially via an additional \mbox{Wi-Fi} interface) in order to extract information about the coexisting \mbox{Wi-Fi} devices~\cite{Sadek2015, Guan2016, Jian2017, Cano2015}.
One major proposal for LTE-U was carrier sense with adaptive transmission (CSAT) by Qualcomm in~\cite{Sadek2015}, which implements adaptive duty cycle and channel selection based on estimating the number of active nodes, and their duty cycle and energy. Additional puncturing (i.e. short off-time during the longer on-time) was introduced to protect \mbox{Wi-Fi} delay-sensitive applications. The authors found that LTE could coexist with \mbox{Wi-Fi} at least as well as \mbox{Wi-Fi} coexisting with itself in terms of throughput. 

Some proposals considering mathematical formulations of different \mbox{Wi-Fi}/LTE fairness coexistence goals included optimizing the LTE network throughput with \mbox{Wi-Fi} access time constraints through a cognitive coexistence scheme that determines dynamically the transmission time, channel selection, and channel width for LTE~\cite{Guan2016}; and achieving proportional fairness in terms of \mbox{Wi-Fi}/LTE channel airtime, while maximizing the overall aggregate throughput~\cite{Cano2015, Jian2017}. 
Although (sub-)optimal solutions were obtained, these works did not compare the performance of their proposed schemes with other adaptive schemes, which are less computationally complex, so it is not clear whether adopting such formal mathematical approaches significantly improve the coexistence performance.
Specifically, in~\cite{Guan2016} it was suggested to implement the proposed algorithm in a server with powerful computing capabilities in the LTE network, which needed information about \mbox{Wi-Fi} obtained via crowd sourcing from LTE users. As such, it is not clear how sensitive this solution is to increased delay due to data computation and transfer to/from the server. Additionally, crowd-sourced data may be unreliable or insufficient in practice, if the number of users is low.
Furthermore, \cite{Cano2015, Jian2017} considered a single coordinated LTE BS/network, so it is not clear how the proposed algorithms perform for multiple uncoordinated LTE networks that apply the algorithms independently.      

Different than previous work, centralized coordination between LTE and \mbox{Wi-Fi} was proposed in~\cite{Sagari2015}, through which the adaptive duty cycle mechanism and transmit power were optimized, such that a similar throughput was achieved for the two technologies. Although algorithms that have information about the entire network result in good network performance in general, they are applicable only for a restricted number of deployments in practice, where a single operator manages all deployments implementing either of the technologies.

Machine learning techniques were also proposed for adapting the LTE duty cycle, i.e. Q-learning for achieving the desired LTE capacity~\cite{RupasingheNewOrleans2015}, and multi-armed bandit machine learning to maximize the LTE average capacity and minimize the LTE transmit power~\cite{Sriyananda2016}. Both schemes were shown to result in considerable gains in the aggregate \mbox{Wi-Fi} and LTE throughput over fixed duty cycling. However, it is not clear how long the learning process takes and whether such complex mechanisms perform better than other adaptive schemes. Furthermore, the proposed learning algorithms do not consider the \mbox{Wi-Fi} performance, but only the LTE target capacity~\cite{RupasingheNewOrleans2015} or LTE minimum capacity~\cite{Sriyananda2016}, for which a single example value was evaluated in the respective works. As such, it is not clear what the \mbox{Wi-Fi} performance would be for other LTE target capacities and \mbox{Wi-Fi}/LTE traffic types.

\subsubsection{\textbf{\mbox{Wi-Fi}/LBT-LTE and \mbox{Wi-Fi}/Duty-Cycle-LTE Coexistence}}
Some work in the literature has investigated  coexistence with both LBT- and duty-cycle-LTE~\cite{Cano2016a, Li2016a, Simic2016, Voicu2016, Voicu2017, Voicu2017a, Jeon2014, Ali2016, Zhang2015a}.
We note that this is an important contribution, since it facilitates the comparison of two major distinct time-sharing approaches. Since duty cycling can be adapted based on the number of active nodes, e.g. CSAT, LBT and adaptive duty cycling can implement the same functionality, i.e. facilitate an equal share of the channel for each device. The following tradeoff is expected: LBT has a higher MAC overhead due to its sensing time, but results in a lower number of collisions, whereas adaptive duty cycling has a lower MAC overhead, but also a higher number of collisions. 
It was reported that, in order to achieve \mbox{Wi-Fi}/LTE coexistence fairness (i.e. LTE degraded the \mbox{Wi-Fi} performance at most as much as \mbox{Wi-Fi} coexisting with itself would do) LBT was preferred to fixed 50\% duty cycling~\cite{Ali2016, Jeon2014}. 
However, the following LTE coexistence mechanisms were found to be equivalent for improving the \mbox{Wi-Fi} performance: (i)~a low fixed duty cycle; or (ii)~LBT with more sensitive sensing thresholds or lower priority than \mbox{Wi-Fi} when contending for the channel through the random backoff procedure~\cite{Li2016a}.
Overall, a similar \mbox{Wi-Fi} performance was obtained when coexisting with LBT-LTE or with adaptive duty-cycle-LTE based on CSAT~\cite{Cano2016a, Simic2016, Voicu2016, Voicu2017, Voicu2017a, Zhang2015a}. 
As an insight, for longer LTE transmission time, the LTE throughput for LBT and CSAT was the same, but this increased the \mbox{Wi-Fi} delay~\cite{Cano2016a}.
Also, adaptive duty cycle was more beneficial in low-density networks, whereas LBT was better in high-density networks~\cite{Voicu2016}. 
For the specific case of \mbox{Wi-Fi}/LTE coexistence in the 5~GHz unlicensed band, it was shown that the choice of time-sharing mechanism for LTE (i.e. LBT or adaptive duty cycling) is irrelevant, due to the large number of available channels~\cite{Simic2016}. This was confirmed when ACI was also modelled~\cite{Voicu2017, Voicu2017a}. We note that \cite{Voicu2017, Voicu2017a} adopted a new evaluation framework, i.e. risk-informed interference assessment, which was relevant for both policy and engineering coexistence goals.

\subsection{Summary \& Insights}
\label{sumSpecComm}

A large number of existing works considered inter-technology coexistence in a spectrum commons and we classified them into works addressing coexistence with low traffic technologies and coexistence among broadband technologies.
In general, it is not straightforward to compare the coexistence performance of different proposed spectrum sharing mechanisms in different works, due to the different considered scenarios, evaluation metrics, and coexistence goals. Moreover, most works compare the coexistence performance of their proposed mechanisms only with the performance of standalone technologies (i.e. not in coexisting deployments) or with the case where the coexisting technologies do not implement any additional sharing mechanism compared to their standard specifications. 

For coexistence with low-traffic technologies, a similar number of works considered spectrum sharing techniques at Layer~1 as at Layer~2, where most of them were distributed, as expected for multiple uncoordinated network deployments in unlicensed bands. 
Furthermore, a large number of works presented experimental results, which is important for capturing the coexistence performance in real deployments. We note that conducting experiments was facilitated by the availability of commercial hardware, especially for cases where only standard features of different technologies were evaluated.
For coexistence of low-traffic technologies, e.g. IEEE 802.15.4, with broadband technologies, e.g. \mbox{Wi-Fi}, it was found that the mismatch in transmit power between the two technologies was dominant, such that at short separation distances, the low-traffic technology was significantly affected, even if the broadband technology also implemented sharing in time at Layer~2, e.g. CSMA/CA.  
Regarding Layer~1 techniques, FHSS was found efficient for Bluetooth when coexisting with \mbox{Wi-Fi}, whereas \mbox{Wi-Fi}'s CSMA/CA did not react fast enough to signals with fast hopping rate. We note, however, that the efficiency of FHSS was facilitated by the availability of some channels that were not occupied by \mbox{Wi-Fi}. Is is thus not clear what the performance of FHSS is for very dense deployments and congested channels, as expected in emerging networks.  
More advanced PHY techniques were also proposed, e.g. interference cancellation and reconfigurable antennas, but were evaluated for only one link of each technology, which suggests that further investigation is needed, in order to determine the efficiency of such techniques in real deployments. 

For coexistence among broadband technologies in a spectrum commons, most of the works considered \mbox{Wi-Fi}/LTE coexistence, as \mbox{Wi-Fi} used to be the only widely deployed broadband technology in such bands. LTE was only recently proposed to operate in the unlicensed bands, where it would thus be the second broadband technology. 
Most works on \mbox{Wi-Fi}/LTE coexistence considered Layer~2 spectrum sharing in time and frequency and only few experimental results were reported, due to the limited availability of testbeds where full LTE stacks are implemented and can be modified in a straightforward manner. 
In general, it was reported that inter-technology sharing in frequency via channel selection is more efficient than time-sharing mechanisms like LBT or duty cycling. However, this requires a sufficient number of channels, which may not always be available. As such, channel selection can only be used to enhance the coexistence performance of time sharing mechanisms.

For time sharing via LBT or duty cycling, it was found that adaptive mechanisms are required to achieve \mbox{Wi-Fi}/LTE coexistence, e.g. LBT with adaptive sensing duration or sensing threshold, or adaptive duty cycling. Fixed variants were not able to fulfil the considered fairness criteria in different works, as it is expected that they cannot take into account variations of device numbers, traffic, mobility, etc.  
The distributed mechanisms of LBT with binary exponential random backoff and adaptive duty cycling based on CSAT for LTE were found to be overall equivalent from the point of view of the resulting \mbox{Wi-Fi} performance. This is due to the fact that LBT causes fewer collisions, but has a higher MAC overhead due to the sensing time, whereas adaptive duty cycling causes more collisions, but has a lower MAC overhead. Further implementation differences are as follows: LBT required significant changes to the 3GPP LTE standard, whereas adaptive duty cycling can be implemented based on the existing LTE specifications for subframe blanking. Nonetheless, adaptive duty cycling has the additional disadvantage that it does not comply with some regional spectrum regulations at Layer~0 and requires channel monitoring (potentially via a colocated \mbox{Wi-Fi} interface) to estimate the number of coexisting devices. As such, equipment cost considerations at Layer~8 may also affect the choice of LBT or adaptive duty cycling as the spectrum sharing mechanism at Layer~2.   

Other adaptive variants for LBT and duty cycling were based on mathematical optimization, which has higher computation complexity. Most of these proposed solutions require coordination among LTE deployments and sometimes also with the \mbox{Wi-Fi} deployments. It is not clear how feasible such approaches are in practice and how the efficiency of such methods varies with Layer~7~and~8 parameters such as uncoordinated LTE deployments of different operators, different traffic types, mobility, and delay in obtaining the required information about the \mbox{Wi-Fi} network. Furthermore, the coexistence performance of the proposed optimized schemes was not compared with that of fully distributed adaptive schemes with low computation complexity. Consequently, it is not yet understood whether such highly optimized solutions, even in ideal conditions, significantly improve the network performance over conventional distributed schemes. 

Furthermore, it was found that power control alone cannot ensure coexistence between different broadband technologies, due to the high data rate requirements and dense deployments, but that it can improve coexistence in conjunction with sharing in time or frequency. 
Although PHY techniques were also proposed for coexistence among broadband technologies, e.g. interference cancellation, and beamforming techniques, they were evaluated with one of the technologies implemented for a single link only, so further investigation is needed to determine the efficiency and impact of such techniques in real network-wide deployments.

\section{Discussion \& Future Research Directions}
\label{sec_discussion}

In this section we summarize the insights from our survey on spectrum sharing mechanisms for wireless inter-technology coexistence and we indicate open challenges and possible future research directions.

\subsection{A System-Level View of Inter-Technology Spectrum Sharing}
The design of spectrum sharing mechanisms is influenced by both technical and non-technical aspects, such as regulatory restrictions, business models, and social practices. 
Due to non-technical aspects, implementing the most efficient spectrum sharing mechanisms may not be straightforward.
For instance, changes in spectrum regulations were required for TVWS before secondary technologies could share underutilized spectrum with TV services. 
Another example is the lack of business agreements among network operators, so that information exchange among e.g. different \mbox{Wi-Fi} hotspots operating in the same band may not be possible. 
Consequently, coordinated spectrum sharing mechanisms cannot be implemented; instead, potentially less efficient, distributed sharing schemes must be used.
It is thus critical to \textbf{consider the design of spectrum sharing mechanisms for inter-technology coexistence from a unified, system-level perspective that includes both technical and non-technical aspects}. 
The technology circle considered in this survey represents such a system-level framework, which incorporates Layers~1--7 of the OSI stack, regulatory restrictions at Layer~0, and business models and social practices at Layer~8.

\subsection{Recent Trends in Spectrum Sharing}
In our literature review in Sections~\ref{litHierFr} and~\ref{litSpecComm}, we identified three major recent technical and regulatory trends in terms of how spectrum is shared: \textbf{(i)}~more broadband technologies operating in a spectrum commons, i.e. \mbox{Wi-Fi}/LTE coexistence in the unlicensed bands; \textbf{(ii)}~introducing multiple primary technologies with equal access rights in the same spectrum band, which is managed by a single entity, e.g. LTE/NB-IoT coexistence; and \textbf{(iii)}~increasingly more bands set to be open for technologies with primary/secondary access rights, where \emph{secondary/secondary} inter-technology coexistence is also an issue, e.g. TVWS, the CBRS band, the 2.3-2.4 GHz band with LSA.
All three of these major developments represent the case of coexisting technologies with \emph{equal} spectrum access rights, which was the focus of this survey. 
We note that, as discussed throughout Section~\ref{litHierFr}, the spectrum sharing mechanisms considered in the existing literature for primary/primary and secondary/secondary coexistence resemble either that of traditional, centrally coordinated cellular networks, or that of distributed networks operating in the unlicensed bands, as an instance of a spectrum commons. 

\subsection{Challenges for Spectrum Sharing in a Spectrum Commons}
Designing spectrum sharing mechanisms for inter-technology coexistence in a \emph{spectrum commons} is the most challenging out of the three identified coexistence cases with equal spectrum access rights, due to the high \emph{heterogeneity} of the coexisting devices. 
A spectrum sharing mechanism for a technology in such bands has to take into account the (intra-technology) spectrum sharing mechanisms of already existing technologies, but also to anticipate the behaviour of future technologies. 
This can be addressed through regulatory limitations at Layer~0 for MAC protocols at Layer~2, e.g. ETSI specifying LBT in the 5~GHz unlicensed band in Europe.
As a result, 3GPP standardized LAA with LBT to coexist with \mbox{Wi-Fi} in the 5~GHz band.
By contrast, no such regulatory limitation on Layer~2 exists in the U.S. for the 5~GHz band, so \mbox{LTE-U} adopted an adaptive duty cycle MAC to facilitate coexistence with \mbox{Wi-Fi}. This was selected due to considerations at Layer~8, i.e. meeting the expectations to protect \mbox{Wi-Fi}~\cite{FCC2015}, while making the minimum changes to the LTE technology, in order to accelerate the time-to-market of commercial deployments. 

Importantly, LAA and \mbox{LTE-U} are examples where \emph{inter-technology} spectrum sharing mechanisms also changed the way that \emph{intra-technology} coexistence is achieved; implementing LBT or duty cycling at the MAC layer for LTE enables spectrum sharing with \mbox{Wi-Fi} devices, but also among LTE devices/operators.
We emphasize that inter-technology coexistence among more than two wireless technologies has largely not been investigated in the literature yet. Thus far, this was not of practical interest, as \mbox{Wi-Fi} was the only widely-deployed broadband technology in the unlicensed bands, whereas low-traffic technologies did not pose major coexistence problems among themselves. However, studying coexistence among more than two dominant technologies may become important in the near future, due to the increasing heterogeneity of broadband technologies operating in a spectrum commons, e.g. \mbox{Wi-Fi}, LAA, and \mbox{LTE-U} operating in the 5~GHz unlicensed band.  
Our survey also showed that properly evaluating inter-technology interactions in dense deployments is already complex, even for only two dominant technologies. This opens another valid research question, of whether current methodologies and modelling tools are sufficient to reliably capture the key interactions among multiple dominant technologies in the variety of coexistence cases that may arise. 

\subsection{Layer~2 Spectrum Sharing in a Spectrum Commons}
Our literature review further revealed that, for inter-technology coexistence in a spectrum commons, the preferred spectrum sharing mechanisms are currently the traditional sharing in frequency and time at Layer~2,  
especially for broadband technologies (\emph{cf.} Section~\ref{litrev_hightraffic}). Most of the works acknowledged that achieving coexistence through such mechanisms is possible, e.g. for \mbox{Wi-Fi}/LTE coexistence in the unlicensed bands via LBT, adaptive duty cycling, and channel selection.
It was also found that, whenever a large number of channels is available (e.g. in the 5~GHz unlicensed band), channel selection is an efficient mechanism to manage inter-technology interference, which results in only marginal performance degradation due to coexistence. 
For the complementary case of inter-technology \emph{co-channel} transmissions, LBT with adaptive sensing time and adaptive duty cycling where found to provide a similar level of coexistence fairness and performance. 
We note that evaluations of dynamic and heterogeneous channel widths across different coexisting technologies, due to advanced features like channel bonding, are largely missing from the literature. 
Studying the impact of dynamic, heterogeneous channel-width selection for distributed deployments as expected in a spectrum commons is thus an important future research direction, as this results in more complex network-wide interference relationships among different technologies and devices.     

Furthermore, as evident from our literature review in Sections~\ref{lbtOptim} and~\ref{dutOptim}, resource allocation algorithms derived from optimization of formal mathematical problems are difficult to implement, as they would require information exchange at a level that may not feasible in practice, due to Layer~8 aspects. 
Moreover, we believe that perfect spectrum sharing optimization for inter-technology coexistence is in general not applicable for a spectrum commons, due to the limited information about other coexisting technologies and devices, the high level of heterogeneity, the large number of network managers, and the dynamics of the deployments; these factors are a direct effect of equal spectrum access rights at Layer~0 and distributed network ownership at Layer~8. 

Nonetheless, many works in the literature studied inter-technology coexistence with respect to a formal optimization goal, e.g. proportional fairness in Sections~\ref{litrev_lbt} and~\ref{litrev_dut}, despite the potential challenges of implementing such solutions in practice. 
We emphasize that the validity of these optimum solutions in real (non-idealized) deployments is still an open question, given the variability of system parameters like traffic demand, hardware performance, and network size.
Furthermore, it is not clear whether the performance of these solutions is better than for conventional distributed adaptive mechanisms with lower computation complexity, like LBT with binary exponential random backoff or adaptive duty cycle based on CSAT. 
An important future research direction is thus performing a thorough sensitivity analysis, to determine the spectrum sharing mechanisms in the design space that are near-optimal yet robust for practical engineering deployments.

\subsection{Layer~1 Spectrum Sharing in a Spectrum Commons}
In the reviewed literature, Layer~1 spectrum sharing mechanisms were found to be efficient for low-traffic coexisting technologies, e.g. FHSS for Bluetooth in Section~\ref{rev_bluetooth}. 
However, Layer~1 techniques such as interference cancellation and beamsteering were found to be less feasible in practice for achieving inter-technology coexistence among broadband technologies in Section~\ref{wifiltecoex}.
Such techniques are based on acquiring information through multiple wireless interfaces that decode signals of other coexisting technologies. Additionally, interference cancellation also requires changes in the MAC layer. 
Further research is needed to determine the practical feasibility of achieving inter-technology coexistence via Layer~1 spectrum sharing mechanisms like interference cancellation and beamforming in large-scale heterogeneous deployments. 

\subsection{Performance Evaluation of Spectrum Sharing Mechanisms}
We found that comparing the coexistence performance of different candidate spectrum sharing mechanisms is not straightforward, especially given the large amount of research work in the literature, with different assumptions and methods, often referring to different coexistence goals (\emph{cf.} Tables~\ref{table_review_3b}, \ref{table_review_4b}, and \ref{table_review_4c}). 
As evident throughout our literature review, most of the works on inter-technology coexistence, and especially those addressing coexistence in a spectrum commons, focus on evaluating only a single or variants of a given main spectrum sharing mechanism (e.g. variants of LBT-LTE, or duty-cycle-LTE).
Moreover, this considered candidate spectrum sharing mechanism is often compared only to the baseline cases where a single technology uses the spectrum, or where newly coexisting technologies do not implement any additional sharing mechanism, e.g. LTE continuously transmitting in the 5 GHz unlicensed band as it traditionally does in dedicated licensed spectrum. Consequently, it is seldom possible to directly compare the coexistence performance reported in different works for different spectrum sharing mechanisms. 
In order to address this issue, coexistence goals should be more clearly and explicitly defined in the first place. 
Also, it is important to study candidate spectrum sharing mechanisms for different coexisting technologies within the same framework. 

Furthermore, only few experimental results were reported for inter-technology coexistence in a spectrum commons, due to the lack of testbeds with fully operational protocol stacks where different coexistence mechanisms can be implemented in a straightforward manner. 
We emphasize that empirical studies are crucial for evaluating the performance of inter-technology coexistence in real deployments and revealing potential implementation issues. Consequently, an important future research direction is developing flexible and accessible software and hardware platforms that can be configured with a moderate amount of effort to implement standard and proposed protocol stacks.

\subsection{Other Open Challenges} 
 
The recent introduction of different LTE variants as broadband technologies in a spectrum commons suggests that, for capacity increase, operating in unlicensed bands is straightforward to adopt from a technical perspective. These LTE variants aggregate unlicensed carriers, i.e. LAA, and LTE-U, or operate exclusively in the unlicensed bands, i.e. MulteFire~\cite{MulteFireAlliance2017}.
This opens an interesting spectrum regulatory research question: whether it may be attractive to open more shared bands for traffic offloading and reserve licensed spectrum only for important signalling traffic and QoS-guaranteed services.

Finally, we note that most of the reviewed spectrum sharing mechanisms for inter-technology coexistence in a spectrum commons are fully distributed, and only a few centralized, as summarized in Section~\ref{litSpecComm} and Tables~\ref{table_review_3} and \ref{table_review_4}. 
Considering more fundamental performance limits of inter-technology spectrum sharing mechanisms is still missing from the literature. Specifically, investigating the impact of different levels of coordination among networks of different technologies is a rich yet largely unexplored research direction.

\section{Conclusions}
\label{sec_conclusions}
In this survey we explored the design space of spectrum sharing mechanisms for wireless inter-technology coexistence from a unified, system-level perspective, i.e. the technology circle, that integrates technical and non-technical aspects at different layers. 
We reviewed the literature on inter-technology coexistence with respect to different layers of the technology circle, where we considered technologies with equal spectrum access rights: (i)~primary/primary; (ii)~secondary/secondary; and (iii)~technologies operating in a spectrum commons. 
Throughout the literature review we identified the following three major trends for inter-technology coexistence: (i)~more broadband technologies operating in a spectrum commons; (ii)~introducing multiple primary technologies with equal access rights in the same spectrum band; and (iii)~increasingly more bands set to be open for technologies with primary/secondary access rights, where secondary/secondary inter-technology coexistence may also become an issue. 

Spectrum sharing mechanisms for primary/primary and secondary/secondary coexistence in the literature were similar to those in centralized coordinated cellular networks, or to those in a spectrum commons.
Out of the three identified cases of inter-technology coexistence with equal spectrum access rights, coexistence in a spectrum commons is the most challenging, due to the high heterogeneity of coexisting devices and technologies. 
For such cases, Layer~2 mechanisms like distributed spectrum sharing in time and frequency (e.g. LBT, adaptive duty cycling, and channel selection) are currently considered efficient for ensuring coexistence, whereas the coexistence performance of advanced PHY layer techniques (e.g. interference cancellation, beamforming) in large, dense deployments has been largely unaddressed. 
Furthermore, our survey revealed that the performance of proposed spectrum sharing mechanisms in different works is difficult to compare directly, due to the different assumptions, baselines, scenarios, coexistence goals, and evaluation methods, where only few works assess multiple spectrum sharing approaches within the same framework. 

The key open challenges that we identified for inter-technology coexistence with equal spectrum access rights are: investigating the coexistence performance of interference cancellation and beamforming in network-wide deployments; evaluating the performance of more than two coexisting broadband technologies; considering heterogeneous channel widths throughout coexisting deployments; performing sensitivity analyses for highly optimized solutions with respect to parameters at different layers of the technology circle in real deployments; developing accessible software and hardware testing platforms; and considering the impact of different levels of coordination among coexisting technologies.

\ifCLASSOPTIONcaptionsoff
  \newpage
\fi



\bibliographystyle{IEEEtran}
\bibliography{IEEEabrv,bibliography}
\end{document}